\documentclass[preprint,5p,10pt,times]{elsarticle}

\usepackage{amssymb}
\usepackage{amsthm}
\usepackage{multirow}
\usepackage{booktabs}
\usepackage{ifthen}
\usepackage[colorlinks=true]{hyperref}
\usepackage{orcidlink}
\usepackage{flushend}
\usepackage{amsmath}
\usepackage{svg}
\usepackage[caption=false, font=footnotesize]{subfig}
\usepackage[linesnumbered,ruled,vlined]{algorithm2e}

\makeatletter
\def\ps@pprintTitle{%
    \let\@oddhead\@empty
    \let\@evenhead\@empty
    \def\@oddfoot{\footnotesize\itshape
        \parbox{\dimexpr\textwidth}{
            \strut\\Accepted Manuscript at Mechanical Systems and Signal Processing (2025).\\\strut\\
            \begin{minipage}{0.89\textwidth}
                \textup{This document contains both the Accepted Manuscript and its Supplementary Material; the latter corresponds to the appendices of this document.\linebreak We refer the reader to the published journal article at \href{https://doi.org/10.1016/j.ymssp.2025.112403}{\texttt{doi:10.1016/j.ymssp.2025.112403}}. © 2025. This manuscript version is made available under the \textup{CC-BY-NC-ND 4.0} license (\href{https://creativecommons.org/licenses/by-nc-nd/4.0/}{\texttt{https://creativecommons.org/licenses/by-nc-nd/4.0/}}) as per Elsevier's article-sharing policy.}
            \end{minipage} \hfill
            \begin{minipage}{0.1\textwidth}
                \includegraphics[width=\textwidth]{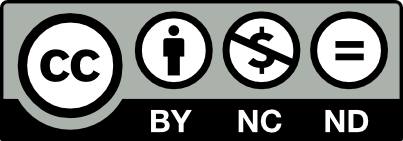}
            \end{minipage}
            } \hfill}}%
    \let\@evenfoot\@oddfoot
\makeatother

\newtheorem{theorem}{Theorem}
\newtheorem{lemma}{Lemma}

\newdefinition{definition}{Definition}
\newdefinition{remark}{Remark}

\makeatletter
\newcommand{\manuallabel}[2]{\def\@currentlabel{#2}\label{#1}}
\makeatother

\newproof{prfthone}{Proof of Theorem~\ref{the:mismodelling-independence-link}}
\newproof{prfthtwo}{Proof of Theorem~\ref{the:strong-consistency}}
\newproof{prfththree}{Proof of Theorem~\ref{the:finite-nominal-time}}
\newproof{prfthfour}{Proof of Theorem~\ref{the:hypothesis-test-metrics}}
\newproof{prfthfive}{Proof of Theorem~\ref{the:gen-mdd-dependency}}
\newproof{prfthsix}{Proof of Theorem~\ref{the:gen-mdd-dependency}}
\newproof{prflemma}{Proof of Lemma~\ref{lem:degenerate-rv}}

\begin{document}

\sloppy
\flushbottom

\begin{frontmatter}

\title{Fault Detection and Monitoring using a Data-Driven Information-Based Strategy:\texorpdfstring{\linebreak}{} Method, Theory, and Application}

\author[ids]{Camilo~Ramírez~\orcidlink{0000-0002-2774-0012}}
\ead{camilo.ramirez@ug.uchile.cl}

\author[ids]{Jorge~F.~Silva~\orcidlink{0000-0002-0256-282X}}
\ead{josilva@ing.uchile.cl}

\author[promes]{Ferhat~Tamssaouet~\orcidlink{0000-0003-2664-3138}}
\ead{ferhat.tamssaouet@univ-perp.fr}

\author[ids]{Tomás~Rojas~\orcidlink{0000-0003-4110-8731}}
\ead{tomas.rojas.c@ug.uchile.cl}

\author[ids]{Marcos~E.~Orchard~\orcidlink{0000-0003-4778-2719}}
\ead{morchard@ing.uchile.cl}

\affiliation[ids]{organization={Information and Decision Systems Group, Department of Electrical Engineering, University of Chile},
            city={Santiago},
            country={Chile}}

\affiliation[promes]{organization={PROMES-CNRS, UPVD},
            city={Perpignan},
            country={France}}

\begin{abstract}
The ability to detect when a system undergoes an incipient fault is of paramount importance in preventing a critical failure. Classic methods for fault detection –~including model-based and data-driven approaches~– rely on thresholding error statistics or simple input-residual dependencies but face difficulties with non-linear or non-Gaussian systems. Behavioral methods –~e.g., those relying on digital twins~– address these difficulties but still face challenges when faulty data is scarce, decision guarantees are required, or working with already-deployed models is required. In this work, we propose an information-driven fault detection method based on a novel concept drift detector, addressing these challenges. The method is tailored to identifying drifts in input-output relationships of additive noise models –~i.e., model drifts~– and is based on a distribution-free mutual information (MI) estimator. Our scheme does not require prior faulty examples and can be applied distribution-free over a large class of system models. Our core contributions are twofold. First, we demonstrate the connection between fault detection, model drift detection, and testing independence between two random variables. Second, we prove several theoretical properties of the proposed MI-based fault detection scheme: (i) strong consistency, (ii) exponentially fast detection of the non-faulty case, and (iii) control of both significance levels and power of the test. To conclude, we validate our theory with synthetic data and the benchmark dataset N-CMAPSS of aircraft turbofan engines. These empirical results support the usefulness of our methodology in many practical and realistic settings, and the theoretical results show performance guarantees that other methods cannot offer.
\end{abstract}


\end{frontmatter}

\section{Introduction}

The objective of Prognostics and Health Management (PHM) is to provide methodologies and tools for creating tailored maintenance plans based on the specific characteristics, operations, and degradation scenarios of a given asset \citep{kordestani2019failure}. This approach aims to achieve optimal system availability while minimizing costs, representing a comprehensive and efficient strategy for system health management. PHM integrates Fault Detection and Isolation (FDI), health management, and prognostic capabilities, including Remaining Useful Life (RUL) prediction \citep{hu2022prognostics}.

Fault Detection and Isolation (FDI) is primarily concerned with identifying when a fault occurs, understanding its characteristics, and pinpointing its location within a system. System degradations, inevitable over time, lead to incipient or critical failures. Incipient failures can be defined as alterations that do not prohibit the system from operating, whereas critical failures occur when degradation levels result in the system not operating as required \cite{song2017integration}. As a consequence, the detection of incipient failures is subject to exhaustive study, especially in complex and multivariate systems \cite{abid2021review}, and is of paramount importance because early detection of incipient failures can enable maintenance plans to be created that prevent the occurrence of a critical failure \cite{atamuradov2017prognostics}.

The literature offers a diverse range of FDI methods \citep{abid2021review}, which can be classified into two approaches: model-based and data-driven. Initially, FDI efforts focused on model-based approaches such as parity-based methods \citep{abbasi2020parity}, parameter identification-based methods \citep{zhou2021identification}, and observer-based methods \citep{perez2023fault}. These methods are based on the idea of comparing system outputs with failure models and comparing this difference with a threshold; then, a fault is detected when the value of the difference exceeds the threshold \citep{tabatabaeipour2015active}. However, in the presence of significant noise or an unknown noise distribution, these methods may have poor performance and make the task of determining the most adequate fault decision threshold difficult \citep{tabatabaeipour2015active, jauberthie2013fault}. Overall, model-based methods are effective for simple systems whose phenomenology can be understood and represented by explicit mathematical models; although providing accurate and interpretable results, they are challenging to implement for complex systems \citep{tamssaouet2023system}.

Data-driven approaches, which contrast with model-based approaches, are also prevalent in the current literature. These phenomenological-agnostic methods involve performing mathematical or statistical operations on measurements or training neural networks using measurements to extract information from the system and predict faults. The information is obtained through measured signals and their conversion into fault-characterizing features \citep{abid2021review, nguyen2022feature}. Machine fault detection and diagnosis, particularly in rotating machinery, employ various methods for data collection, such as vibration monitoring and thermal imaging \citep{abid2021review}. The collected data undergoes processing using methods like spectral analysis, wavelet analysis, short-term Fourier transform, and others. The processed data can then be directly used for fault detection by setting fixed or adaptive thresholds \citep{wang2023fingerprint}. As the system state is generally not directly measurable from data, two approaches are used for its estimation: statistical and non-statistical. Statistical methods, such as Kalman filter \citep{long2023fault}, principal component analysis \citep{elsamanty2023principal}, among others, excel in rapid fault detection for linear systems but may not be optimal for diagnosis and for detection in non-linear or non-Gaussian systems. The non-statistical data-driven methods involve using mathematical classification models from supervised learning methods. For instance, Support Vector Machines are sensitive to initial parameters, necessitating a parameter-tuning process for each signal dataset. Artificial Neural Networks (ANNs) are widely utilized for their self-learning capabilities and automatic feature extraction; however, they may tend to over-fit the training set \citep{mohd2020neural}. Recent advancements in ANNs and the adoption of deep learning algorithms have led to the development of novel classification models for fault detection and diagnosis \citep{iqbal2019fault}. Various architectures, such as Convolutional Neural Networks, have performed successfully in various industrial applications \citep{zhu2023review}. These deep learning models offer the ability to learn complex structures from datasets, although they require larger samples and longer processing times to achieve higher accuracy.

Compared to model-based methods, data-driven approaches are easier to implement but face challenges like limited interpretability, poor uncertainty management, and difficulty obtaining comprehensive data from faulty scenarios. Indeed, run-to-failure experiments are often infeasible for complex systems due to safety and cost concerns. To address this, behavioral models, including AI-based and data-driven models, are increasingly used for monitoring and control, with digital twins emerging as a notable solution \citep{tamssaouet2023system}.

Behavioral models capture the statistical dynamics of systems, which change under faults, manifesting as an alteration --~namely, a drift~-- in statistical relationships among system variables \cite{bi2017detection}. As a consequence, a link between detecting faults and statistical drifts arises \cite{gama2004learning, ardakani2017toward}. Concept drift refers to changes in the distribution of a stochastic phenomenon over time \citep{gama2014survey, lu2018learning} and is studied under various terms, such as data shift \cite{quinonero2022dataset} and anomaly detection \cite{yamanaka2019autoencoding}. Drift detection methods can be supervised, relying on anomalous data \citep{yamanaka2019autoencoding}, or unsupervised, using error rates and $p$-values as measures of abnormality \citep{lu2018learning}. Notable error-based methods include DMM \citep{gama2004learning}, ADWIN \citep{bifet2007learning}, and STEPD \citep{nishida2007detecting}, though they assume discrete targets, limiting their use with continuous variables. Regression-suited methods, such as those in \cite{lima2022learning}, exist but are used in adaptive scenarios and are constrained by algorithm-specific limitations \citep[Table~6]{lima2022learning}, making them unsuitable for already deployed models or in problems with modeling more complex than their base algorithm.

In the current landscape, behavioral models offer real advantages for FDI. This approach is particularly attractive as it does not assume a history, making it applicable to new or critical systems. However, there are several open challenges, which include (i) the nonexistence or scarcity of faulty data, which makes the supervised methods inoperable, (ii) the lack of decision guarantees for the majority of existing methods, particularly in complex multi-variate non-linear and non-Gaussian systems, and (iii) the inability to work with already-deployed behavioral models not necessarily dedicated to FDI. In addition, one can observe a link in the literature between concept drift detection and FDI, but we do not observe a formal connection between results found in concept drift literature and its applications to FDI.

\subsection{Contributions of this Work}
We propose a novel information- and data-driven fault detection strategy leveraged by a concept drift theory tailored to capture input-output relationship (model) drifts within the system. Remarkably, our method does not require the availability of previous faulty data and is agnostic to learning algorithms, expert modeling, and data distributions, which makes it applicable to a large family of scenarios. Our key contributions are as follows:
\begin{itemize}
    \item We establish a formal framework that links the fault detection (FD) task with a kind of concept drift (model drift) detection that is especially suited to address alterations in input-output relationships. We develop a theoretical proof demonstrating the equivalence between testing model drift and testing independence between a regression input and its residual for the rich class of additive noise models.
    \item We introduce the use of a non-parametric mutual information estimator \cite{cover_2006, silva2012complexity} to perform fault detection. We prove that the resulting scheme is strongly consistent (asymptotic expressiveness for model drift detection), has finite-length performance guarantees, and has a specific vanishing error convergence rate.
    \item We complement our theoretical and methodological contributions with numerical analyses on controlled (synthetic) scenarios and in the realistic benchmark dataset of turbofan engines N-CMAPSS \cite{arias2021aircraft}. Our empirical findings support the practical capabilities of our method in complex fault detection settings.
\end{itemize}

\subsection{Related Works}
Some related works use mutual information (MI) in their fault detection pipeline, but most use MI to select features (e.g., \cite{hassani2021unsupervised, zhong2020quality, hu2023mutual}) or to circumvent the high-dimensionality of the process (e.g., \cite{rashid2012new, jiang2018information}). Exceptions are the works of Lv et al.\! \cite{lv2021interpretable} and Kumar et al.\! \cite{kumar2021optimization}. In \cite{lv2021interpretable}, Lv et al.\! proposed to monitor the statistics of a component-pairwise MI estimation matrix of all variables involved in the system to detect faults and identify the variables associated with the fault by checking deviations, in testing time, from the values of the statistics observed in a healthy scenario. In \cite{kumar2021optimization}, Kumar et al.\! used a kernel-based MI estimation to build a fitness function of a data-driven genetic algorithm that optimizes the hyperparameters of their fault detection method.

From the basic fact that MI is used, all the previously mentioned works (\cite{hassani2021unsupervised, zhong2020quality, hu2023mutual, rashid2012new, jiang2018information, lv2021interpretable, kumar2021optimization}) might appear to be related to the method presented in this paper; However, there are essential differences worth mentioning. From a formulation perspective, our work is constructed from a novel theory --~developed in this paper~-- that connects model drift detection with independence testing. Notably, this formal path justifies the adoption of MI to detect input-output deviations within a system and, consequently, the usage of MI features as health indicators, which strongly quantify the fault severity; this is a guarantee that is not present in any of the mentioned methods. Moreover, our methodology offers unique performance guarantees that are not offered by other methods.

\subsection{Paper Outline}
The outline of the remainder of our work is as follows. In Sec.~\ref{sec:preliminaries}, we introduce the model drift detection task, state its link with fault detection, and introduce definitions useful for the rest of the work. In Sec.~\ref{sec:md_formalization}, we formalize model drift and show the equivalence between testing model drift and testing independence (Theorem~\ref{the:mismodelling-independence-link}). In Sec.~\ref{sec:md_method}, our new methodology to perform model drift detection and, consequently, fault detection is introduced. In Sec.~\ref{sec:performance}, we show a regime of parameters in which our methodology has the following desirable properties: strong consistency (Theorem~\ref{the:strong-consistency}), exponentially-fast decision convergence on healthy systems (Theorem~\ref{the:finite-nominal-time}), and error convergence guarantees (Theorem~\ref{the:hypothesis-test-metrics}). In Sec.~\ref{sec:method-results-discussion}, we discuss our methodology and its properties. In Sec.~\ref{sec:experimental}, we apply our method to a diversity of synthetic-defined systems, visualizing and validating our theoretical development, and in Sec.~\ref{sec:n-cmapss}, we apply it to the N-CMAPSS dataset, a benchmark dataset of realistic simulations of turbofans; with this, we demonstrate our method's capability to detect faults in a real-life application. Finally, in Sec.~\ref{sec:final-summary}, we summarize our contributions and propose directions for future research.

\section{Preliminaries}
\label{sec:preliminaries}
In this section, we describe our main decision problem related to fault detection, state general definitions, and introduce the concept of mutual information.

\subsection{Fault Detection as a Model Drift Detection Task}
\label{sec:faults-as-drift}
In this paper, we deal with system monitoring from a behavioral data-driven approach. This means that data obtained from a system is treated as realizations of the random variables that compose the system. In this setting, each variable of the system corresponds to a random variable, and the system as a whole is described by the joint distribution of the random vector built upon all the system's variables.

Concept drift corresponds to an arbitrary change in the statistical properties of a certain phenomenon over time \citep{lu2018learning}. Moreover, when considering an explanatory-response relationship between two random vectors $X$ and $Y$, their joint distribution ($P_{X,Y}$) can be decomposed into the product of the explanatory-marginal distribution ($P_X$) and the predictive distribution ($P_{Y|X}$), i.e., $P_{X,Y} = P_X\cdot P_{Y|X}$. Then, we can distinguish between \textit{virtual drifts} (changes in $P_X$) and \textit{actual drifts} (changes in $P_{Y|X}$) \citep{bayram2022concept}. 

A fault event within a system implies an alteration in its inner dynamics, which leads to a disturbance in the predictive distribution of the response given the explanatory variables ($P_{Y|X}$); i.e., a fault is an event that leads to an actual drift. In contrast, a model drift can have different origins, which include changes in operational parameters, replacement of sub-components of a system, and faults. In this context, we can define any unwanted alteration of the inner dynamics of a system as a fault. Therefore, in a setting where no desired alterations are being made to a system, we can establish an equivalence between the existence of a fault and the existence of an actual drift.

In Sec.~\ref{sec:md_formalization}, we show how this drift can be decoupled into drifts on the underlying deterministic explanatory-response relationship (\textit{model drift}) and drifts on the noise model (\textit{noise drift}), and consequently, how the fault-drift equivalence reduces to the equivalence between fault existence and model drift existence. Taking this equivalence into consideration, we propose an information- and data-driven model drift detection method. 

\subsection{Model Drift Detection}
\label{sec:md-test-def}
Let us consider two random vectors (r.v.s) $X$ and $Y$, whose distributions might change over time, taking values in $\mathcal{X}$ and $\mathcal{Y}$, respectively. Formally, we say that $(X,Y)$ has a \textit{nominal} distribution $P_{X,Y}\in\mathcal{P}(\mathcal{X}\times\mathcal{Y})$ and an \textit{actual} distribution $P_{X,Y}'\in\mathcal{P}(\mathcal{X}\times\mathcal{Y})$.\footnote{The nominal distribution is the one describing the phenomenon in its nominal, healthy, or desired behavior; the actual distribution describes the phenomenon in its current and potentially unknown behavior.} As we are focusing on model drift (MD) detection, we need to distinguish between two statistical hypotheses: The null hypothesis ($\mathcal{H}_0:$ the absence of MD), under which the actual distribution has the same underlying model as the nominal distribution, and the alternative hypothesis ($\mathcal{H}_1$: the presence of MD), under which the actual distribution has a different underlying model than the nominal distribution.

To test MD, we require i.i.d. samples from the actual distribution (the evidence) of $(X,Y)$ and a data-driven decision rule. We denote by $\mathbf{Z}_n=(Z_j)_{j=1}^n$ the $n$ i.i.d. realizations from $(X,Y)\sim P_{X,Y}'$. A \textit{decision rule} of length $n$, corresponds to a function $\phi_n:(\mathcal{X}\times\mathcal{Y})^n\rightarrow\{0,1\}$ from the $n$-size sample space ($(\mathcal{X}\times\mathcal{Y})^n$) to the decision space ($\{0,1\}$), where 0 means accepting $\mathcal{H}_0$ and $1$ means rejecting $\mathcal{H}_0$. The collection of the $n$-size decision rules is denoted by $\Pi_n$. Finally, $\Phi=(\phi_n(\cdot))_{n\in\mathbb{N}}$ is said to be a \textit{decision scheme} if $\phi_n(\cdot) \in\Pi_n,\forall n\in\mathbb{N}$.

\subsection{Performance Metrics for MD and Fault Detection}
We introduce three standard concepts --~strong consistency, detection time, and power and significance level~-- that are widely used by the decision community to measure the quality of decision rules and schemes in the context of hypothesis testing. These are of special relevance since, in our problem formalization, the null hypothesis (i.e., $\mathcal{H}_0$) corresponds with the system being in a healthy condition, and the alternative hypothesis (i.e., $\mathcal{H}_1$) corresponds with the system being subjected to some fault.

\subsubsection{Strong Consistency}
An important concept in hypothesis testing is \textit{strong consistency} \citep{gretton2010consistent}. This requirement means that eventually, in the sample size, a scheme converges to the correct decision (almost-surely).
\begin{definition}[{Adapted from \cite{gretton2010consistent} and \cite[Definition~1]{gonzalez2021indtest}}]
\label{def:strong-consistency}
\hfill A decision scheme $\Phi=(\phi_n(\cdot))_{n\in\mathbb{N}}$ is said to be \textit{strongly consistent} if for any nominal and actual distributions $P_{X,Y}\in\mathcal{P}(\mathcal{X}\times\mathcal{Y})$ and $P_{X,Y}'\in\mathcal{P}(\mathcal{X}\times\mathcal{Y})$, the following almost-surely convergence holds true:
\begin{equation}
\label{eqn:strong-consistency}
\mathbb{P}\left(\left.\lim_{n\rightarrow\infty}\phi_n(\mathbf{Z}_n)=i\,\right|\,\mathcal{H}_i\right)=1,\forall i\in\{0,1\},
\end{equation}
where $\mathbb{P}$ represents the process distribution of $(Z_n)_{n\in\mathbb{N}}$, being $Z_n\sim P_{X,Y}',\forall n\in\mathbb{N}$. 
\end{definition}

\subsubsection{Finite-Sample Analysis}
A refined (non-asymptotic) metric to evaluate a consistent scheme is the number of samples required to converge to the right decision \citep{lehmann1986testing}. For this purpose, given a sequence of binary values $\mathbf{s}=(s_n)_{n=1}^{\infty}\in\{0,1\}^{\infty}$ (decisions) and $i\in\{0,1\}$ (the right decision), the collapsing time of $\mathbf{s}$ is expressed by ${\textrm{col}(\mathbf{s},i) \equiv \sup\{n\in\mathbb{N}:s_n=1-i\}}$.\footnote{For any $i\in\{0,1\}$, if $\mathbf{s}=(i)_{n\in\mathbb{N}}$, then $\textup{col}(\mathbf{s},i)=0$.} If $\textrm{col}(\mathbf{s},i)<\infty$, we say that $\mathbf{s}$ \textit{collapses} to $i$ after $\textup{col}(\mathbf{s},i)$ observations.

Given a sampling sequence $\mathbf{z}=(z_j)_{j=1}^{\infty}\in(\mathcal{X}\times\mathcal{Y})^{\infty}$, the sequence of decisions obtained by applying a scheme $\Phi$ in $\mathbf{z}$ is denoted as $\mathbf{s}_{\Phi,\mathbf{z}} \equiv (\phi_n((z_j)_{j=1}^n))_{n=1}^{\infty}$. Then, for $i\in\{0,1\}$, we are interested in the following set: ${\mathcal{S}_i^{\Phi} \equiv  \left\{\mathbf{z}\in(\mathcal{X}\times\mathcal{Y})^{\infty}:\text{col}(\mathbf{s}_{\Phi,\mathbf{z}},i) < \infty \right\}}$.

\begin{definition}[{Adapted from \cite{lehmann1986testing} and \cite[Definition~3]{gonzalez2021indtest}}]
\label{def:detection-time}
\hfill Let $\Phi=(\phi_n(\cdot))_{n\in\mathbb{N}}$ be a decision scheme and   $i\in\{0,1\}$. For any $\mathbf{z}=(z_j)_{j=1}^{\infty}\in \mathcal{S}_i^{\Phi}$, the $i$-\textit{detection time} of $\Phi$ is
\begin{equation}
\mathcal{T}_i^{\Phi}(\mathbf{z})=\textup{col}(\mathbf{s}_{\Phi,\mathbf{z}},i)<\infty,
\end{equation}
otherwise, i.e., if $\mathbf{z}\notin \mathcal{S}_i^{\Phi}$, we have that $\mathcal{T}_i^{\Phi}(\mathbf{z})=\infty$.
\end{definition}

\subsubsection{Power and Significance Level}
Finally, we introduce the significance level of the test and the power of the test \citep{lehmann1986testing}. These are the probability of incurring a type I error (erroneously reject $\mathcal{H}_0$) and the probability of not incurring a type II error (successfully reject $\mathcal{H}_0$ when $\mathcal{H}_1$ holds true), respectively.

\begin{definition}[{Adapted from \cite[Equations (3.1) and (3.2)]{lehmann1986testing}}]
\label{def:hypothesis-test-metrics}
Let $\phi_n(\cdot)$ be a decision rule of length $n$. 
The \textit{significance level} $(\alpha_{\phi_n})$ and \textit{power} $(1-\beta_{\phi_n})$ of $\phi_n(\cdot)$ are such that
\begin{align}
\alpha_{\phi_n} &= \mathbb{P}\left(\phi_n(\mathbf{Z}_n)=1\,|\,\mathcal{H}_0\right),\\
\beta_{\phi_n} &= \mathbb{P}\left(\phi_n(\mathbf{Z}_n)=0\,|\,\mathcal{H}_1\right).
\end{align}
\end{definition}

\subsection{Mutual Information}
Our work proposes an information-driven method for fault detection; in particular, we base our method on the celebrated \textit{Mutual Information} (MI), which was proposed by Claude Shannon \cite{shannon1948mathematical} to quantify a higher order dependency between two r.v.s. \citep{czyz2024beyond}. We focus on the MI between two continuous r.v.s equipped with a joint density.\footnote{A probability measure on $(\mathbb{R}^r,\mathcal{B}(\mathbb{R}^r))$ has a density if it is absolutely continuous with respect to the Lebesgue measure in $(\mathbb{R}^r,\mathcal{B}(\mathbb{R}^r))$. For $\mathbb{R}^r$ with arbitrary $r\in\mathbb{N}$, $\mathcal{B}(\mathbb{R}^r)$ denotes the Borel $\sigma$-field of $\mathbb{R}^r$.} Let $X$ and $Y$ be two continuous r.v.s taking values in $\mathcal{X}$ and $\mathcal{Y}$, with a joint density $f_{X,Y}:\mathcal{X}\times\mathcal{Y}\rightarrow\mathbb{R}$. The MI between $X$ and $Y$ is
\begin{equation}
\label{eqn:mi}
    I(X;Y) \equiv \int_{\mathcal{X}}\int_{\mathcal{Y}}f_{X,Y}(x,y)\log\left(\frac{f_{X,Y}(x,y)}{f_X(x)\cdot f_Y(y)}\right)\mathrm{d}y\hspace{0.5mm}\mathrm{d}x,
\end{equation}
where $f_X(\cdot)$ and $f_Y(\cdot)$ denote the marginal densities of $X$ and $Y$, respectively, induced by $f_{X,Y}(\cdot)$.

\section{Problem Formalization}
\label{sec:md_formalization}
We address fault detection via the detection of deviations of input-output relationships -- i.e., model drifts (MDs) -- as presented in Sec.~\ref{sec:faults-as-drift}. In this section, we formalize MD using the concept of noise outsourcing \citep{austin2015exchangeable}. By that means, we define a framework of fault and MD detection suitable for monitoring a system in a regression context. The steps of this formalization are shown in Figure~\ref{fig:formal-diagram}.

\subsection{General Model}
\begin{figure*}[ht!]
    \centering
    \subfloat[\label{sfig:abstract-plant}Abstract system.]{\hspace{2mm}\includegraphics[width=3.3cm]{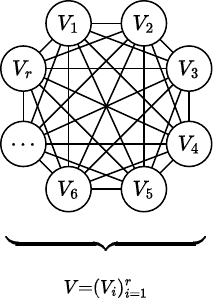}\hspace{2mm}}\hspace{0.8cm}
    \subfloat[\label{sfig:abstract-in-out}Input-output selection.]{\hspace{2mm}\includegraphics[width=3.3cm]{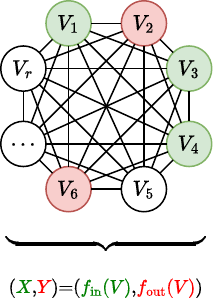}\hspace{2mm}}\hspace{0.8cm}
    \subfloat[\label{sfig:general-model}Functional generative model.]{\hspace{2mm}\includegraphics[width=3.3cm]{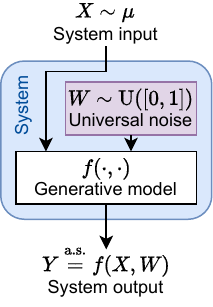}\hspace{2mm}}\hspace{0.8cm}
    \subfloat[\label{sfig:additive-model}The additive noise model (ANM).]{\hspace{2mm}\includegraphics[width=3.3cm]{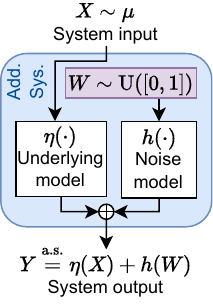}\hspace{2mm}}
    \caption{Diagrams of the abstraction stages of our formal framework.}
    \label{fig:formal-diagram}
\end{figure*}

Let $V$ be a continuous r.v. taking values in $\mathbb{R}^r$ (with $r\in\mathbb{N}$). The system (or phenomenon of interest) is represented by $V$ and is illustrated in Figure~\ref{sfig:abstract-plant}. In the context of input-output system modeling, it is common to identify an explanatory (input) subset of $V$ and a response (output) subset of $V$ \citep{draper1998applied}. Formally, the input variables are denoted by $X\equiv f_{\mathrm{in}}(V)$, and the output variables are denoted by $Y\equiv f_{\mathrm{out}}(V)$, taking values in $\mathcal{X}=\mathbb{R}^p$ and $\mathcal{Y}=\mathbb{R}^q$, respectively.\footnote{$p\in\mathbb{N}$, $q\in\mathbb{N}$, and $r\in\mathbb{N}$ are arbitrary natural numbers.}  Then, we move from the description of the distribution of $V$ to the description of the input-output joint distribution of $(X,Y)$ that we consider as our observable system. A special case of this setting is illustrated in Figure~\ref{sfig:abstract-in-out}, where a system $V$ is a collection of variables, and we partition a subset of this set into input ($X$) and output ($Y$) variables. 

\subsection{Functional Generative Model via Noise Outsourcing}
The following result offers a functional description of the distribution of $(X,Y)$.

\begin{lemma}[see {\cite[Lemma~3.1]{austin2015exchangeable}}\footnote{The result shown in \cite[Lemma~3.1]{austin2015exchangeable} can be traced back as a straightforward application of \cite[Proposition~5.13]{kallenberg1997foundations} by noting that $X$ and $Y$ are trivially conditionally independent given $X$.}]
\label{lem:noise-outsourcing}
Let $X$ and $Y$ be r.v.s taking values in $\mathbb{R}^p$ and $\mathbb{R}^q$, respectively. There exists a random variable $W\sim\mathrm{Uniform}([0,1])$, independent of $X$, and a measurable function $f:\mathbb{R}^p\times[0,1]\rightarrow\mathbb{R}^q$, such that $(X,Y)\overset{\textup{a.s.}}{=}(X,f(X,W))$.\footnote{``a.s.'' stands for ``almost-surely.'' $A\overset{\text{a.s.}}{=}B$ means that $\mathbb{P}(A=B)=1$.}
\end{lemma}
 
Importantly, this lemma states that any system $(X,Y)$ admits the following functional predictive structure: $Y|X\overset{\text{a.s.}}{=}f(X,W)$. Therefore, if the distribution of the input ($X$) is known or controlled, knowing $f(\cdot,\cdot)$ is sufficient to fully describe the predictive distribution of $Y|X$ and the joint distribution of $(X,Y)$. In particular, $f(X,W)$ can be seen as a \textit{generative model} for $Y|X$. We illustrate this in Figure~\ref{sfig:general-model}.

\subsection{The Class of Additive Noise Models}
\label{sec:general-additive-models}
Using Lemma~\ref{lem:noise-outsourcing}, we focus on the expressive family of additive noise models (ANMs; see Figure~\ref{sfig:additive-model}), which are widely used in system theory and system monitoring applications \citep{draper1998applied}.

\begin{definition}
\label{def:additive-model}
A system $(X,Y)$ is said to follow an \textup{additive noise model}, denoted by $(X,Y)\sim\textup{add}(\eta,h;\mu)$, if both following properties hold true:
\begin{itemize}
    \item[(i)] $\mathbb{P}(X\in A)=\mu(A),\forall A\in\mathcal{B}(\mathbb{R}^p)$,
    \item[(ii)] $(X,Y)\overset{\textup{a.s.}}{=}(X,\eta(X)+h(W))$,
\end{itemize}
where $\mu$ is a probability measure over $(\mathbb{R}^p,\mathcal{B}(\mathbb{R}^p))$, ${W\sim\textup{Uniform}([0,1])}$ a r.v. independent of $X$, and $\eta:\mathbb{R}^p\rightarrow\mathbb{R}^q$ and $h:[0,1]\rightarrow\mathbb{R}^q$ two measurable functions, which are denominated as the system's \textit{underlying model} and \textit{noise model}, respectively.
\end{definition}

Some observations regarding Definition~\ref{def:additive-model}: 
\begin{itemize}
\item  For the rest of the exposition, we focus on the rich case where $\mathbb{E}[h(W)]=\mathbf{0}\in\mathbb{R}^q$. This assumption induces no loss of generality, as it is simple to embed any noise bias in $\eta(\cdot)$.\footnote{If $\mathbb{E}[h(W)]=\mathbf{b}\neq\mathbf{0}$, then we can define $\bar{\eta}(\cdot)$ and $\bar{h}(\cdot)$ such that ${\bar{\eta}(x) = \eta(x)+\mathbf{b},\forall x\in\mathbb{R}^p}$, and $\bar{h}(w) = h(w)-\mathbf{b}, \forall w\in[0,1]$, with the consequence that $\mathbb{E}[\bar{h}(W)]=\mathbb{E}[h(W)]-\mathbf{b}=\mathbf{0}\in\mathbb{R}^q$.} This is further justified in \ref{sec:unbias-noise}.

\item A system following an ANM is fully defined by its tuple $(\eta,h,\mu)$, and its predictive distribution is fully described by the tuple $(\eta,h)$; hence, an actual drift can be decoupled into drifts in $\eta(\cdot)$ and $h(\cdot)$. We denominate these as \textit{model drifts} and \textit{noise drifts}, respectively.

\item In system monitoring, we can assume that $\mu$ is known or controlled, as this is the marginal distribution of the explanatory (input) variable; note that drifts in $\mu$ are virtual drifts. In addition, $h(\cdot)$ is the noise model, and its drifts do not alter the underlying deterministic input-output relationship; hence, they cannot be attributed to unwanted alterations of the inner dynamics of a system (faults) but to external disturbances, such as alterations in the measurement procedures. As neither virtual nor noise drifts can be attributed to system faults and as faults do not necessarily imply either of those drifts, they are out of the scope of our study.

\item As a consequence of the above, the focus of our study is to detect model drifts, i.e., disturbances on the underlying deterministic input-output relationship $\eta(\cdot)$. As disturbances in the underlying model are a proxy for unwanted alterations in the system's inner dynamics (faults), and faults induce model drifts, we establish an equivalence between fault detection and model drift detection.

\end{itemize}

\subsection{Model Equivalence}
\label{sec:a-s-equivalence}
For the class of ANMs (Def.~\ref{def:additive-model}), it is crucial to have an equivalence relationship that can be used to represent the idea of MD and, in consequence, the idea of the existence of faults.

\begin{definition}
\label{def:model-almost-equivalence}
Let $\eta_1:\mathbb{R}^p\rightarrow\mathbb{R}^q$ and $\eta_2:\mathbb{R}^p\rightarrow\mathbb{R}^q$ be measurable functions. $\eta_1(\cdot)$ and $\eta_2(\cdot)$ are said to be \textit{almost-surely equivalent w.r.t. $\mu$}, denoted by $\eta_1\simeq_{\mu}\eta_2$, if for every r.v. $X\sim\mu$ it holds that $\eta_1(X)\overset{\textup{a.s.}}{=}\eta_2(X)$. If $\eta_1\simeq_{\mu}\eta_2$ does not hold, it is denoted as $\eta_1\not\simeq_{\mu}\eta_2$.
\end{definition}

\begin{remark} 
\label{rmk:mmse}
It is worth noting that for an arbitrary system $(X,Y)\sim\textup{add}(\eta,h;\mu)$, the minimum mean square error (MMSE) estimation of $Y$ given $X$ is $\mathbb{E}[Y|X]=\eta(X)+\mathbb{E}[h(W)]=\eta(X)$ \citep[Theorem~13-1]{mendel1995lessons}. Hence, it follows for the optimal (MMSE) estimator that $\hat{Y}\simeq_{\mu}\eta$, for any input distribution $\mu\in\mathcal{P}(\mathbb{R}^p)$. Equipped with Def.~\ref{def:model-almost-equivalence}, in the following sections, we show that MD over this class of ANMs can be observed by using the MMSE estimator of the reference (nominal) scenario $\mathcal{H}_0$.
\end{remark}

\subsection{Model Drift Detection --- Formal Hypotheses for Fault\texorpdfstring{\linebreak}{} Detection}
\label{sec:mismodelling-test-def}
Within our class of models from Def.~\ref{def:additive-model} and using the almost-surely equivalence definition introduced in Def.~\ref{def:model-almost-equivalence}, we can state the hypothesis test required for dealing with model drift. Let us consider a system with nominal distribution $\text{add}(\eta,h;\mu)$ and actual distribution $\text{add}(\tilde{\eta},\tilde{h};\mu)$. We define a healthy system -- i.e., the absence of model drift -- as our null hypothesis $\mathcal{H}_0$, and the presence of unwanted system deviations --~i.e., faults: the presence of model drift~-- as our alternative hypothesis $\mathcal{H}_1$. Then, formally, we have that
\begin{equation}
    \label{eqn:hypothesis-test}
    \begin{aligned}
    \mathcal{H}_0 &: \eta\simeq_{\mu}\tilde{\eta},\\
    \mathcal{H}_1 &: \eta\not\simeq_{\mu}\tilde{\eta}.
    \end{aligned}
\end{equation}

As a consequence of Remark~\ref{rmk:mmse}, if $\eta_{\text{nominal}}(\cdot)$ is the MMSE estimator built from the nominal distribution, then ${\eta\simeq_{\mu}\tilde{\eta} \Leftrightarrow \eta_{\text{nominal}}\simeq_{\mu}\tilde\eta}$ holds true for every $\mu\in\mathcal{P}(\mathbb{R}^p)$. Hence, $\eta_{\text{nominal}}(\cdot)$ can be used as a snapshot of the nominal system to be tested with the actual underlying model via samples $\mathbf{Z}_n = (Z_j)_{j=1}^n$ collected from the actual system with distribution $\text{add}(\tilde{\eta},\tilde{h};\mu)$.

\subsection{Fault Existence and Input-Residual Dependency} 
\label{sec:md-ir-dependency}
The first main contribution of our work formalizes a link between fault existence and input-residual dependency. From this point on, the following assumptions will be considered:\footnote{Although assumptions (i) and (ii) might seem equivalent and both r.v.s in assumption (iii) might seem the same, neither of those statements is necessarily true; this is because $Y-\eta(X)$ is a.s. equal to $\tilde{\eta}(X)-\eta(X)+\tilde{h}(W)$ and not $\tilde{h}(W)$ when $(X,Y)\sim\text{add}(\tilde{\eta},\tilde{h};\mu)$.}
\begin{itemize}
    \item[(i)] $\mathbb{E}[\tilde{h}(W)]=\mathbf{0}\in\mathbb{R}^q$.
    \item[(ii)] $\mathbb{E}[Y-\eta(X)]=\mathbf{0}\in\mathbb{R}^q$.
    \item[(iii)] Both r.v.s $(X,Y-\eta(X))$ and $(X,\tilde{h}(W))$ have densities.
\end{itemize}
We name these as our ``\textit{standard assumptions}.''\,\footnote{Assumption (i) was already discussed in the observations regarding Def.~\ref{def:additive-model}. Both assumptions (i) and (ii) are formally justified in \ref{sec:unbias-noise} and \ref{sec:unbias-model}, respectively. Assumption (iii) emerges from the continuous nature of our setup. All these assumptions can be relaxed but are taken into consideration for the sake of clarity and simplicity of our presentation.}

\begin{theorem}
    \label{the:mismodelling-independence-link}
    Let $X$ and $Y$ be r.v.s with values in $\mathbb{R}^p$ and $\mathbb{R}^q$, respectively, such that $(X,Y)\sim\textup{add}(\tilde{\eta},\tilde{h};\mu)$, and let ${\eta:\mathbb{R}^p\rightarrow\mathbb{R}^q}$ be a measurable function. Under the \textup{standard assumptions}, the following statements are equivalent:
    \begin{itemize}
        \item[(i)] $\tilde{\eta}\simeq_{\mu}\eta$, i.e., $\mathcal{H}_0$ in (\ref{eqn:hypothesis-test}).
        \item[(ii)] $X$ and $Y-\eta(X)$ are independent.
    \end{itemize}
\end{theorem}

The proof of Theorem~\ref{the:mismodelling-independence-link} is presented in \ref{sec:proof-mismodelling-independence-link}. As shown in the proof, Theorem~\ref{the:mismodelling-independence-link} is a consequence of a generalized version of this result --~Theorem~\ref{the:gen-mdd-dependency}, see \ref{sec:general-theorem}~-- which relaxes assumptions (i) and (ii).

Importantly, Theorem \ref{the:mismodelling-independence-link} shows that testing the existence of faults (see Sec.~\ref{sec:mismodelling-test-def}) reduces to analyzing the statistical dependency (or higher order dependency) between the input $X$ and $Y-\eta(X)$. This last variable, which we denominate as \textit{residual}, can be seen as the regression error obtained from the MMSE estimator built from $\mathcal{H}_0$ (the nominal model).

\section{Information- and Data-Driven Model Drift Detection}
\label{sec:md_method}
In this section, we show our core methodological contribution: a novel MI-based pipeline for testing the existence of faults. Mutual information can be used to determine if two arbitrary r.v.s. ($A$ and $B$) are independent or not. In fact, $I(A;B)=0$ if, and only if, $A$ and $B$ are independent \citep{cover_2006}. Using this expressive property and driven by Theorem~\ref{the:mismodelling-independence-link}, we propose the adoption of a data-driven consistent estimation of the MI between $Y-\eta(X)$ and $X$ to inform the decision about $\mathcal{H}_0$ and $\mathcal{H}_1$. To implement this idea (in the form of a decision scheme --~see Sec.~\ref{sec:md-test-def}~--), we adopt a distribution-free MI estimator \citep{silva2012complexity} and extend its statistical properties into our fault detection problem expressed in~(\ref{eqn:hypothesis-test}).

\subsection{Data-Driven Decision Strategy --- The Residual Information Value (RIV) Pipeline}
\label{sec:the_detector}
\begin{figure*}[ht!]
\centering
\includegraphics[width=0.95\textwidth]{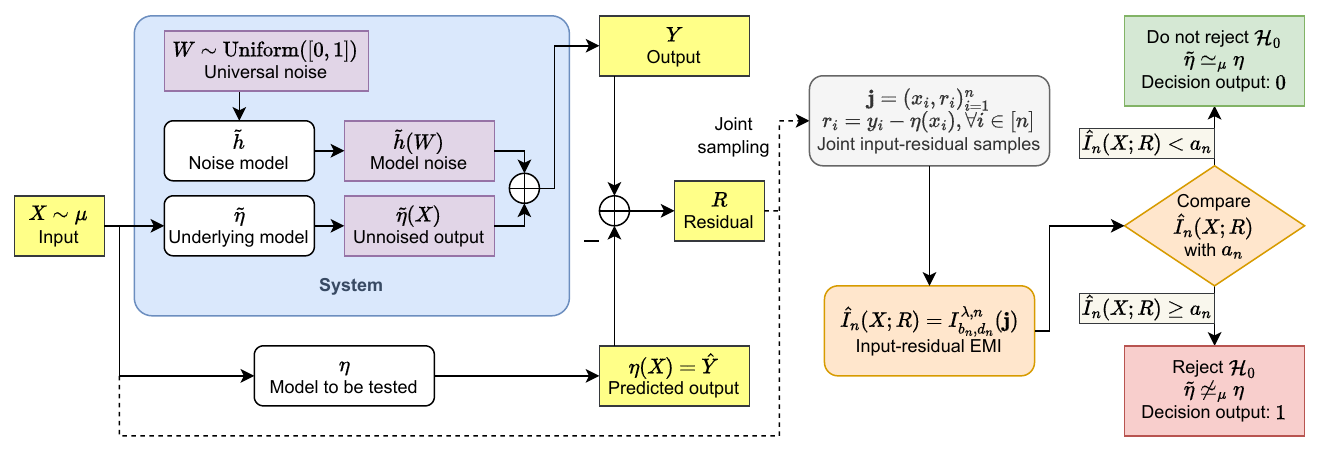}
\caption{Block diagram for the pipeline induced by our family of decision rules: $\psi_{b_n,d_n,a_n}^{\lambda, n}(\cdot)$.}
\label{fig:pipeline}
\end{figure*}

Returning to the problem formulated in Sec.~\ref{sec:mismodelling-test-def}, let ${(X,Y)\sim\text{add}(\tilde{\eta},\tilde{h};\mu)}$ be a system and $\eta:\mathbb{R}^p\rightarrow\mathbb{R}^q$ be the function that models its reference (nominal) scenario ($\mathcal{H}_0$; in other words, under $\mathcal{H}_0$: $\tilde{\eta}\simeq_{\mu}\eta$ and under $\mathcal{H}_1$: $\tilde{\eta}\not\simeq_{\mu}\eta$), and let $\mathbf{Z}_n = (X_j,Y_j)_{j=1}^n$ be $n$ i.i.d. samples of $(X,Y)\sim\text{add}(\tilde{\eta},\tilde{h};\mu)$. Our MI-based decision rule of length $n$, which we denote by $\psi_{b_n,d_n,a_n}^{\lambda,n}:(\mathbb{R}^{p+q})^n\rightarrow\{0,1\}$, is computed as follows: 
\begin{enumerate}
    \item From $\mathbf{Z}_n$, a sampling of the residual $R\equiv Y-\eta(X)$ is built: $\mathbf{R}_n \equiv (R_j)_{j=1}^n \equiv (Y_j-\eta(X_j))_{j=1}^n$.
    \item The value of $I(X;R)$ --~see~(\ref{eqn:mi})~-- is estimated using an MI estimator presented in Sec.~\ref{sec:mi-estimator}. In particular, we denote $\hat{I}_n(X;R) \equiv I_{b_n,d_n}^{\lambda,n}(\mathbf{J}_n)$ as the estimated MI (EMI) between $X$ and $R$ --~which we denominate the \textit{residual information value} (RIV)~--, being $\mathbf{J}_n\equiv(X_j,R_j)_{j=1}^n$ the data and $I_{b_n,d_n}^{\lambda,n}(\cdot)$ the MI estimator. The estimator parameters are $(\lambda,b_n,d_n)$, and their roles are explained in Sec.~\ref{sec:mi-estimator}.  
    \item Reject $\mathcal{H}_0$ if the RIV, $\hat{I}_n(X;R)$, is above or equal to a threshold $a_n>0$; otherwise, do not reject $\mathcal{H}_0$. 
\end{enumerate}
The pipeline induced by our family of decision rules is illustrated in Figure~\ref{fig:pipeline} and presented as a pseudocode in Algorithm~\ref{alg:method}. 

\begin{algorithm}[ht!]
    \caption{Residual Information Value (RIV) Method for Detecting Faults\label{alg:method}}
    \KwIn{Nominal model $\eta(\cdot)$, input-output $n$-size i.i.d. sample from the system $\mathbf{Z}_n=(X_j,Y_j)_{j=1}^n$, MI estimator parameters for $n$ samples $(\lambda,b_n,d_n)$, decision threshold for $n$ samples $a_n$}
    \KwOut{Binary decision --- $0$ (system is healthy) or $1$ (system is undergoing a failure)}

    \SetKwFunction{FRIV}{RIVDecision}
    \SetKwProg{Fn}{Function}{:}{}
    \Fn{\FRIV{$\eta(\cdot)$, $\mathbf{Z}_n$, $\lambda$, $b_n$, $d_n$, $a_n$}}{
        \For{$j\in\{1,2,\ldots,n\}$}{
            \tcp{Compute the output estimation}
            $\hat{Y}_j \leftarrow \eta(X_j)$\;
            \tcp{Compute the residual}
            $R_j \leftarrow Y_j - \hat{Y}_j$\;
        }
        \tcp{Build the joint input-residual dataset}
        $\mathbf{J}_n \leftarrow (X_j,R_j)_{j=1}^n$\;
        \tcp{Estimate the input-residual MI}
        $\texttt{RIV}\leftarrow I_{b_n,d_n}^{\lambda,n}(\mathbf{J}_n)$\;
        \If{$\textup{\texttt{RIV}} < a_n$}{
            \tcp{Do not reject $\mathcal{H}_0$, i.e., system is healthy}
            $\texttt{decision}\leftarrow 0$\;
        }
        \Else{
            \tcp{Reject $\mathcal{H}_0$ in favor of $\mathcal{H}_1$, i.e., system is faulty}
            $\texttt{decision}\leftarrow 1$\;
        }
        \Return \texttt{decision}\:\;
    }
\end{algorithm}

In summary, the decision rule corresponds to ${\psi_{b_n,d_n,a_n}^{\lambda,n}(\mathbf{Z}_n)=1_{[a_n,\infty)}(\hat{I}_n(X;R))}$, where all $(b_n)_{n\in\mathbb{N}}\equiv\mathbf{b}$, $(d_n)_{n\in\mathbb{N}}\equiv\mathbf{d}$ and $(a_n)_{n\in\mathbb{N}}\equiv\mathbf{a}$ are sequences of positive numbers, and $\lambda\in(0,\infty)$. Then, we can define our proposed decision scheme as a sequence of decision rules $\Psi_{\mathbf{b},\mathbf{d},\mathbf{a}}^{\lambda} \equiv \left(\psi_{b_n,d_n,a_n}^{\lambda, n}(\cdot)\right)_{n\in\mathbb{N}}$ parametrized by $(\lambda, \mathbf{b},\mathbf{d},\mathbf{a})$.

\subsection{The Distribution-Free Mutual Information Estimator}
\label{sec:mi-estimator}
In this subsection, we show the basic structure of the MI estimator adopted for our scheme, i.e., $I_{b_n,d_n}^{\lambda,n}:(\mathbb{R}^{p+q})^n\rightarrow[0,\infty)$, introduced in Sec.~\ref{sec:the_detector}. This non-parametric estimator is presented in full detail in \cite{silva2012complexity}.

The adopted MI estimator uses the data in three stages:

{\bf Stage one} (\textit{Data-driven partition}): it builds a partition of the sample space ($\mathbb{R}^{p+q}$) using axis-parallel hyperplanes, which perform a statistical equivalent division of the data \cite[Sec.~III.A]{silva2012complexity}. This produces a set of (data-driven) cells ${\mathcal{A}\equiv\{A_{\ell}\}_{\ell\in\Lambda}\subseteq\mathcal{B}(\mathbb{R}^{p+q})}$, indexed by the nodes of a binary tree. This tree is grown until each cell has at most $n\cdot b_n$ samples \cite[Equation~(8)]{silva2012complexity}. Then, the resulting tree is pruned to achieve a complexity regularized optimum with a factor $\lambda$. This pruning (regularization) stage is designed to ensure that the estimation is within a confidence interval shown in \cite[Corollary~2]{silva2012complexity} with a probability of at least~$1-\delta_n$.\footnote{What we denote as $d_n$ is denoted as $\delta$ and $\delta_n$ in \cite{silva2012complexity}.}

{\bf Stage two} (\textit{Distribution estimation}): it estimates (from data) both the joint distribution and the marginal distributions over the cells of $\mathcal{A}$ (the empirical probability), exploiting the fact that the axis-parallel construction implies that for every $\ell\in\Lambda$, there exist $A^{(1)}_{\ell}\in\mathcal{B}(\mathbb{R}^p)$ and $A^{(2)}_{\ell}\in\mathcal{B}(\mathbb{R}^q)$ such that $A_{\ell}=A_{\ell}^{(1)}\times A_{\ell}^{(2)}$. The empirical probability is defined as $P_n(A)\equiv\frac{1}{n}\sum_{j=1}^n1_A(J_j),\forall A\in\mathcal{B}(\mathbb{R}^{p+q})$.

{\bf Stage three} (\textit{MI estimation}): it estimates the MI using an empirical expression of (\ref{eqn:mi}). This corresponds~to 
\begin{equation}
    \label{eqn:mi-estimation}
    \hat{I}_n(X;R) = I_{b_n,d_n}^{\lambda,n}(\mathbf{J}_n) \equiv \sum_{\ell\in\Lambda}P_n(A_{\ell})\cdot\log\left(\frac{P_n(A_{\ell})}{Q_n(A_{\ell})}\right),
\end{equation}
where $Q_n(A_{\ell})\equiv P_n(A_{\ell}^{(1)}\times\mathbb{R}^q)\cdot P_n(\mathbb{R}^p\times A_{\ell}^{(2)})$.

\section{Performance Results}
\label{sec:performance}
In this section, we state the theoretical properties of our method illustrated in Figure~\ref{fig:pipeline}. We show concrete performance results, from asymptotic to important finite-sample properties. These results support the capacity of our scheme to reach the right decision once sufficient evidence is collected from the input-output data observed in the system: $\mathbf{Z}_n = (X_j,Y_j)_{j=1}^n$. 

Before stating our results, we need to introduce the following notation: Let $\mathbf{u}=(u_n)_{n\in\mathbb{N}}$ and $\mathbf{v}=(v_n)_{n\in\mathbb{N}}$ be two arbitrary sequences of non-negative numbers; first, we define ${1/\mathbf{u}\equiv(1/u_n)_{n\in\mathbb{N}}}$; second, ${\mathbf{u}\approx\mathbf{v}\Leftrightarrow\exists C\in(0,\infty):\lim_{n\rightarrow\infty}u_n/v_n=C}$; third, $\mathbf{u}\in O(v_n)\Leftrightarrow\exists C\in(0,\infty),\exists n_0\in\mathbb{N}:\forall n\geq n_0,u_n\leq Cv_n$; fourth, $\mathbf{u}\in\ell_1(\mathbb{N})\Leftrightarrow\sum_{n\in\mathbb{N}}|u_n|<\infty$; and lastly, that $\mathbf{u}\in o(1)\Leftrightarrow\lim_{n\rightarrow\infty}u_n=0$.

\subsection{Strong Consistency}
\begin{theorem}
\label{the:strong-consistency}
    Under the \textup{standard assumptions} (see Sec.~\ref{sec:md-ir-dependency}), any decision scheme (see Sec.~\ref{sec:the_detector}) in the family $\Psi_{\textup{SC}}$ expressed by
    \begin{equation}
    \label{eqn:strong_family}
        \Psi_{\textup{SC}} \equiv \left\{\Psi_{\mathbf{b},\mathbf{d},\mathbf{a}}^{\lambda}:\begin{array}{l}\mathbf{b}\approx(n^{-l})_{n\in\mathbb{N}},\\
        1/\mathbf{d}\in O(\textup{exp}(n^{1/3})),\mathbf{d}\in\ell_1(\mathbb{N}),\\
        \mathbf{a}\in o(1),\lambda\in(0,\infty),l\in(0,1/3)\end{array}\right\}
    \end{equation}
    is strongly consistent for detecting faults in the sense of Def.~\ref{def:strong-consistency}.
\end{theorem}

The proof of Theorem~\ref{the:strong-consistency} is presented in \ref{sec:proof-strong-consistency}. This result shows a large regime of parameters in~(\ref{eqn:strong_family}), where our scheme converges (with the number of samples) to the right decision almost-surely with respect to the process distribution of $(Z_n)_{n\in\mathbb{N}}$.

\subsection{Exponentially-Fast Decision for Healthy Systems}
\begin{theorem}
\label{the:finite-nominal-time}
    Under the \textup{standard assumptions} (see Sec.~\ref{sec:md-ir-dependency}), for any decision scheme $\Psi$ in the family $\Psi_{\textup{FL}}$ expressed by
    \begin{equation}
    \label{eqn:fast-family}
        \Psi_{\textup{FL}} \equiv \left\{\Psi_{\mathbf{b},\mathbf{d},\mathbf{a}}^{\lambda}:\begin{array}{l}\mathbf{b}\approx(n^{-l})_{n\in\mathbb{N}},\mathbf{d}\approx(\textup{exp}(n^{-1/3})),\\
        \mathbf{a}\in o(1), \lambda\in(0,\infty),l\in(0,1/3)\end{array}\right\},
    \end{equation}
    we have that $\forall m\in\mathbb{N},$
    \begin{equation}
    \label{eqn:finite-nominal-time}
    \mathbb{P}(\mathcal{T}_0^{\Psi}((Z_n)_{n=1}^{\infty})<m\,|\,\mathcal{H}_0)\geq1-K\textup{exp}(-m^{1/3}),
    \end{equation}
    for a universal (distribution-free) constant $K>0$.\footnote{See Def.~\ref{def:detection-time} for further details on the notation $\mathcal{T}_0^{\Psi}(\cdot)$.}
\end{theorem}

The proof of Theorem~\ref{the:finite-nominal-time} is presented in \ref{sec:proof-finite-nominal-time}. This result refines Theorem \ref{the:strong-consistency} and says that under $\mathcal{H}_0$, the probability of arriving to the right decision (with less than $m$ samples) converges exponentially fast to 1.

\subsection{Power and Significance Level}
\begin{theorem}
\label{the:hypothesis-test-metrics}
Under the \textup{standard assumptions} (see Sec.~\ref{sec:md-ir-dependency}), for any decision scheme $\Psi=(\psi_n(\cdot))_{n\in\mathbb{N}}$ in the family expressed in (\ref{eqn:fast-family}), i.e., $\forall\Psi\in\Psi_{\textup{FL}}$, it follows that
\begin{align}
    &\lim_{n\rightarrow\infty}(1-\beta_{\psi_n}) = 1,\label{eqn:power-convergence}\\
    &\alpha_{\psi_n}\leq K\cdot\textup{exp}(-n^{1/3}),\forall n\in\mathbb{N},\label{eqn:significance-vanish}
\end{align}
with $K>0$, a universal (distribution-free) constant.
\end{theorem}

The proof of Theorem~\ref{the:hypothesis-test-metrics} is presented in \ref{sec:proof-hypothesis-test-metrics}.

\section{Discussion of the Methodology and Results}
\label{sec:method-results-discussion}

Regarding our methodological contribution and the significance of our results, we can highlight the following: 

\begin{itemize}

    \item Our method, illustrated in Figure~\ref{fig:pipeline}, works over the rich family of ANMs. Importantly, beyond our standard assumptions, we do not require any special distribution for the r.v.s that determine the system, and we do not impose any restriction on the underlying model (i.e., on $\eta(\cdot)$) and the noise distribution (induced by $h(\cdot)$ in the ANM).

    \item From the adoption of a distribution-free MI estimator, we have the capability to detect any input-residual statistical dependency; in consequence, we are able to detect any kind of fault --~i.e., deviations from $\eta(\cdot)$, determined by $\tilde{\eta}(\cdot)$~-- with the only restriction of these functional deviations occurring within the input's marginal distribution ($\mu$) support.

    \item From $\Psi_{\text{FL}}$, we provide a concrete parameter regime to practitioners --~see (\ref{eqn:fast-family})~-- to ensure strong consistency, an exponentially-fast decision on $\mathcal{H}_0$, and error convergence to $0$ in the sample size. 

    \item Any decision scheme $\Psi\in\Psi_{\text{FL}}$ has an exponentially-fast decision convergence for healthy systems, i.e., under $\mathcal{H}_0$ (see Theorem~\ref{the:finite-nominal-time}). This finite-sample size behavior is remarkable and distinctive from other existing methods. To the best of our knowledge, this exponential velocity for detecting $\mathcal{H}_0$ is not observed in other methods for fault detection.

    \item Our method is unsupervised in the sense that it does not require examples from failure scenarios to be implemented. In practice, we only need healthy data or a phenomenological insight about the system to start monitoring it and detect its faults via model drift detections.
\end{itemize}

In the following two sections, we apply our method to practical scenarios, starting with systems that have synthetic distributions (see Sec.~\ref{sec:experimental}) and finishing with a benchmark dataset of turbofan engines (see Sec.~\ref{sec:n-cmapss}).

\section{Numerical Analysis on Synthetic Distributions}
\label{sec:experimental}

In this section, we perform a numerical analysis to validate our pipeline by applying our methodology to several synthetic parametric systems. In these systems, we fix nominal parameters that are disturbed by a change in the parameters (i.e., a model drift). Consequently, drifts represent an induced fault in our synthetic system, and their magnitudes are proxies for fault severities -- see Sec.~\ref{sec:faults-as-drift} and Sec.~\ref{sec:general-additive-models} where the fault-drift relationship is presented for general systems and systems within the rich ANM family (see Def.~\ref{def:additive-model}), respectively. Experiments in this section are subjected to an error-bar analysis in\linebreak {\ref{sec:synthetic-error-bar}}.

We consider systems within the family of ANMs determined by a parametric function expressed by $\eta_{\theta}:\mathbb{R}^p\rightarrow\mathbb{R}^q$ and $h:[0,1]\rightarrow\mathbb{R}^q$. The parameter for $\eta_{\theta}(\cdot)$ is $\theta\in\Theta\subseteq\mathbb{R}^{\nu}$, where $\nu\in\mathbb{N}$ is the dimensionality of $\Theta$, which varies according to the system in consideration. For these parametric constructions, we consider a nominal system to be determined by its nominal parameters $\theta$, i.e., $(X,Y)\sim\text{add}(\eta_{\theta},h;\mu)$. On the other hand, a faulty -- i.e., drifted -- system is described by its nominal parameters ($\theta$) and its deviation; the latter is expressed as a perturbation $\delta=(\delta_{\ell})_{\ell=1}^{\nu}\in\mathbb{R}^{\nu}$. In consequence, a faulty system is expressed by $(X,Y)\sim\text{add}(\eta_{\theta,\delta},h;\mu)$. In particular, $\delta=\mathbf{0}\in\mathbb{R}^{\nu}$ implies that the system is non-drifted from its reference scenario, i.e., $\text{add}(\eta_{\theta,\mathbf{0}},h;\mu) = \text{add}(\eta_{\theta},h;\mu)$.

\subsection{Experimental Setup}
\label{sec:experimental-setup}

Our setup for analyzing fault detection considers a parametric system $(X,Y)$ with an un-drifted nominal distribution $\text{add}(\eta_{\theta},h;\mu)$ and a potentially drifted distribution $\text{add}(\eta_{\theta,\delta},h;\mu)$. In this setting, the nominal model for $Y$ given $X$ corresponds to $\eta_{\theta}(\cdot)$, and the value of $\delta$ describes and quantifies the fault. The fault detector uses actual data $\mathbf{Z}_n$, i.e., $n$ i.i.d. samples from the actual distribution $(X,Y)\sim\text{add}(\eta_{\theta,\delta},h;\mu)$ to test the actual underlying model ($\eta_{\theta,\delta}(\cdot)$) with the nominal model ($\eta_{\theta}(\cdot)$).

For simplicity, in this section, we restrict our analysis to a 2-dimensional input space, a univariate output space, and a 2-dimensional $\delta$-space, i.e., $p=2$, $q=1$, and $\nu=2$. We explore four systems, denominated as \textit{Linear}, \textit{Polynomial}, \textit{Trigonometric} and \textit{MLP}; their expressions for $\eta_{\theta,\delta}(\cdot)$ are shown in Table~\ref{tab:fn_forward}.\footnote{$f^{\text{MLP}}_{\theta,\delta}$ is the function defined by a simple multi-layer perceptron with a single hidden layer of two units and parameters $\theta$ disturbed by $\delta$. In particular, $\delta_1$ and $\delta_2$ disturb additively the weights from the first and second input unit to the first hidden unit. We fix $\delta_\ell=0,\forall\ell>2$, in our experiments over the MLP system.} Additional details for all the distributions and functions used in our experimental setup, including marginal distributions, noise models, and the values of the coefficients that make up the nominal parameters ($\theta$), are contained in \ref{sec:exp-desc}.

\begin{table}[ht!]
\begin{center}
\begin{tabular}{lll}\toprule
     \textbf{System} & \textbf{Value of $\eta_{\theta,\delta}(x_1,x_2)$}\\\midrule
     Linear & $(c_1+\delta_1)x_1 + (c_2+\delta_2)x_2$\\
     Polynomial & $(c_1+\delta_1)x_1^2 + (c_2+\delta_2)x_2^3$\\
     Trigonometric & $(A+\delta_1)\sin(x_1x_2+\phi+\delta_2)$\\
     MLP & $f_{\theta,\delta}^{\text{MLP}}(x_1,x_2)$\\\bottomrule
\end{tabular}
\end{center}
\caption{Expressions for $\eta_{\theta,\delta}(x_1,x_2)$ in the explored systems.}
\label{tab:fn_forward}
\end{table}

For all cases, the nominal system ($\mathcal{H}_0$) corresponds to $\eta_{\theta,\mathbf{0}}(\cdot)$. In our experiments, we range $\delta_1$ and $\delta_2$ from $-0.15$ to $0.15$ with a step of $0.0015$. Within this range, it holds that $\delta=\mathbf{0}\Leftrightarrow\eta_{\theta}\simeq_{\mu}\eta_{\theta,\delta}$. Consequently, applying our method to observations from $(X,Y)\sim\text{add}(\eta_{\theta,\delta},h;\mu)$ will raise an MD detection ($\mathcal{H}_1$) if, and only if, $\delta\neq(0,0)$.

To better visualize our method, we compute the RIV, i.e., the input-residual EMI ($\hat{I}_n(X;R)$), and show its value prior to thresholding it (see Sec.~\ref{sec:threshold-analysis} for an in-depth thresholding analysis). This is done for every $\delta$ within the aforementioned range. These MI estimations are made using parameters from the range expressed in (\ref{eqn:fast-family}); therefore, each decision scheme $\Psi_{\mathbf{b},\mathbf{d},\mathbf{a}}^{\lambda}\in\Psi_{\text{FL}}\subseteq\Psi_{\text{SC}}$ satisfies the hypotheses of Theorem~\ref{the:strong-consistency},~\ref{the:finite-nominal-time},~and~\ref{the:hypothesis-test-metrics}. In particular, we choose ${\lambda=2.3\cdot10^{-5}\in(0,\infty)}$, ${\mathbf{d}=(\text{exp}(n^{-1/3}))_{n\in\mathbb{N}}\approx(\text{exp}(n^{-1/3}))_{n\in\mathbb{N}}}$, and ${\mathbf{b}=(wn^{-l})_{n\in\mathbb{N}}\approx(n^{-l})_{n\in\mathbb{N}}}$ with $w=5\cdot10^{-2}$ and ${l=0.167\in(0,1/3)}$. As we simulated a fixed amount of ${n=2000}$ samples, the reported EMIs are computed with ${b_n\approx0.014051}$ and $d_n\approx1.082605$.

We consider two baseline strategies with which to compare our method. The first, which we denote as the \textit{Correlation method}, is the maximum absolute value of the Pearson correlation coefficients between the residual and each input coordinate;\footnote{We denote this as $\text{MAPC}(X,R) \equiv \max\{|\text{corr}(X_1,R)|,|\text{corr}(X_2,R)|\}$; MAPC stands for \textit{Maximum Absolute Pearson Correlation}.} we select this method, as the input-error correlation is used as a common validation measure for system identification \cite[Ch.~9]{keesman2011system}. The second strategy, which we denote as the \textit{RMSE method}, is the root mean squared error computed from the residual; supporting this selection, there are different works based on thresholding RMSE, both in FDI (e.g., \cite{perez2023fault, dash2024bond, chowdhury2006fault}) and concept drift detection applied to regression~\cite{lima2022learning}.

Our method fundamentally improves upon the widely used baselines found in the literature. Firstly, correlation is defined as the variance-normalized covariance of $(X,R)$, which is a second-order statistic of the joint input-residual distribution. Then, the correlation method cannot capture drift complexities beyond a second-order relationship, such as faults that induce non-linear dependencies. Additionally,  a non-zero correlation implies a non-zero mutual information (MI); this means that all fault events detectable with the Correlation method are also detectable with our approach. Secondly, RMSE is a statistic of the residual-marginal distribution that fails to retain any information regarding the input-residual interaction. Therefore, unlike our method, it is vulnerable to noise disturbances that are independent of the system process.

\subsection{Numerical Results and Discussions about our Method}
\label{sec:numerical-discussion}

\begin{figure}[ht!]
\centering
\includegraphics[width=0.45\textwidth]{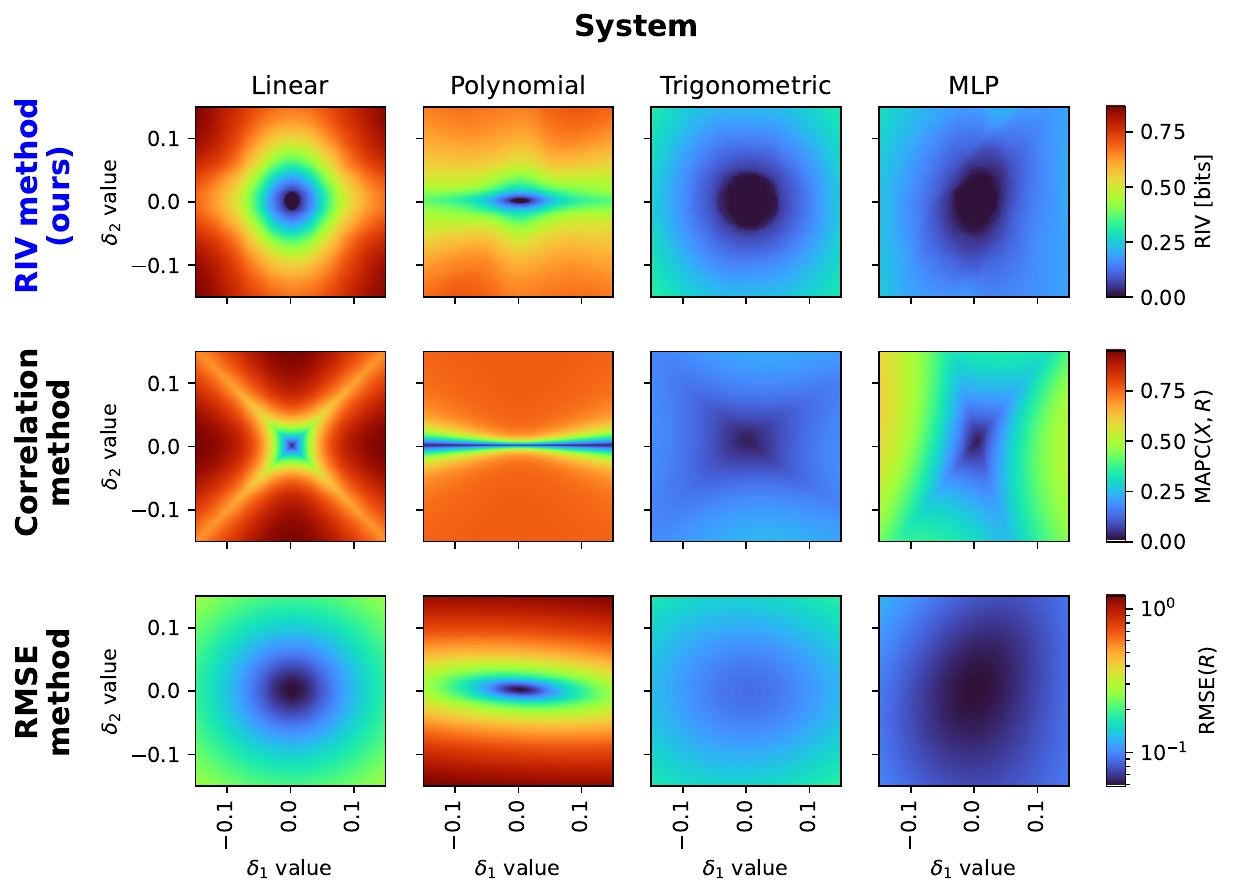}
\caption{Numerical results for our MI-based model drift detection method and baselines on parametrized drifts.}
\label{fig:experiment}
\end{figure}

Figure~\ref{fig:experiment} shows the values obtained for all four systems using our method and both baseline methods. Every single cell of every colormap has the average value over 10 different seeds for that specific $\delta$ value, method, and system. We can see that for all systems, the residual information value (RIV) reaches $0.0\pm0.0$ (i.e., empirical $0$ across all random seeds) when $\delta=(0,0)$. In consequence, for any threshold sequence $\mathbf{a}\in o(1)$, our method returns the right decision on $\mathcal{H}_0$. In addition, we can see in every model a region of $\delta$ values around the $\mathcal{H}_0$ point, which would raise false negatives as their RIVs are $0$; this is due to the difficulty of detecting dependency in slightly non-independent r.v.s. In general, as $\delta$ is farther from $(0,0)$, it is easier to identify input-residual dependencies.

In addition, we can see that each RIV colormap gives us an insight into the underlying model subjected to the drift. For example, the Linear and Trigonometric systems have colormaps where the RIV increases monotonically with $||\delta||$ following no privileged directions (isotropic monotonicity); hence, if we weight equally drifts in $\delta_1$ and $\delta_2$, the RIV is a clear indicator of the drift's magnitude. The greater RIV values in the Linear system, in contrast with the Trigonometric system, express that it is simpler to detect model drift in the Linear system. In regard to the Polynomial system, we can see a privileged direction of RIV increment (anisotropic monotonicity); this is the $\delta_1=0$ axis. The reason for this privilege is the cubic relationship of $Y$ with $X_2$ (whose weight is disturbed by $\delta_2$) in contrast with the quadratic dependence of $Y$ with $X_1$ (whose weight is disturbed by $\delta_1$). This implies that $\delta_2$-disturbances induce stronger input-residual dependencies (cubic) than $\delta_1$-disturbances (quadratic). Finally, regarding the MLP system, we can see oblique privilege directions; this pattern is due to asymmetric non-linearities induced by the activation function of the system. We remark that these colormaps, or high-dimensional versions of them, can be built for any system that can be identified with a white-box model to have a clear phenomenological interpretation of its possible drifts.

\subsection{Comparison with Baselines}
\label{sec:numerical-comparison}
Contrasting our method with the baselines, we can make the following observations. Regarding correlation, for the Polynomial system, the MAPC approximates zero across all points defined by $\delta_2=0$. This means that this method is unable to detect a fault caused by $\delta_1\neq0$ when $\delta_2=0$ -- e.g., for $\delta=(0.15,0)$ and $\delta=(-0.15,0)$, which represent non-subtle faults. This lack of fault observability is a consequence of the quadratic dependency between the input and the residuals when $\delta_2=0$. This is in clear contrast with the capacity of our MI-based residual analysis, as the MI captures dependencies among all complexities.

As for the RMSE, we can see a local minimum at $\delta=(0,0)$ for all colormaps. However, the value of the local minimum is system-dependent --~e.g., MLP and Trigonometric systems minima are $(5.79\pm0.05)\cdot10^{-3}$ and $(8.68\pm0.08)\cdot10^{-3}$, respectively~-- which implies the necessity of a system-dependent threshold. This design is not straightforward and is a limitation of this error-based strategy; in this line, there exists a systematic study for error-based drift detection methods \cite{lima2022learning}. In contrast, our RIVs reach zero under $\mathcal{H}_0$ independently of the system; this is expressed in the exponentially-fast convergence of the decision under $\mathcal{H}_0$ as shown in Theorem~\ref{the:finite-nominal-time}. As a consequence, the RMSE method requires samples from faulty systems to set the decision threshold.

\subsection{Complementary Analysis}
\label{sec:complementary-analysis}

In this subsection, we complement our analysis with experiments that address challenges present in applied systems. First, we apply our methodology to autoregressive systems, which are relevant in controlled systems since their inputs (i.e., controller's outputs) depend on the previous system's outputs.  In contrast with the previous analysis, the main challenge is that input-output samples are not i.i.d. Second, we perform a thresholding analysis to derive the receiving operating characteristic (ROC) curves for different scenarios for the alternative hypothesis ($\mathcal{H}_1$).  This ROC  analysis is useful for systems with a known set of faults or degradation profiles where the analysis of the trade-off between test power and significance level is desired; this trade-off is achieved via a fixed threshold. Lastly, we perform a systematic analysis of the scenario where samples from faulty systems are available; this is done via an analysis of the area under the curve (AUC) of ROC curves in different settings for our method and both baselines.

\subsubsection{Autoregressive Systems}
\label{sec:ar-analysis}

In these experiments, we apply our method and both baselines on autoregressive (AR) systems. These AR systems are described by
\begin{align}
    D_j &= \eta_{\theta,\delta}(D_{j-1},U_j) + H_j,\\
    Y_j &= D_j + W_j,
\end{align}
where all $U$, $H$, and $W$ are independent univariate r.v.s. representing the exogenous input, the model noise, and the measurement noise, respectively. In these systems, the nominal model is still $\eta_{\theta,\mathbf{0}}(\cdot,\cdot)$. We perform experiments on a linear AR system (\textit{ARX}) and a non-linear AR system (\textit{NARX}); their expressions for $\eta_{\theta,\delta}(\cdot,\cdot)$ are presented in Table~\ref{tab:eta_autoregressive}.\footnote{In-depth details for these constructions are in \ref{sec:exp-desc}.} For these AR systems, an input sampling corresponds to $(X_j)_{j=1}^n = (Y_{j-1},U_j)_{j=1}^n$; hence, the temporal dependency between instances of the input implies that $\mathbf{Z}_{n}$ is not an i.i.d. sequence. We experimented with these systems to analyze the capabilities of our method beyond its formal hypotheses.

\begin{table}[ht!]
    \begin{center}
        \begin{tabular}{ll}\toprule
            \textbf{System}  &\textbf{Value of $\eta_{\theta,\delta}(d,u)$} \\\midrule
            ARX & $(c_1+\delta_1)d + (c_2+\delta_2)u$\\
            NARX & $(c_3+\delta_1+c_4\exp(-d^2))\cdot d + (c_5+\delta_2)\cdot u^2$\\\bottomrule
        \end{tabular}
    \end{center}
    \caption{Expressions for $\eta_{\theta,\delta}(d,u)$ in autoregressive systems.}
    \label{tab:eta_autoregressive}
\end{table}

\begin{figure}[ht!]
\centering
\includegraphics[height=6cm]{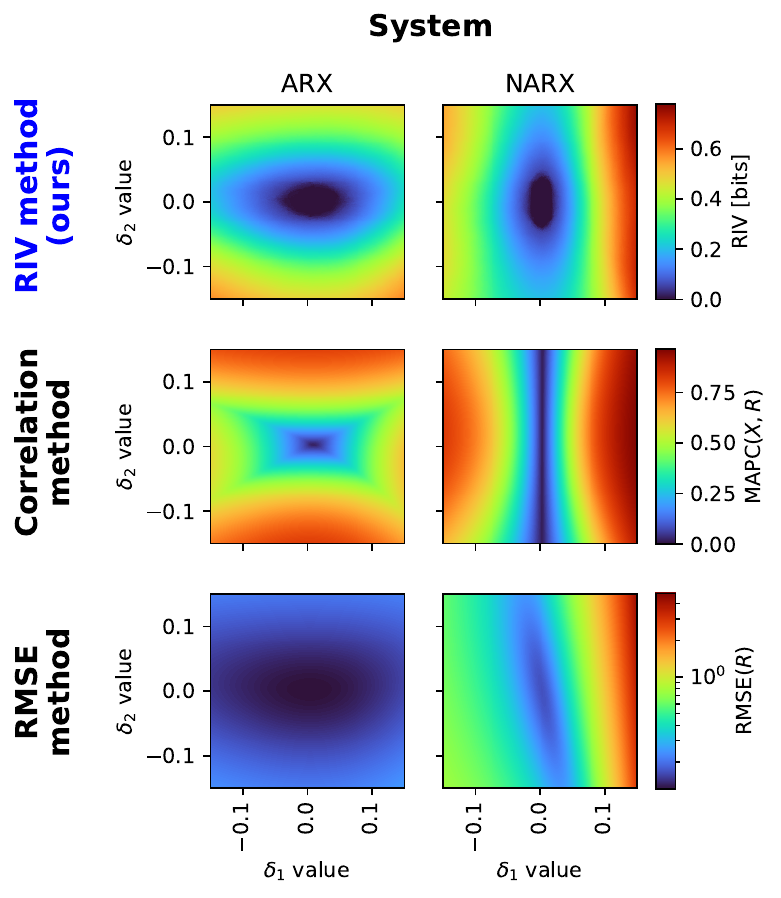}
\caption{Numerical results for the RIV method and baselines on parametrized drifts over AR systems.}
\label{fig:ar_experiment}
\end{figure}

In Figure~\ref{fig:ar_experiment}, we show the results of applying our method and both baselines using the same amount of samples ($n$) and parameters of the MI estimator ($\lambda$, $\mathbf{b}$, and $\mathbf{d}$; shown in Sec.~\ref{sec:experimental-setup}) to both ARX and NARX systems. We can see that the main observations discussed in Sec.~\ref{sec:numerical-discussion} remain valid for both systems. In particular, we get zero RIV under $\mathcal{H}_0$ and an increase of the RIV when moving away from $\delta=(0,0)$. We also observe an approximately isotropic monotonicity of the RIV for the ARX system; this is in contrast with the clear asymmetric pattern for the NARX system. This last asymmetric information pattern can be attributed to the non-linearities of the system.

Regarding the baselines, we can still observe the inability of the Correlation method to identify non-trivial drifts, which can be attributed to quadratic input-residual dependencies --~see MAPC values for the NARX system at $\delta_1=0$, e.g., $\delta=(0,-0.15)$ and $\delta=(0,0.15)$. Additionally, we can still observe the system-dependent non-zero minima of the RMSE method; these are $0.121\pm0.002$ and $0.168\pm0.002$ for the ARX and NARX systems, respectively.

Our results for the AR systems demonstrate the practical potential of our method in realistic scenarios. This capability is especially useful for controlled systems, as they have an inherent temporal dependence embedded in their signals due to the control loop. Moreover, our results show how the baselines' limitations we mentioned in Sec.~\ref{sec:numerical-comparison} still hold on AR systems.

\subsubsection{Thresholding Analysis}
\label{sec:threshold-analysis}

In these experiments, we analyze receiving operating characteristic (ROC) curves to visualize the effect of the threshold in our decision rules. Due to the high discrimination capacity of our framework when $n$ becomes large --~justified by our formal results and our previous analysis in Sec.~\ref{sec:numerical-discussion}~-- to obtain non-trivial ROC curves, we focus on challenging scenarios. In particular, we consider small sample sizes and alternative hypotheses that drift subtly from the null hypothesis (incipient faults).

We perform this ROC analysis on the same systems described in Sec.~\ref{sec:experimental-setup} and Sec.~\ref{sec:ar-analysis}. We keep the same parameters of the MI estimator ($\lambda$, $\mathbf{b}$, and $\mathbf{d}$ shown in Sec.~\ref{sec:experimental-setup}) and add the threshold sequence $\mathbf{a} = c/\mathbf{n} \equiv (c/n)_{n\in\mathbb{N}}\in o(1)$ to analyze the decision outputs -- instead of the RIV values shown in the previous experiments -- where $c>0$ is the threshold factor to be explored. The threshold sequence ($c/\mathbf{n}$) is such that $\Psi_{\mathbf{b},\mathbf{d},c/\mathbf{n}}^{\lambda}\in\Psi_{\text{FL}}\subseteq\Psi_{\text{SC}},\forall c>0$, ensuring that Theorems \ref{the:strong-consistency}, \ref{the:finite-nominal-time}, and~\ref{the:hypothesis-test-metrics} hypotheses are satisfied. Through the rest of this section, we denote $\Psi_{\mathbf{b},\mathbf{d},c/n}^{\lambda} \equiv (\psi_n^c(\cdot))_{n\in\mathbb{N}}$ to remark the dependency of our decision rules on the design parameter $c$.

Since our fault detector's hypothesis test is a composite one, see (\ref{eqn:hypothesis-test}), it is convenient and informative for this ROC performance analysis to fix alternative hypotheses. We have shown that the RIV is a clear indicator of the discriminability difficulty between $\mathcal{H}_0$ and $\mathcal{H}_1$ --~see Sec.~\ref{sec:numerical-discussion}; then, we index the alternative hypothesis in terms of their RIV. With this convention,  the null hypothesis and the (fixed) alternative hypotheses are reduced to
\begin{equation}
\begin{aligned}
    \label{eqn:restricted-HT}
    \mathcal{H}'_0 &: \delta=(0,0),\\
    \mathcal{H}_1^{\gamma} &: \delta = \varrho(\gamma),
\end{aligned}
\end{equation}
where $\varrho(\gamma)$ is an arbitrary system-dependent $\delta$ value such that the system's RIV is $\gamma$. Since there is no analytic expression for the RIV for most of our explored systems, we select $\varrho(\gamma)$  numerically from the RIV maps shown in Figures~\ref{fig:experiment} and \ref{fig:ar_experiment}; this means that $\varrho(\gamma)$ is determined as the $\delta$ value within our explored range whose estimated RIV approaches $\gamma$.

The ROC curves show the trade-off between the test's significance level and power \cite{bradley1997use} under a defined alternative hypothesis when varying the decision threshold. As we are using (\ref{eqn:restricted-HT}) instead of (\ref{eqn:hypothesis-test}), the two errors $\alpha_{\phi_n}$ and $\beta_{\phi_n}$ must be adapted to $\alpha_{n,c}^{\gamma} \equiv\mathbb{P}(\psi_n^c(\mathbf{Z}_n)=1\,|\,\mathcal{H}_0')$ and ${\beta_{n,c}^{\gamma} \equiv \mathbb{P}(\psi_n^c(\mathbf{Z}_n)=0\,|\,\mathcal{H}_1^{\gamma})}$, respectively. Hence,  for a certain amount of samples ($n$) and target RIV ($\gamma$), the ROC curve is represented by the pair $(\alpha_{n,c}^{\gamma},1-\beta_{n,c}^{\gamma})$ when $c\in(0,\infty)$.\footnote{Better ROCs are those that are farther above the diagonal ${\{(x,x):x\in[0,1]\}}$. A \textit{perfect} ROC reaches point $(0,1)$, which means that there exists a threshold for zero error rate (perfect discriminability).} We estimated ROCs for each system by simulating 1000 random seeds from its $\mathcal{H}_0'$ scenario and $\mathcal{H}_1^{\gamma}$ for each $\gamma$ value of interest and then computing the false positive rate (FPR; approximation of $\alpha_{n,c}^{\gamma}$) and true positive rate (TPR; approximation of $1-\beta_{n,c}^{\gamma}$) as we range $c$.

\begin{figure*}[ht!]
    \centering
    \includegraphics[width=\linewidth]{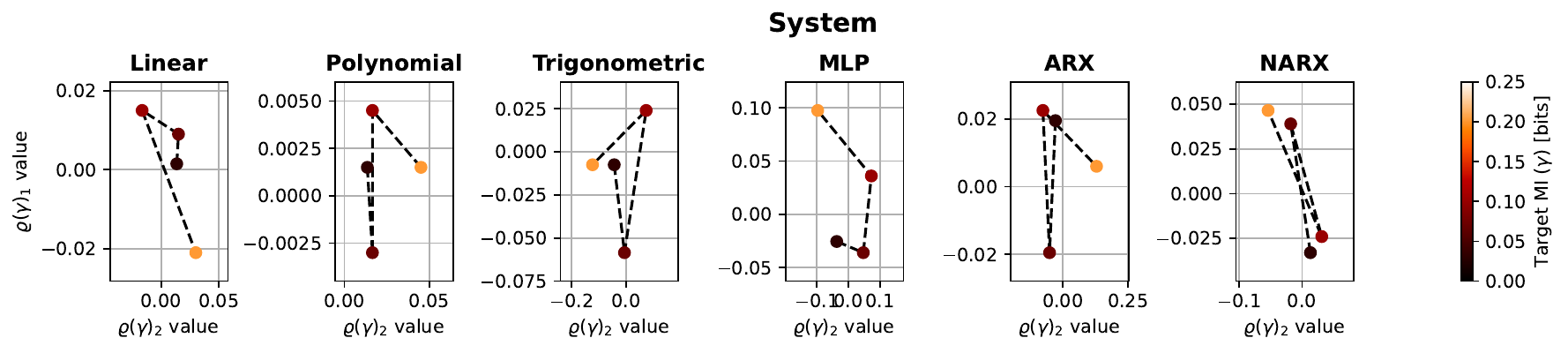}
    \caption{Values of $\varrho(\gamma)$ for different systems for target RIVs $\gamma\in\{0.03, 0.07,0.1,0.2\}$ bits.}
    \label{fig:approx-deltas}
\end{figure*}

We present the ROC performance curves for different sample sizes (in the mentioned low sampling regime) and drifts' discriminability (measured with the RIVs). In Figure~\ref{fig:roc-var-n}, we show the ROC curves for the sample sizes ${n\in\{50,100,250,500\}}$ with a fixed target RIV of $\gamma=0.1\:\text{bits}$, and in Figure~\ref{fig:roc-var-mi}, we show the ROC curves for the following target RIVs $\gamma\in\{0.03,0.07,0.1,0.2\}$ (measured in bits) with a fixed sample size of $n=100$ (which, as a reference, is $5\%$ of the sample size used in Figures~\ref{fig:experiment} and \ref{fig:ar_experiment}). In both figures, the value of $c$ that corresponds to each point in the ROC curve is shown as the point color's brightness. In Figure~\ref{fig:approx-deltas}, we show the values of $\varrho(\gamma)\equiv(\varrho(\gamma)_1,\varrho(\gamma)_2)\in\mathbb{R}^2$ for the explored target RIVs, i.e., for $\gamma\in\{0.03, 0.07, 0.1, 0.2\}$.

\begin{figure*}[ht!]
    \centering
    \includegraphics[width=\linewidth]{ROC_var_n.png}
    \caption{ROC curves for different systems with a fixed target RIV of $\gamma=0.1\:\text{bits}$ and $n\in\{50,100,250,500\}$.}
    \label{fig:roc-var-n}
\end{figure*}

\begin{figure*}[ht!]
    \centering
    \includegraphics[width=\linewidth]{ROC_var_mi.png}
    \caption{ROC curves for different systems with a fixed amount of samples $n=100$ and $\gamma\in\{0.03,0.07,0.1,0.2\}$ bits.}
    \label{fig:roc-var-mi}
\end{figure*}

We can see in Figure~\ref{fig:roc-var-n} that as $n$ increases, we get monotonically better ROC curves for all systems; this means that for all significance levels ($\alpha_{n,c}^{\gamma}$), its associated power of the test ($1-\beta_{n,c}^{\gamma}$) increases as $n$ increases (which, in turn, means that more data is available for decision making). In this regard, the ROC curves approximate the perfect ROC curve, which is an empirical validation of the strong consistency of our method --~see Theorem~\ref{the:strong-consistency}~-- and error rates convergence to zero --~see Theorem~\ref{the:hypothesis-test-metrics}. On the other hand, we can see in Figure~\ref{fig:roc-var-mi} that as $\gamma$ increases, we get monotonically better ROCs through all systems. This performance trend validates the interpretation of the proposed RIV indicator as a measure of the task difficulty.

\subsubsection{Comparison with Baselines on Availability of Prior Faulty Data}

We conclude our numerical analysis by evaluating scenarios in which data from faulty systems is available prior to the decision-making process. We note that these scenarios overlook our method's capability to function without relying on the existence of faulty data; hence, it is expected for our method to behave similarly to those that require prior data from faulty systems -- i.e., regarding our baselines, the RMSE~method.

We report the area under the curve (AUC) of the different (estimated) ROC curves.  To compute the AUC in each case, we treat a ROC trajectory as a function ${\text{ROC}_{n,\gamma}:[0,1]\rightarrow[0,1]}$ that maps significance levels to test's powers, i.e., ${\text{ROC}_{n,\gamma}(\alpha_{n,c}^{\gamma}) = 1-\beta_{n,c}^{\gamma}},$ $\forall(n,c,\gamma)\in\mathbb{N}\times(0,\infty)^2$.\footnote{Although $\alpha_{n,c}^{\gamma}$ and $\beta_{n,c}^{\gamma}$ are defined in terms of the decision rules of our method (i.e., $\psi_n^c$), these can be easily adapted to the Correlation and RMSE methods by thresholding estimations of $\text{MAPC}(X,R)$ and $\text{RMSE}(R)$.} Hence, the AUC of $\text{ROC}_{n,\gamma}$ is  $\text{AUC}_{n,\gamma} = \int_0^1\text{ROC}_{n,\gamma}(\alpha)\:\mathrm{d}\alpha$, which can be estimated with simple numerical approximations. In this global analysis, higher AUC values correspond to better ROC curves (i.e., a better performance tradeoff).\footnote{A perfect ROC has an AUC value of $1.0$, and a ROC generated from a random binary decision rule has an AUC value of $0.5$.}

Figure~\ref{fig:auc} shows the AUC values for $(n,\gamma)\in\{50,100,250,500,2000\}\times\{0.03,0.07,0.1,0.15,0.2\}$ (where 
$\gamma$ is measured in bits).\footnote{AUC values for the ROCs presented in Figures~\ref{fig:roc-var-n} and \ref{fig:roc-var-mi} are shown in the third row ($\gamma=0.1$) and second column ($n=100$), respectively, of each system's matrix of the RIV method} Consistently with the results presented in Figures \ref{fig:roc-var-n} and \ref{fig:roc-var-mi}, we observe improvements in the ROC curves as either $n$ or $\gamma$ increases for all studied systems; this further validates the strong consistency and error convergence to zero of our method. Notably, the achievement of a perfect empirical AUC value of $1.0$ means that the distributions of the 1000 RIV values generated from $\mathcal{H}_0'$ and $\mathcal{H}_1^{\gamma}$, respectively, are perfectly separable; this, again, can be explained by the exponentially fast detection and vanishing significance levels of our method -- see Theorems~\ref{the:finite-nominal-time} and \ref{the:hypothesis-test-metrics} -- since the exponential decay of $a_{n,c}^{\gamma}$ on $n$ is independent of $(c,\gamma)$ -- see (\ref{eqn:significance-vanish}) -- which means that the ROC curves (built from all $c>0$) converge exponentially-fast on $n$ to the axis of zero significance level ($\alpha_{n,c}^{\gamma}=0$).

\begin{figure*}[ht!]
    \centering
    \includegraphics[width=\linewidth]{AUC.png}
    \caption{Values of the AUC for ROC curves generated with different values of $(n,\gamma)$ on different systems using different methods.}
    \label{fig:auc}
\end{figure*}

Looking at both baseline results, we can see that the correlation method converges fast to an error-free ROC for the Linear and ARX systems. In contrast, this method is challenged --~reflected in low AUC values~-- by the non-linear systems. In particular, it faces difficulties in discriminating faults on the Polynomial system when $\gamma=0.2\:\text{bits}$ and the Trigonometric system across all scenarios. This is remarkable, in particular, for the Trigonometric system and can be explained by the nature of the trigonometric functions, which raise input-residual dependencies that are not correlated when faults are induced. For the RMSE method, on the other hand, we can see a fast convergence to error-free ROCs on all systems but the ARX system. Due to the system's autoregressive nature, this result arises from how noise propagates within the system, hindering the ability to differentiate between faulty and faultless signatures based solely on error magnitude.

We note that our AUC experiment, shown in Figure~\ref{fig:auc}, did not consider regimes where the Correlation method is rendered useless (i.e., $\delta_2=0$ on the Polynomial system or $\delta_1=0$ on the NARX system) nor regimes where the RMSE can be easily confused (e.g., with varying noises). These results show the competitiveness of our method even when its capacity to detect faults without prior faulty examples is neglected.

In summary, the analyses presented in Sec.~\ref{sec:experimental} offer an empirical validation of the power of our method, which is consistent with the Theorems presented in Sec.~\ref{sec:performance}. We have obtained consistent results for all the explored scenarios -- including the AR systems, where the i.i.d. hypothesis does not hold -- strengthening the universality of our method in regard to the underlying model -- determined by $\eta(\cdot)$.

\section{Study Case --- N-CMAPSS Dataset}
\label{sec:n-cmapss}

In this section, we apply our data-driven fault detection strategy to the N-CMAPSS dataset \cite{arias2021aircraft}. This dataset corresponds to aircraft engine (turbofan) run-to-failure simulations under realistic flight scenarios; these simulations are run under the \textit{Commercial Modular Areo-Propulsion System Simulation} (CMAPSS) turbofan model developed at NASA \cite{frederick2007user} using as input real recordings of environmental variables obtained in real commercial flights. The N-CMAPSS dataset is widely used as a benchmark in the PHM community.

\subsection{Dataset Description}
The N-CMAPSS dataset provides 8 sub-datasets denominated from DS01 to DS08; our work focuses on DS04. Each sub-dataset contains simulations for a number of turbofan units that varies according to the sub-dataset. These units are already divided and specified as train or test units. Each turbofan unit is simulated in a sequence of flights of the same class --- short-length, medium-length, and long-length flights; in DS04, there are only units with medium and long-length flights. Each complete flight is referred to as a \textit{cycle} of the unit, which corresponds with the recording of a simulated flight. Each recording is a multivariate series sampled at 1~Hz during the time interval where the altitude of the unit is above 10~000~ft (3048 m). The attributes of these recordings are divided into 4 scenario descriptors, 14 sensor measurements, and 14 virtual measurements. In addition, the dataset provides the ground-truth model health parameters, the remaining useful life (RUL), and auxiliary data such as the flight class and health state.

We discarded the virtual measurements to ensure a realistic application of our methodology; consequently, we built our nominal models and computed the RIVs only with the scenario descriptors and the sensor measurements. In Table~\ref{tab:variables}, we detail the variables we consider in our pipeline.\footnote{LPC, HPC, LPT, and HPT stand for \textit{low pressure compressor}, \textit{high pressure compressor}, \textit{low pressure turbine}, and \textit{high pressure turbine}, respectively.} To describe the application of our methodology, we refer to each variable by either its number ID or its symbol as a subscript of $V$; e.g., we refer to the fuel flow by either $V_{\text{Wf}}$ or $V_5$.

\begin{table*}[ht!]
    \caption{Variables of interest in the N-CMAPSS dataset.}
    \label{tab:variables}
    \begin{center}
    \begin{tabular}{lllll}\toprule
        \textbf{Category} & \textbf{Number ID} & \textbf{Symbol} & \textbf{Description} & \textbf{Unit}\\\midrule
        \multirow{4}{10em}{Scenario descriptor} & 1 & alt & Altitude & ft\\
        & 2 & Mach & Flight Mach number & --\\
        & 3 & TRA & Throttle-resolver angle & \%\\
        & 4 & T2 & Total temperature at fan inlet & $^\circ$R\\\midrule
        \multirow{14}{10em}{Sensor measurements} & 5 & Wf & Fuel flow & pps\\
        & 6 & Nf & Physical fan speed & rpm\\
        & 7 & Nc & Physical core speed & rpm\\
        & 8 & T24 & Total temperature at LPC outlet & $^{\circ}$R\\
        & 9 & T30 & Total temperature at HPC outlet & $^{\circ}$R\\
        & 10 & T48 & Total temperature at HPT outlet & $^{\circ}$R\\
        & 11 & T50 & Total temperature at LPT outlet & $^{\circ}$R\\
        & 12 & P15 & Total pressure in bypass-duct & psia\\
        & 13 & P2 & Total pressure at fan inlet & psia\\
        & 14 & P21 & Total pressure at fan outlet & psia\\
        & 15 & P24 & Total pressure at LPC outlet & psia\\
        & 16 & Ps30 & Static pressure at HPC outlet & psia\\
        & 17 & P40 & Total pressure at burner outlet & psia\\
        & 18 & P50 & Total pressure at LPT outlet & psia\\\bottomrule
    \end{tabular}
    \end{center}
\end{table*}

All units start with a random small degradation and are subjected to increasing noisy degradation throughout their cycles. This degradation can evolve in a normal or abnormal way, following a linear or exponential fashion, respectively. The latter starts at some number of cycles determined by the simulation following criteria regarding energy usage of the unit's subcomponent. In Table~\ref{tab:ds-summary}, we summarize the characteristics of the units simulated in DS04. We refer to the number of cycles occurring prior to the onset of the exponential evolution for the degradation as \textit{healthy flights}.

\begin{table*}[ht!]
    \caption{Summary of the units simulated in sub-dataset DS04.}
    \label{tab:ds-summary}
    \begin{center}
    \begin{tabular}{lllll}\toprule
        \textbf{Division}\hspace{5mm} & \textbf{Flight class}\hspace{5mm} & \textbf{Unit ID}\hspace{5mm} & \textbf{Healthy flights}\hspace{5mm} & \textbf{Total flights}\hspace{5mm}\\\midrule
        \multirow{6}{4em}{Train} & \multirow{2}{7em}{Medium-length} & 1 & 21 & 87\\
        & & 3 & 21 & 100\\\cmidrule{2-5}
        & \multirow{4}{7em}{Long-length} & 2 & 16 & 73\\
        & & 4 & 15 & 69\\
        & & 5 & 16 & 100\\
        & & 6 & 16 & 83\\\midrule
        \multirow{4}{4em}{Test} & \multirow{3}{7em}{Medium-length} & 7 & 21 & 87\\
        & & 8 & 21 & 99\\
        & & 9 & 21 & 73\\\cmidrule{2-5}
        & Long-length & 10 & 16 & 85\\\bottomrule
    \end{tabular}
    \end{center}
\end{table*}

The degradation is parametrized with the \textit{model health parameters} (MHPs). In Table~\ref{tab:degradation}, we detail the available MHPs in the N-CMAPSS dataset (all MHPs are adimensional). We can observe that these MHPs are analogous to the fault parameters ($\delta$) described in Sec.~\ref{sec:experimental}. Hence, we refer to each MHP by its number ID as a subscript of $\delta$, e.g., we refer to the LPT efficiency modifier~as~$\delta_9$.

\begin{table}[ht!]
    \caption{Model health parameters in the N-CMAPSS dataset.}
    \label{tab:degradation}
    \begin{center}
    \begin{tabular}{lll}\toprule
        \textbf{\# ID} & \textbf{Symbol} & \textbf{Description} \\\midrule
        1 & fan\_eff\_mod & Fan efficiency modifier\\
        2 & fan\_flow\_mod & Fan flow modifier\\
        3 & LPC\_eff\_mod & LPC efficiency modifier\\
        4 & LPC\_flow\_mod & LPC flow modifier\\
        5 & HPC\_eff\_mod & HPC efficiency modifier\\
        6 & HPC\_flow\_mod & HPC flow modifier\\
        7 & HPT\_eff\_mod & HPT efficiency modifier\\
        8 & HPT\_flow\_mod & HPT flow modifier\\
        9 & LPT\_eff\_mod & LPT efficiency modifier\\
        10 & LPT\_flow\_mod & LPT flow modifier\\\bottomrule
    \end{tabular}
    \end{center}
\end{table}

\begin{figure}[ht!]
    \centering
    \subfloat[Trajectories for the complete usage of the units.]{\includegraphics[width=8cm]{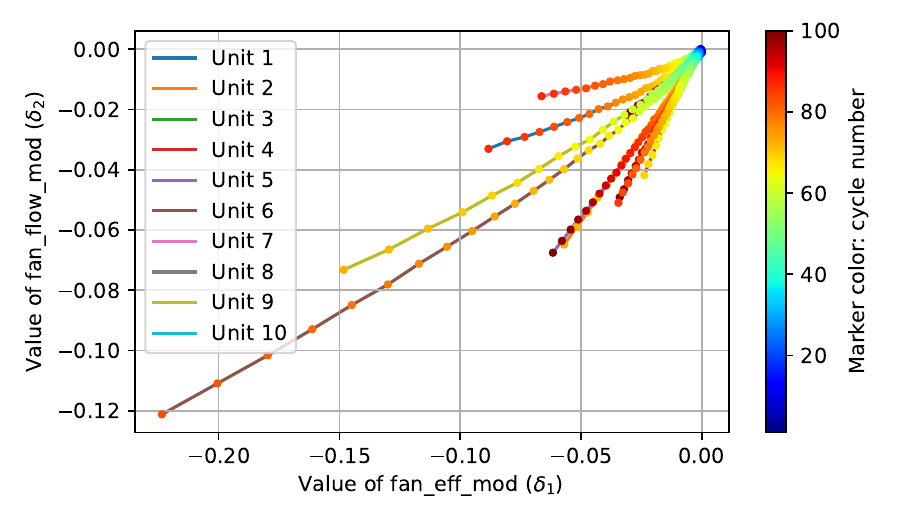}}\\
    \subfloat[Trajectories for the first 50 cycles.]{\includegraphics[width=8cm]{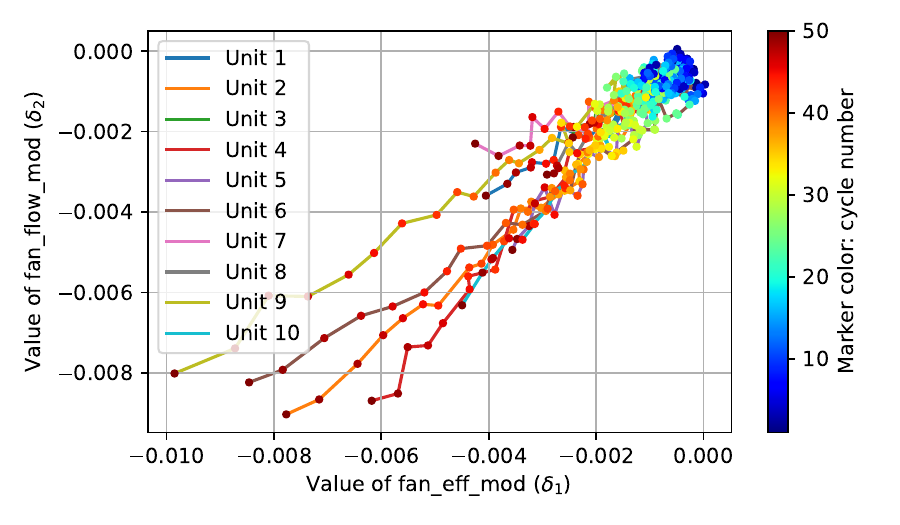}}
    \caption{Trajectories of the fan modifiers ($\delta_1$ and $\delta_2$) over units' usage.}
    \label{fig:delta-trajectory}
\end{figure}

The units of a particular sub-dataset are only subjected to a particular kind of degradation. Units in DS04 are subjected only to fan degradations, i.e., alterations to only fan efficiency and fan flow modifiers. This means that in DS04, all from $\delta_3$ to $\delta_{10}$ are fixed to $0$, and only $\delta_1$ and $\delta_2$ are able to have non-zero values, implying that $(\delta_1,\delta_2)=(0,0)$ represents a perfectly healthy turbofan. In Figure~\ref{fig:delta-trajectory}, we show the trajectory of fan modifiers as the unit's number of cycles increases. We can see a noisy degradation close to $\delta=(0,0)$ over the first flights and a drift away from point $(0,0)$ as the number of cycles increases. In all sub-datasets, the degradation remains constant during the course of a cycle. For the sake of realism, we note that each unit follows its own trajectory, that the degradation is noisy, and that there is no perfectly healthy turbofan in any case.

\subsection{Methodology Specifics for N-CMAPSS}
The first step in applying our methodology is to define the system. We will consider that a turbofan is a system composed of all variables shown in Table~\ref{tab:variables}, i.e., $V\equiv (V_{\ell})_{\ell=1}^{18}$ (taking values in $\mathbb{R}^{18}$) is the r.v. representing the system (see Figure~\ref{sfig:abstract-plant}). Next, we define the input and output. As this selection is left to the user, we will choose an arbitrary variable from Table~\ref{tab:variables} with ID number $k\in\{1,2,\ldots,18\}$ as the target (response or output) and will consider all the rest as explanatory (input). This means that considering $V_k$ as the target, the output is the r.v. $Y_{[k]}\equiv V_k$ (taking values in $\mathbb{R}$), and the input is the r.v. ${X_{[k]}\equiv (X_{j,[k]})_{j=1}^{17} \equiv (V_{\ell})_{\ell\in\{1,2,\ldots,18\}\setminus\{k\}}}$ (taking values in $\mathbb{R}^{17}$) --~see Figure~\ref{sfig:abstract-in-out}.

The next step is to build a nominal model for $Y_{[k]}|X_{[k]}$. For that, we need to obtain an approximation of the MMSE estimator of $Y_{[k]}$ given $X_{[k]}$; we denominate the optimal model as $\eta_{[k]}:\mathbb{R}^{17}\rightarrow\mathbb{R}$. We approximate $\eta_{[k]}(\cdot)$ with a multilayer perceptron (MLP) using data from the first flights of the train units, specifically, the first 15 cycles of units with medium-length flights (IDs 1 and 3) and the first 10 cycles of units with long-length flights (IDs 2, 4, 5, and 6). We used the first cycles to ensure data came from non-abnormal degraded units (see Figure~\ref{fig:delta-trajectory}).

The 18 MLPs used to model each possible target share the same properties and training hyperparameters: each MLP has 4 hidden layers with 1024 hidden units per layer. The input attributes are normalized with a standard scaler. The loss function is the MSE loss. The optimizer is ADAM \cite{KingBa15} with a learning rate of $10^{-6}$ and values of $\beta_1=0.9$ and $\beta_2=0.999$. The training is done with a batch size of 256 samples and early stopping with 32 tolerance epochs and a maximum of 256 epochs using $30\%$ of the nominal data as validation.

Building a nominal model is the only prerequisite of our method. This nominal model is built only with healthy flights of train units; hence, the unsupervised capability of our methodology to detect faults is clear, as there is no data coming from degraded units involved prior to the actual test. This is especially useful in scenarios with a scarcity of faulty observations.

\subsection{Numerical Results --- RIV and Degradation}
\label{sec:riv_results}

We are interested in analyzing how the residual information value (RIV) evolves as the degradation of the units evolves. For this reason, we associate a RIV to each cycle of each unit. This can be done by using the whole recording of that particular cycle and unit to obtain the input and output (which depend on the targeted variable) as a sampling of the system in that particular state. The sampled input is fed to the MLP model to compute the residuals and then to obtain the input-residual EMI, i.e., the RIV (see Figure~\ref{fig:pipeline}) of that particular unit and cycle.

\begin{figure}[ht!]
    \centering
    \subfloat[Unit ID 7.]{\includegraphics[width=4.1cm]{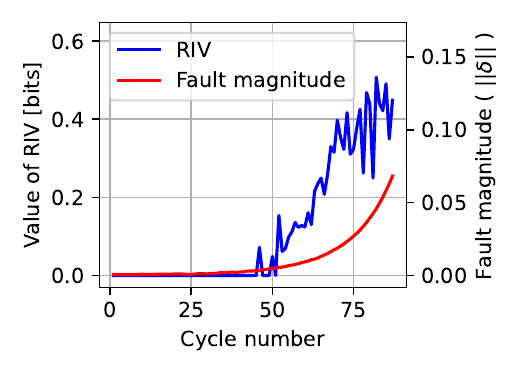}}
    \subfloat[Unit ID 8.]{\includegraphics[width=4.1cm]{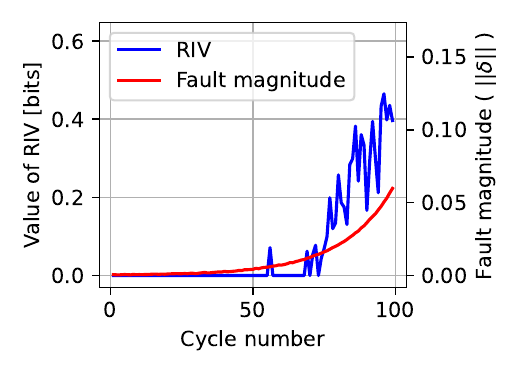}}\\
    \subfloat[Unit ID 9.]{\includegraphics[width=4.1cm]{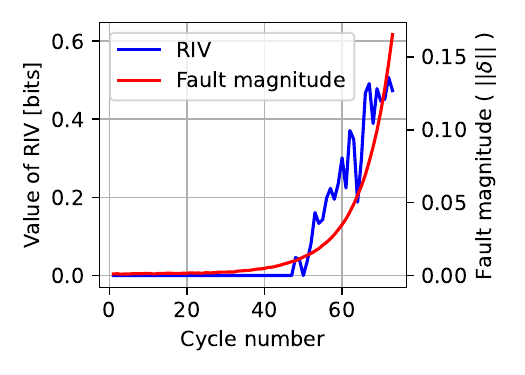}}
    \subfloat[Unit ID 10.]{\includegraphics[width=4.1cm]{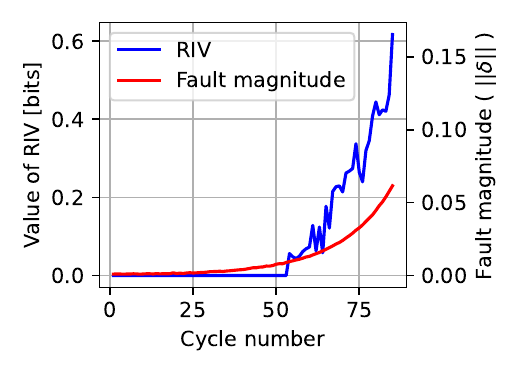}}
    \caption{Degradation magnitude ($||\delta||$) and RIV ($k=15$) evolution over usage for test units.}
    \label{fig:riv-evolution}
\end{figure}

In Figure~\ref{fig:riv-evolution}, we show how the RIV evolves in each test unit for a model with P24 ($k=15$) as its target. We also show how the fault magnitude ($||\delta||$) evolves in comparison. In these examples, a detection is raised when the RIV goes above $0$. This occurs in the neighborhood of cycle number 50 of each unit when the fault magnitude goes above approximately 0.01. Importantly, we observe a clear correlation between the RIV and $||\delta||$.

\begin{figure}[ht!]
    \centering
    \subfloat[Detailed correlation for all models and units.\label{sfig:riv-corr-matrix}]{\includegraphics[width=7.75cm]{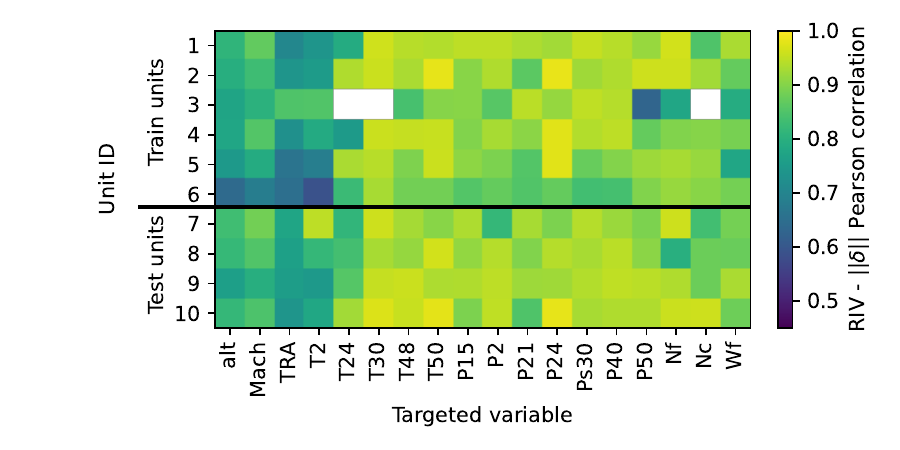}}\\
    \subfloat[Train and test units' average correlation.\label{sfig:riv-corr-bar}]{\includegraphics[width=7.75cm]{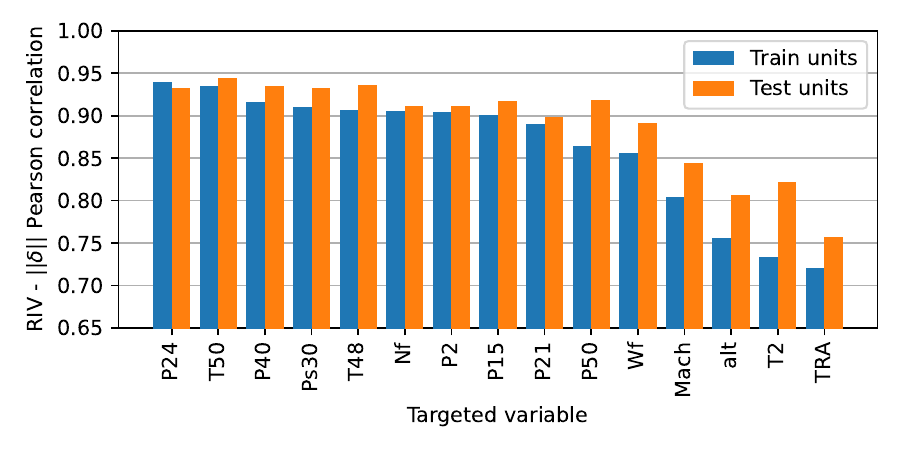}}
    \caption{Pearson correlation between the RIVs and the fault magnitude ($||\delta||$).}
    \label{fig:riv-corr}
\end{figure}

To complement this analysis, in Figure~\ref{fig:riv-corr}, we show how RIVs correlate with fault magnitudes for the different possible models and different units. In Figure~\ref{sfig:riv-corr-matrix}, we show the RIV-degradation correlation for each unit and each possible targeted variable. The white cells in Figure~\ref{sfig:riv-corr-matrix} are undetermined correlations caused by a 0 RIV signal for all cycles of that specific targeted variable and unit. In Figure~\ref{sfig:riv-corr-bar}, we group the correlations by train and test units and show their sorted average; we remove the models for T24, T30, and Nc of this analysis due to their undetermined correlation at unit ID~3. It is easy to see that the election of the target variable is crucial to obtain better RIVs in terms of their correlation with the fault magnitude; in this case, P24 ($k=15$) is the best variable to target as an output. These correlations show the potential of the RIV values to estimate the health state of the system.

\subsection{Beyond Detection --- Residual Information Features (RIFs)}

Although the results shown in Sec.~\ref{sec:riv_results} validate our methodology in a realistic benchmark dataset, these results can be improved. The RIVs are estimations of the MI between $X_{[k]}$ and $R_{[k]} \equiv Y_{[k]}-\hat{\eta}_{[k]}(X_{[k]})$; i.e., an EMI between a 17-dimensional ($X_{[k]}$) and a unidimensional ($R_{[k]}$) r.v. The estimator we are using is tailored for low dimensionalities (see \cite{gonzalez2021indtest}); hence, a straightforward way to enrich our analysis is not to compute $\hat{I}_n(X_{[k]};R_{[k]})$ but to compute $\hat{I}_n(X_{j,[k]};R_{[k]})$ for each ${j\in\{1,2,\ldots,17\}}$. This means computing 17 estimations of MI of univariate pairs instead of a single value of EMI involving a high-dimensionality variable. In practical terms, instead of obtaining a single value, we are obtaining a rich vector; we name this vector as the \textit{residual information feature} (RIF).

\begin{figure*}[ht!]
    \centering
    \subfloat[Unit ID 7.]{\includegraphics[width=8.5cm]{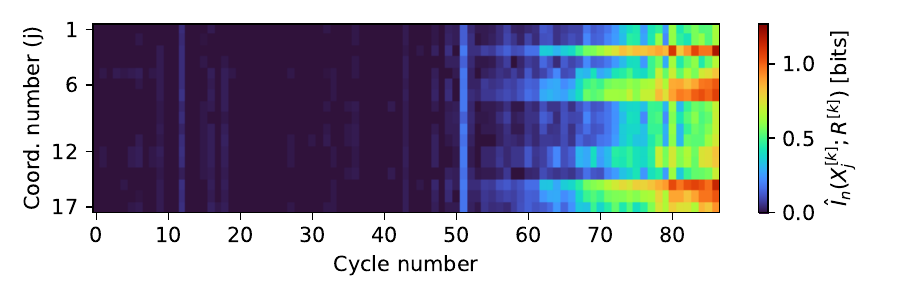}}\hspace{5mm}
    \subfloat[Unit ID 8.]{\includegraphics[width=8.5cm]{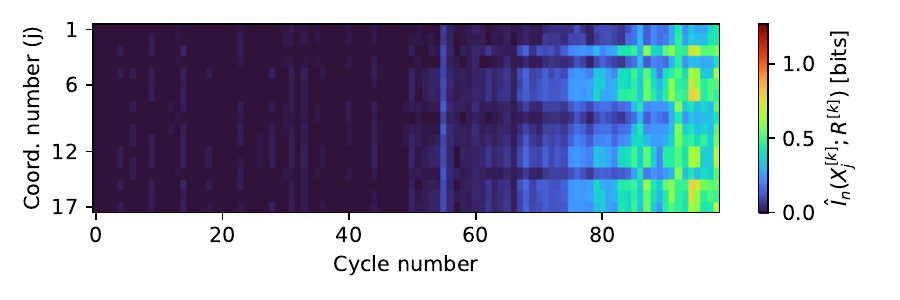}}\\
    \subfloat[Unit ID 9.]{\includegraphics[width=8.5cm]{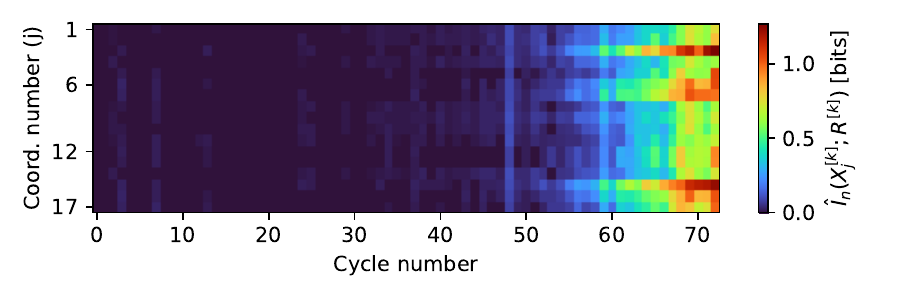}}\hspace{5mm}
    \subfloat[Unit ID 10.]{\includegraphics[width=8.5cm]{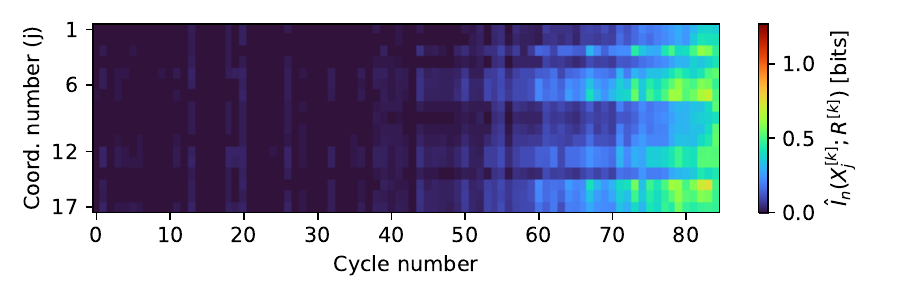}}
    \caption{RIF $(\hat{I}_n(X_{j,[k]};R_{[k]}))_{j=1}^{17}$ evolution for $k=11$ over usage for test units.}
    \label{fig:rif-evolution}
\end{figure*}

\begin{figure}[ht!]
    \centering
    \subfloat[Detailed correlation for all models and units\label{sfig:rif-corr-matrix}.]{\includegraphics[width=7.75cm]{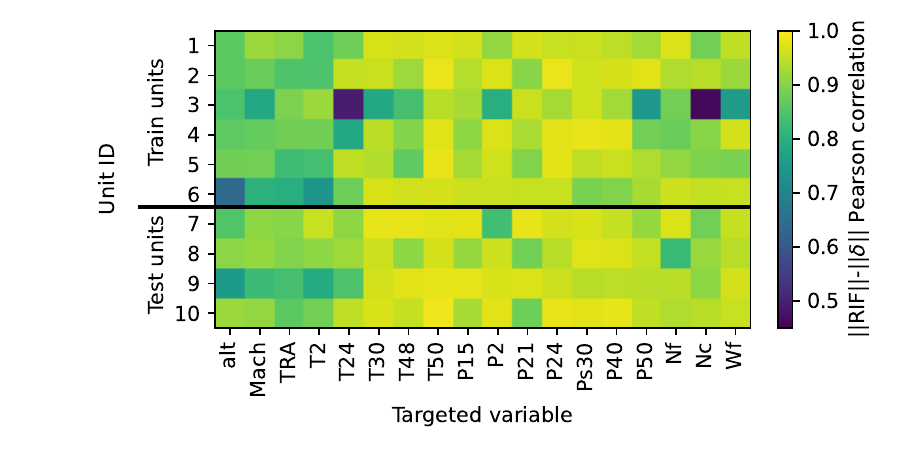}}\\
    \subfloat[Train and test units' average correlation\label{sfig:rif-corr-bar}.]{\includegraphics[width=7.75cm]{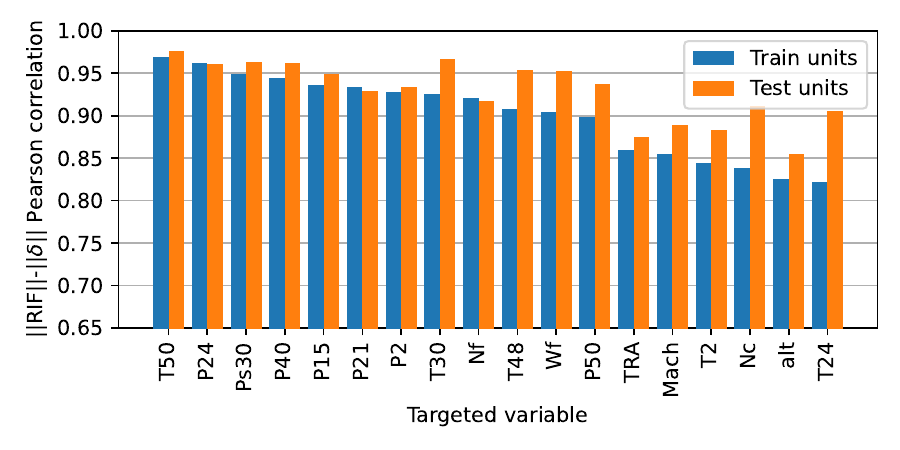}}
    \caption{Pearson correlation between RIF magnitudes and the fault\linebreak magnitudes.}
    \label{fig:rif-corr}
\end{figure}

In Figure~\ref{fig:rif-evolution}, we show how the RIFs evolve in each test unit for a model with T50 ($k=11$) as its target. We can see a clear correlation between the RIF coordinates and the degradation. In Figure~\ref{fig:rif-corr}, we show how RIF magnitudes ($||\text{RIF}||$) correlate with fault magnitudes ($||\delta||$) for the different possible models and different units. In Figure~\ref{sfig:rif-corr-matrix}, we show the $||\text{RIF}||$-$||\delta||$ correlation for each unit and each possible targeted variable, and in Figure~\ref{sfig:rif-corr-bar}, we show the same correlation grouped by train and test units sorted by average. We can make the same observations about these results as in Sec.~\ref{sec:riv_results} with two main differences: We are getting better correlations when computing component-wise EMIs, and we obtain a more detailed description of the fault when computing a feature (RIF) instead of a single value (RIV).

It should be noted that for each particular unit and cycle, we have 18 (one per targeted variable) possible RIFs, i.e., 18 vectors of 17 components; if we concatenate these RIFs, we obtain a 306-dimensional vector describing the fault, i.e., a signature vector. To evaluate the signature capability to describe the fault, we train an MLP regressor using the vector of concatenated RIFs as the input and $||\delta||$ as the target; for this objective, we use all data available from training units. The MLP used in the signature-$||\delta||$ regression has 4 hidden layers with 128 hidden units per layer. The signature coordinates are normalized with a min-max scaler. The loss function is the MSE loss. The optimizer is ADAM with a learning rate of $10^{-6}$, and values of $\beta_1=0.9$ and $\beta_2=0.999$. The training is done with a batch size of 256 samples and early stopping with 1024 tolerance epochs and a maximum of 32~768 epochs using $30\%$ of the data from the training units as validation.

\begin{figure}[ht!]
    \centering
    \subfloat[Unit ID 7.]{\includegraphics[width=4.1cm]{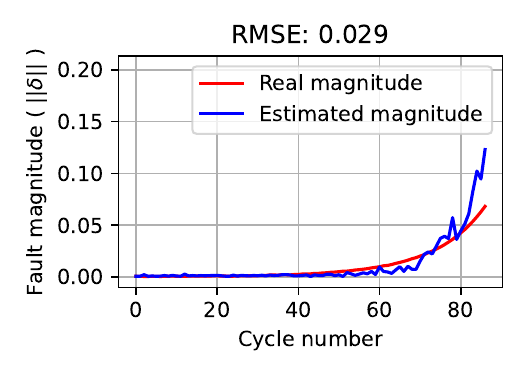}}
    \subfloat[Unit ID 8.]{\includegraphics[width=4.1cm]{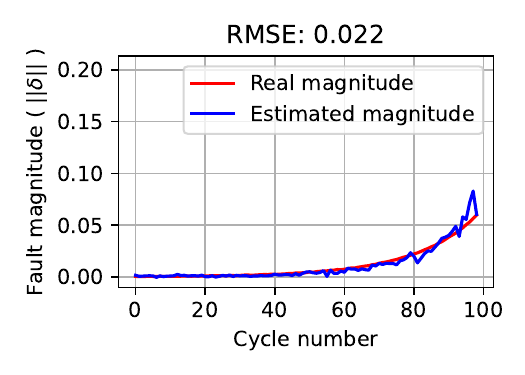}}\\
    \subfloat[Unit ID 9.]{\includegraphics[width=4.1cm]{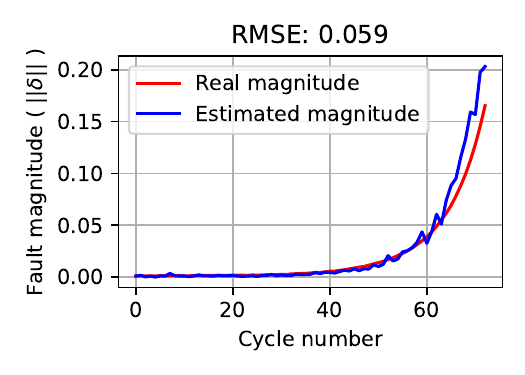}}
    \subfloat[Unit ID 10.]{\includegraphics[width=4.1cm]{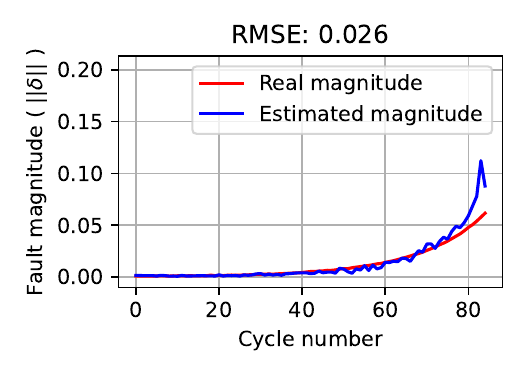}}
    \caption{Real and estimated values of $||\delta||$ using RIF's signatures.}
    \label{fig:delta-estimations}
\end{figure}

In Figure~\ref{fig:delta-estimations}, we show the true (real) and estimated fault magnitude ($||\delta||$) over each test unit usage and the root mean squared error (RMSE) of the estimation. These estimations are made with the signature-to-magnitude MLP regressor mentioned above. Remarkably, here, we can observe the predictive power of our information signatures to describe and estimate the hidden degradation profile of a test turbofan.

\subsection{Final Discussion}

The results in this section validate our formal contributions and synthetic analysis in a realistic benchmark scenario: N-CMAPSS. Moreover, we have built an observable value to describe hidden degradation and incipient faults in turbofans. Better models, such as physics-informed ones, a better tuning of hyperparameters, and a consideration of the time dynamics of the fault could significantly improve these results. This means that even with simple data-driven models and no knowledge of the system dynamics or degradation mechanisms, it is possible to build a successful fault descriptor. To summarize, in addition to fault detection, our methodology can be used to develop diagnosis and prognostic approaches for complex input-output systems.

\section{Summary and Future Perspectives}
\label{sec:final-summary}
In this work, we introduce a novel information- and data-driven fault detection (FD) method and a new theory for model drift detection tailored to systems with continuous input-output relationships within the ANM family. On the theoretical side, we formally state the link between faults and model drift and prove the equivalence between the latter and an independence test. On the practical side, our proposed FD method does not require prior knowledge of faulty data to make accurate decisions and can be applied to continuous phenomena with no restrictions on the regression learning algorithm or expert knowledge 
used to capture its healthy dynamics. The following points summarize the key contributions of our work.

\begin{itemize}
    \item We show that the fault detection task over the rich class of additive noise models can be equivalent to the machine learning problem of testing independence between a regression input and its correspondent regression residual.

    \item We propose a new decision scheme for fault detection using the residual information values (RIVs), which are mutual information estimations.

    \item We propose the use of a distribution-free MI estimator that enables our method to detect any fault --~which corresponds to an unwanted deviation of the inner dynamics (input-output relationship) of the system that affects the data~-- in disregard of its nature and complexity.

    \item We state a range of parameters for the MI estimator to make our method strongly consistent, able to achieve exponentially fast decision under $\mathcal{H}_0$, and able to achieve both test power convergence to 1 and significance level exponential convergence to 0.

    \item We perform numerous experiments on synthetic systems for which we show the discrimination ability of our method for fault detection. Importantly, we observe that our method even works for systems outside the i.i.d. sampling assumption.

    \item We validate our method in the benchmark N-CMAPSS dataset, evidencing its usability in realistic scenarios. In this context, we also show our method's capability of describing the state of health (diagnosis) of the turbofans in terms of information-driven features (RIFs). 
\end{itemize}

These formal, numerical, and empirical results lead us to explore further practical applications for our method, such as benchmarking its capacity to estimate the state of health of other realistic time series systems and working in real-world scenarios. The ability of the residual information features (RIFs) to describe the state of health suggests a variety of next steps to study their practical usage in fault diagnosis and system prognostics. A concrete step is to study our method on systems controlled with adaptive strategies, which are known for their robustness to faults \cite{chowdhury2006fault, wu2024ts}; in particular, we will focus on how our performance results behave in a distributed hypothesis testing scenario where system's data sources are subjected to communication constraints \cite{espinosa2024survey}.

\appendix

\section{Appendix Outline}

In these appendices, we present the proofs for all our theoretical claims, describe the details of the systems involved in the numerical analysis performed on synthetic distributions, and provide further insight into our theoretical assumptions and the experimental results on the synthetic scenarios. On the theoretics, in \ref{sec:general-theorem}, we show and prove Theorem~\ref{the:gen-mdd-dependency}, a generalized version of Theorem~\ref{the:mismodelling-independence-link}, which establishes the relationship between model drift and input-residual dependency; this proof is supported with the aid of Lemma~\ref{lem:degenerate-rv}, whose proof is shown in \ref{sec:lemma-proof}. In \ref{sec:proof-mismodelling-independence-link}, we prove Theorem~\ref{the:mismodelling-independence-link} as a corollary of its extended version. In \ref{sec:proof-strong-consistency}, \ref{sec:proof-finite-nominal-time}, and \ref{sec:proof-hypothesis-test-metrics}, we prove Theorem~\ref{the:strong-consistency} (strong consistency), \ref{the:finite-nominal-time} (exponentially-fast decision), and \ref{the:hypothesis-test-metrics} (error rate guarantees), respectively, and in \ref{sec:generality}, we justify the generality of our assumptions. On the numerical analysis, in \ref{sec:exp-desc}, we provide a full description of the systems and methods employed in the results shown in Sec.~\ref{sec:experimental}; and in \ref{sec:synthetic-error-bar}, to enhance our analysis, we provide an error-bar analysis that showcases the consistency of the asymptotic and finite-length properties proved for our method.

\section{Generalized Link between Fault Existence and Input-Residual Dependency}
\label{sec:general-theorem}

In this appendix, we show a generalized version of Theorem~\ref{the:mismodelling-independence-link}. This generalized theorem links fault existence and input-residual dependency even outside our assumptions of the absence of bias, i.e., assumptions (i) and (ii) of our \textit{standard assumptions} (see Sec.~\ref{sec:md-ir-dependency}). Before making the statement of this generalized theorem, we need to introduce a broader version of Def.~\ref{def:model-almost-equivalence}, which enables us to work with the class of equivalence we denote as \textit{perfectly biased almost-surely equivalent} models. This definition captures the idea of models that are equivalent in the sense of Def.~\ref{def:model-almost-equivalence} but have a \textit{perfect bias} between them, i.e., an offset.

\begin{definition}
    \label{def:model-b-equivalence}
    Let $\eta_1:\mathbb{R}^p\rightarrow\mathbb{R}^q$ and $\eta_2:\mathbb{R}^p\rightarrow\mathbb{R}^q$ be measurable functions, and $\mathbf{c}\in\mathbb{R}^q$. $\eta_1(\cdot)$ and $\eta_2(\cdot)$ are said to be \textit{almost-surely equivalent w.r.t. $\mu$ with a bias $\mathbf{c}$}, denoted by $\eta_1\approx_{\mu}^\mathbf{c}\eta_2$, if for every r.v. $X\sim\mu$ it holds that $\eta_1(X)\overset{\text{a.s.}}{=}\eta_2(X)+\mathbf{c}$. If there does not exist any $\mathbf{c}\in\mathbb{R}^q$ such that $\eta_1\approx_{\mu}^\mathbf{c}\eta_2$, it is denoted as $\eta_1\not\approx_{\mu}\eta_2$.
\end{definition}

Now, equipped with Def.~\ref{def:model-b-equivalence}, we can state the general version of Theorem~\ref{the:mismodelling-independence-link} as follows.

\begin{theorem}
    \label{the:gen-mdd-dependency}
    Let $X$ and $Y$ be r.v.s with values in $\mathbb{R}^p$ and $\mathbb{R}^q$, respectively, such that $(X,Y)\sim\textup{add}(\tilde{\eta},\tilde{h};\mu)$, and let ${\eta:\mathbb{R}^p\rightarrow\mathbb{R}^q}$ be a measurable function. The following properties are true:
    \begin{itemize}
        \item[(i)] If there exists $\mathbf{c}\in\mathbb{R}^q$ such that $\tilde{\eta}\approx_{\mu}^\mathbf{c}\eta$, then $X$ and $Y-\eta(X)$ are independent.
        \item[(ii)] If $\tilde{\eta}\not\approx_{\mu}\eta$, and both r.v.s. $(X,Y-\eta(X))$ and $(X,\tilde{h}(W))$ have densities, then $X$ and $Y-\eta(X)$ \textup{are not} independent.
    \end{itemize}
\end{theorem}

\begin{remark}
    We highlight that property (i) of Theorem~\ref{the:gen-mdd-dependency} does not require any r.v. to have densities. Then, it would remain true even in a classification context where $Y$ is a discrete random variable. The extension of our theory to discrete-output systems would require rethinking the properties of the MMSE estimator we exploited in this work to consider the properties of a classifier that optimizes a classification loss, such as the cross-entropy. The idea is to use this classifier to capture the nominal dynamics of the discrete-output system.
\end{remark}

\begin{prfthfive}
We divide this proof into proving each property (i) and (ii) that constitute Theorem~\ref{the:gen-mdd-dependency}.

\begin{itemize}
    \item \textsc{Proof of Th.~\ref{the:gen-mdd-dependency}} (i): This proof relies on the definition of independence between r.v.s, this means, that we prove independence between $X$ and $Y-\eta(X)$ by proving that $\forall A\in\mathcal{B}(\mathbb{R}^p),\forall B\in\mathcal{B}(\mathbb{R}^q)$,
    \begin{align}
        \label{eqn:th2-i-initial}
        &\mathbb{P}(X\in A,Y-\eta(X)\in B)\nonumber\\
        =\:&\mathbb{P}(X\in A)\cdot\mathbb{P}(Y-\eta(X)\in B).
    \end{align}
    
    Let us start this proof by noting that $\tilde{\eta}\approx_{\mu}^\mathbf{c}\eta$, from Def.~\ref{def:model-b-equivalence}, implies that for every random variable $\tilde{X}\sim\mu$: ${\tilde{\eta}(\tilde{X})\overset{\text{a.s.}}{=}\eta(\tilde{X})+\mathbf{c}}$. This, in particular, is true for the marginal input of $(X,Y)\sim\text{add}(\tilde{\eta},\tilde{h};\mu)$; i.e., $X\sim\mu$ satisfies that ${\tilde{\eta}(X)\overset{\text{a.s.}}{=}\eta(X)+\mathbf{c}}$.
    
    To continue this proof, we define the following events in~$\Sigma$:\footnote{We are denoting the probability space of all the r.v.s involved in this work as $(\Omega,\Sigma,\mathbb{P})$.}\textsuperscript{,}\footnote{Note that, for the events of interest, ${(X,Y)\sim\text{add}(\tilde{\eta},\tilde{h};\mu) \Leftrightarrow \mathbb{P}(\mathcal{E}^{\text{add}})=1}$ and ${\tilde{\eta}\approx_{\mu}^\mathbf{c}\eta \Leftrightarrow \mathbb{P}(\mathcal{E}_\mathbf{c}^{\text{eq}})=1}$.}
    \begin{align}
        \mathcal{E}^{\text{add}} &\equiv \{\omega\in\Omega:Y(\omega) = \tilde{\eta}(X(\omega)) + \tilde{h}(W(\omega))\},\label{eqn:event-add}\\
        \mathcal{E}_\mathbf{c}^{\text{eq}} &\equiv \{\omega\in\Omega:\tilde{\eta}(X(\omega)) = \eta(X(\omega))+\mathbf{c}\},\label{eqn:event-as-eq}
    \end{align} 
    and then, we have the following sequence of equalities for any arbitrary sets $A\in\mathcal{B}(\mathbb{R}^p)$ and $B\in\mathcal{B}(\mathbb{R}^q)$:
    \begin{align}
            &\mathbb{P}(X\in A,Y-\eta(X)\in B)\nonumber\\
            =\:&\mathbb{P}(X\in A,Y-\eta(X)\in B,\mathcal{E}^{\text{add}})\label{eqn:prf-gen-i-1-eq-1}\\
            =\:&\mathbb{P}(X\in A,\tilde{\eta}(X)+\tilde{h}(W)-\eta(X)\in B,\mathcal{E}^{\text{add}})\label{eqn:prf-gen-i-1-eq-2}\\
            =\:&\mathbb{P}(X\in A,\tilde{\eta}(X)+\tilde{h}(W)-\eta(X)\in B)\label{eqn:prf-gen-i-1-eq-3}\\
            =\:&\mathbb{P}(X\in A,\tilde{\eta}(X)+\tilde{h}(W)-\eta(X)\in B,\mathcal{E}_\mathbf{c}^{\text{eq}})\label{eqn:prf-gen-i-1-eq-4}\\
            =\:&\mathbb{P}(X\in A,\eta(X)+\mathbf{c}+\tilde{h}(W)-\eta(X)\in B,\mathcal{E}_\mathbf{c}^{\text{eq}})\label{eqn:prf-gen-i-1-eq-5}\\
            =\:&\mathbb{P}(X\in A,\tilde{h}(W)+\mathbf{c}\in B)\label{eqn:prf-gen-i-1-eq-6},
    \end{align}
    where (\ref{eqn:prf-gen-i-1-eq-1}), (\ref{eqn:prf-gen-i-1-eq-3}), (\ref{eqn:prf-gen-i-1-eq-4}), and (\ref{eqn:prf-gen-i-1-eq-6}) are a consequence of the following property, $\forall\mathcal{E}_1\in\Sigma,\forall\mathcal{E}_2\in\Sigma$:
    \begin{align}
        \label{eqn:as-event-property}
        \mathbb{P}(\mathcal{E}_1)=1\Rightarrow\mathbb{P}(\mathcal{E}_1,\mathcal{E}_2)=\mathbb{P}(\mathcal{E}_2);
    \end{align}
    Equation (\ref{eqn:prf-gen-i-1-eq-2}) and (\ref{eqn:prf-gen-i-1-eq-5}) are consequences of (\ref{eqn:event-add}) and (\ref{eqn:event-as-eq}), respectively. As $X$ and $W$ are independent, we have that $X$ and $\tilde{h}(W)$ are independent. Moreover, if we define the set $B_{-\mathbf{c}}\equiv\{\mathbf{x}\in\mathbb{R}^q:\mathbf{x}+\mathbf{c}\in B\}$, we can see that
    \begin{align}
        &\mathbb{P}(X\in A,Y-\eta(X)\in B)\nonumber\\
        =\:&\mathbb{P}(X\in A,\tilde{h}(W)+\mathbf{c}\in B)\label{eqn:th2-1-1st-eq-prev-1}\\
        =\:&\mathbb{P}(X\in A,\tilde{h}(W)\in B_{-\mathbf{c}})\label{eqn:th2-1-1st-eq-prev-2}\\
        =\:&\mathbb{P}(X\in A)\cdot\mathbb{P}(\tilde{h}(W)\in B_{-\mathbf{c}}),\label{eqn:th2-1-1st-eq}
    \end{align}
    where (\ref{eqn:th2-1-1st-eq-prev-1}), (\ref{eqn:th2-1-1st-eq-prev-2}), and (\ref{eqn:th2-1-1st-eq}) come from (\ref{eqn:prf-gen-i-1-eq-6}), the definition of $B_{-\mathbf{c}}$, and the independence between $X$ and $\tilde{h}(W)$, respectively. Developing further the term $\mathbb{P}(\tilde{h}(W)\in B_{-\mathbf{c}})$, we can see that
    \begin{align}
        &\mathbb{P}(\tilde{h}(W)\in B_{-\mathbf{c}})\nonumber\\
        =\:&\mathbb{P}(\tilde{h}(W)+\mathbf{c}\in B)\label{eqn:th2-1-2nd-eq-prev-1}\\
        =\:&\mathbb{P}(\tilde{h}(W)+\mathbf{c}\in B,\mathcal{E}_\mathbf{c}^{\text{eq}})\label{eqn:th2-1-2nd-eq-prev-2}\\
        =\:&\mathbb{P}(\tilde{h}(W)+\tilde{\eta}(X)-\eta(X)\in B,\mathcal{E}_\mathbf{c}^{\text{eq}})\label{eqn:th2-1-2nd-eq-prev-3}\\
        =\:&\mathbb{P}(\tilde{\eta}(X)+\tilde{h}(W)-\eta(X)\in B)\label{eqn:th2-1-2nd-eq-prev-4}\\
        =\:&\mathbb{P}(\tilde{\eta}(X)+\tilde{h}(W)-\eta(X)\in B,\mathcal{E}^{\text{add}})\label{eqn:th2-1-2nd-eq-prev-5}\\
        =\:&\mathbb{P}(Y-\eta(X)\in B,\mathcal{E}^{\text{add}})\label{eqn:th2-1-2nd-eq-prev-6}\\
        =\:&\mathbb{P}(Y-\eta(X)\in B),\label{eqn:th2-1-2nd-eq}
    \end{align}
    where (\ref{eqn:th2-1-2nd-eq-prev-2}), (\ref{eqn:th2-1-2nd-eq-prev-4}), (\ref{eqn:th2-1-2nd-eq-prev-5}), and (\ref{eqn:th2-1-2nd-eq}) are a consequence of (\ref{eqn:as-event-property}); Equation (\ref{eqn:th2-1-2nd-eq-prev-3}) and (\ref{eqn:th2-1-2nd-eq-prev-6}) come from (\ref{eqn:event-as-eq}) and (\ref{eqn:event-add}), respectively; and (\ref{eqn:th2-1-2nd-eq-prev-1}) is a consequence of the definition of $B_{-\mathbf{c}}$. Then, from (\ref{eqn:th2-1-1st-eq}) and (\ref{eqn:th2-1-2nd-eq}), we get that
    \begin{align}
        &\mathbb{P}(X\in A,Y-\tilde{\eta}(X)\in B)\nonumber\\
        =\:&\mathbb{P}(X\in A)\cdot\mathbb{P}(\tilde{h}(W)\in B_{-\mathbf{c}})\\
        =\:&\mathbb{P}(X\in A)\cdot\mathbb{P}(Y-\eta(X)\in B).\label{eqn:th2-i-final}
    \end{align}

    Finally, we can see that (\ref{eqn:th2-i-final}) is precisely what we wanted to show, as stated in (\ref{eqn:th2-i-initial}), proving in this way the property (i) of Theorem~\ref{the:gen-mdd-dependency}.

    \item \textsc{Proof of Th.~\ref{the:gen-mdd-dependency}} (ii): We prove point (ii) by contradiction; in particular, we show how assuming both $\tilde{\eta}\not\approx_{\mu}\eta$ and the independence between $X$ and $Y-\eta(X)$ leads to a contradiction. This contradiction has to do with the impossibility of the existence of probability density functions (pdfs) for the continuous r.v.s $(X,Y-\eta(X))$ and $(X,\tilde{h}(W))$ under the stated assumptions.

    Let us start this proof by considering ${f_{X,R}:\mathbb{R}^{p+q}\rightarrow[0,\infty)}$ and $f_{X,\tilde{h}(W)}:\mathbb{R}^{p+q}\rightarrow[0,\infty]$ as pdfs of $(X,Y-\eta(X))$ and $(X,\tilde{h}(W))$, respectively, and let us consider $f_R:\mathbb{R}^q\rightarrow[0,\infty)$ as the marginal pdf of $Y-\eta(X)$ induced from $f_{X,R}(\cdot)$ and $f_{\tilde{h}(W)}:\mathbb{R}^q\rightarrow[0,\infty)$ as the marginal pdf of $\tilde{h}(W)$ induced from $f_{X,\tilde{h}(W)}(\cdot,\cdot)$.

    Let us define the function $D_{\eta,\tilde{\eta}}:\mathbb{R}^p\rightarrow\mathbb{R}^q$ such that $D_{\eta,\tilde{\eta}}(\mathbf{x}) = \eta(\mathbf{x})-\tilde{\eta}(\mathbf{x}),\forall\mathbf{x}\in\mathbb{R}^p$. We can observe that $(X,Y)\sim\text{add}(\tilde{\eta},\tilde{h};\mu)$ implies that ${(X,Y)\overset{\text{a.s.}}{=}(X,\tilde{\eta}(X)+\tilde{h}(W))}$, which in turn, implies that $(X,Y-\eta(X))\overset{\text{a.s.}}{=}(X,\tilde{\eta}(X)+\tilde{h}(W)-\eta(X)) =$ ${(X,\tilde{h}(W)-D_{\eta,\tilde{\eta}}(X))}$. Then, as $f_{X,R}(\cdot,\cdot)$ is a pdf of ${(X,Y-\eta(X))}$, it also is a pdf of $(X,\tilde{h}(W)-D_{\eta,\tilde{\eta}}(X))$.

    Let us consider an arbitrary set $C\in\mathcal{B}(\mathbb{R}^{p+q})$, then we can define a set $\tilde{C}$ expressed by
    \begin{equation}
        \tilde{C} \equiv \{(\mathbf{x},\mathbf{r}+D_{\eta,\tilde{\eta}}(\mathbf{x})):\mathbf{x}\in\mathbb{R}^p,\mathbf{r}\in\mathbb{R}^q,(\mathbf{x},\mathbf{r})\in C\}.
    \end{equation}
    This definition satisfies that $\forall\mathbf{x}\in\mathbb{R}^p,\forall\mathbf{r}\in\mathbb{R}^q$, ${[(\mathbf{x},\mathbf{r})\in C\Leftrightarrow(\mathbf{x},\mathbf{r}+D_{\eta,\tilde{\eta}}(\mathbf{x}))\in\tilde{C}]}$; then, we have the following event equality:
    \begin{align}
        &\{\omega\in\Omega:(X(\omega),\tilde{h}(W(\omega))-D_{\eta,\tilde{\eta}}(X(\omega)))\in C\}\nonumber\\
        =\:&\{\omega\in\Omega:(X(\omega),\tilde{h}(W(\omega)))\in\tilde{C}\},
    \end{align}
    and in consequence
    \begin{equation}
        \label{eqn:prf-5-ii-prob-eq}
        \mathbb{P}((X,\tilde{h}(W)-D_{\eta,\tilde{\eta}}(X))\in C) = \mathbb{P}((X,\tilde{h}(W))\in\tilde{C}).
    \end{equation}
    Hence, if we define $\tilde{f}:\mathbb{R}^{p+q}\rightarrow[0,\infty)$ as a function such that $\tilde{f}(\mathbf{x},\mathbf{r})=f_{X,\tilde{h}(W)}(\mathbf{x},\mathbf{r}+D_{\eta,\tilde{\eta}}(\mathbf{x}))$, $\forall\mathbf{x}\in\mathbb{R}^p$, $\forall\mathbf{r}\in\mathbb{R}^q$, we can see that
    \begin{align}
        &\mathbb{P}((X,\tilde{h}(W)-D_{\eta,\tilde{\eta}}(X))\in C)\nonumber\\
        =\:&\mathbb{P}((X,\tilde{h}(W))\in\tilde{C})\label{eqn:pdf-def-1}\\
        =\:&\int_{\tilde{C}}f_{X,\tilde{h}(W)}(\mathbf{x},\mathbf{h})\:\mathrm{d}(\mathbf{x},\mathbf{h})\label{eqn:pdf-def-2}\\
        =\:&\int_{\tilde{C}}\tilde{f}(\mathbf{x},\mathbf{h}-D_{\eta,\tilde{\eta}}(\mathbf{x}))\:\mathrm{d}(\mathbf{x},\mathbf{h})\label{eqn:pdf-def-3}\\
        =\:&\int_C\tilde{f}(\mathbf{x},\mathbf{r})\:\mathrm{d}(\mathbf{x},\mathbf{r}),\label{eqn:pdf-def-4}
    \end{align}
    where (\ref{eqn:pdf-def-1}) and (\ref{eqn:pdf-def-3}) come from (\ref{eqn:prf-5-ii-prob-eq}) and the definition of $\tilde{f}(\cdot,\cdot)$, respectively; hence, $\tilde{f}(\cdot,\cdot)$ is a pdf of $(X,\tilde{h}(W)-D_{\eta,\tilde{\eta}}(X))$, and in consequence, a pdf of ${(X,Y-\eta(X))}$. This lets us state that the following equalities hold $\lambda^{p+q}$-almost everywhere:\footnote{$\lambda^{p+q}$ is the Lebesgue measure of $(\mathbb{R}^{p+q},\mathcal{B}(\mathbb{R}^{p+q}))$.}
    \begin{align}
        &f_{X,R}(\mathbf{x},\mathbf{r})\nonumber\\
        =\:&\tilde{f}(\mathbf{x},\mathbf{r})\label{eqn:proof-th2-ii-pdfs-prev-1}\\
        =\:&f_{X,\tilde{h}(W)}(\mathbf{x},\mathbf{r}+D_{\eta,\tilde{\eta}}(\mathbf{x})),\label{eqn:proof-th2-ii-pdfs}
    \end{align}
    where (\ref{eqn:proof-th2-ii-pdfs-prev-1}) and (\ref{eqn:proof-th2-ii-pdfs}) come from both $f_{X,R}(\cdot,\cdot)$ and $\tilde{f}(\cdot,\cdot)$ being pdfs of $(X,Y-\eta(X))$, and the definition of $\tilde{f}(\cdot,\cdot)$, respectively. This means that if we define ${E\subseteq\mathbb{R}^{p+q}}$ as the set of points where the pdfs are equal, this is,
    \begin{equation}
        E \equiv \left\{(\mathbf{x},\mathbf{r})\in\mathbb{R}^{p+q}:\hspace{-2mm}\begin{array}{l}
        \mathbf{x}\in\mathbb{R}^p,\mathbf{r}\in\mathbb{R}^q,\\
        f_{X,R}(\mathbf{x},\mathbf{r})=f_{X,\tilde{h}(W)}(\mathbf{x},\mathbf{r}+D_{\eta,\tilde{\eta}}(\mathbf{x}))
        \end{array}\hspace{-2mm}\right\},
    \end{equation}
    we get that $E^{\mathsf{c}}\equiv\mathbb{R}^{p+q}\setminus E\in\mathcal{B}(\mathbb{R}^{p+q})$ is a $\lambda^{p+q}$-null set, i.e., $\lambda^{p+q}(E^{\mathsf{c}}) = 0$.

    One of the assumptions that lead to a contradiction is the independence between $X$ and $Y-\eta(X)$; this independence implies that
    \begin{equation}
        \label{eqn:proof-th2-x-res-pdf}
        \forall\mathbf{x}\in\mathbb{R}^p,\forall\mathbf{r}\in\mathbb{R}^q,f_{X,R}(\mathbf{x},\mathbf{r}) = f_X(\mathbf{x})\cdot f_R(\mathbf{r}).
    \end{equation}
    Then, we get the following $\forall(\mathbf{x},\mathbf{r})\in E$:
    \begin{align}
        &f_X(\mathbf{x})\cdot f_R(\mathbf{r})\nonumber\\
        =\:&f_{X,R}(\mathbf{x},\mathbf{r})\label{eqn:proof-th2-pdf-gen-after-ind-prev-1}\\
        =\:&f_{X,\tilde{h}(W)}(\mathbf{x},\mathbf{r}+D_{\eta,\tilde{\eta}}(\mathbf{x}))\label{eqn:proof-th2-pdf-gen-after-ind-prev-2}\\
        =\:&f_X(\mathbf{x})\cdot f_{\tilde{h}(W)}(\mathbf{r}+D_{\eta,\tilde{\eta}}(\mathbf{x})),\label{eqn:proof-th2-pdf-gen-after-ind}
    \end{align}
    where (\ref{eqn:proof-th2-pdf-gen-after-ind-prev-1}), (\ref{eqn:proof-th2-pdf-gen-after-ind-prev-2}), and (\ref{eqn:proof-th2-pdf-gen-after-ind}) are a consequence of (\ref{eqn:proof-th2-x-res-pdf}), (\ref{eqn:proof-th2-ii-pdfs}), and the independence between $X$ and $\tilde{h}(W)$, respectively. Moreover, let us define $\forall\mathbf{x}\in\mathbb{R}^p$, the sets $E_{\mathbf{x}}\equiv\{\mathbf{r}\in\mathbb{R}^q:(\mathbf{x},\mathbf{r})\in E\}$ and $E_{\mathbf{x}}^{\mathsf{c}}\equiv\mathbb{R}^q\setminus E_{\mathbf{x}}$. We can see that $E_{\mathbf{x}}^{\mathsf{c}} = \{\mathbf{r}\in\mathbb{R}^q:(\mathbf{x},\mathbf{r})\in E^{\mathsf{c}}\},\forall\mathbf{x}\in\mathbb{R}^p$. Then, as $E^{\mathsf{c}}$ is a $\lambda^{p+q}$-measurable set, from Fubini's theorem (see \cite[Th.~3.4.1.]{bogachev2007measure}), we get that $\lambda^p$-almost every set $E_{\mathbf{x}}^{\mathsf{c}}$ is $\lambda^q$-measurable and that
    \begin{equation}
        \lambda^{p+q}(E^{\mathsf{c}}) = \int_{\mathbb{R}^p}\lambda^q(E_{\mathbf{x}}^{\mathsf{c}})\:\mathrm{d}\mathbf{x},\label{eqn:prf-fubinni}
    \end{equation}
    where $\lambda^p$ and $\lambda^q$ are the Lebesgue measures of $(\mathbb{R}^p,\mathcal{B}(\mathbb{R}^p))$ and $(\mathbb{R}^q,\mathcal{B}(\mathbb{R}^q))$, respectively.
    
    In addition, as $\lambda^{p+q}(E^{\mathsf{c}})=0$, if we define a set
    \begin{equation}
        \label{eqn:pdf-eq-marginal-set}
        M\equiv\{\mathbf{x}\in\mathbb{R}^p:\lambda^q(E_{\mathbf{x}}^{\mathsf{c}})=0\}\in\mathcal{B}(\mathbb{R}^p),
    \end{equation}
    we get from (\ref{eqn:prf-fubinni}) that $\lambda^{p}(M^{\mathsf{c}})=0$, where ${M^{\mathsf{c}}\equiv\mathbb{R}^p\setminus M=\{\mathbf{x}\in\mathbb{R}^p:\lambda^q(E_{\mathbf{x}}^{\mathsf{c}})>0\}\in\mathcal{B}(\mathbb{R}^p)}$. This means that there are $\lambda^p$-almost none points $\mathbf{x}\in\mathbb{R}^p$ such that the set of points $\mathbf{r}\in\mathbb{R}^q$ where (\ref{eqn:proof-th2-pdf-gen-after-ind}) does not hold has a $\lambda^q$-measure greater than 0.

    Before continuing this proof, we need to state the following lemma about degenerated random variable distributions. This lemma is related to the concept of \textit{perfectly biased} equivalence we introduce in Def.~\ref{def:model-b-equivalence}.

    \begin{lemma}\footnote{The proof of Lemma~\ref{lem:degenerate-rv} is presented in \ref{sec:lemma-proof}.}
        \label{lem:degenerate-rv}
        Let $U$ be a r.v. taking values in $\mathbb{R}^r$ with ${r\in\mathbb{N}}$. If for every $S\in\mathcal{B}(\mathbb{R}^r)$ it is satisfied that ${\mathbb{P}(U\in S)\in\{0,1\}}$, then there exists $\mathbf{c}\in\mathbb{R}^r$ such that $\mathbb{P}(U=\mathbf{c})=1$.
    \end{lemma}

    Regarding the other assumption that leads to a contradiction, let us note that $\tilde{\eta}\not\approx_{\mu}\eta$, from Def.~\ref{def:model-b-equivalence},
    corresponds to the non-existence of $\mathbf{c}$ such that $\tilde{\eta}\approx_{\mu}^\mathbf{c}\eta$, or which is equivalent, that $\forall\mathbf{c}\in\mathbb{R}^q$, there exists a r.v. $\tilde{X}\sim\mu$ such that $\mathbb{P}(\tilde{\eta}(\tilde{X})=\eta(\tilde{X})+\mathbf{c})<1$. Then, since $X$ has the same distribution as $\tilde{X}$, we have that $\forall \mathbf{c}\in\mathbb{R}^q,\mathbb{P}(\tilde{\eta}(X)=\eta(X)+\mathbf{c})<1$, or equivalently, that $\nexists\mathbf{c}\in\mathbb{R}^q:D_{\eta,\tilde{\eta}}(X)\overset{\text{a.s.}}{=}\mathbf{c}$. Due to the contrapositive of Lemma~\ref{lem:degenerate-rv}, the nonexistence of $\mathbf{c}\in\mathbb{R}^q$ such that $\mathbb{P}(D_{\eta,\tilde{\eta}}(X)=\mathbf{c})=1$ implies that the following proposition is false: $[\mathbb{P}(D_{\eta,\tilde{\eta}}(X)\in S)\in\{0,1\},\forall S\in\mathcal{B}(\mathbb{R}^q)]$; or which is equivalent, that there exists a set $G\in\mathcal{B}(\mathbb{R}^q)$ such that $\mathbb{P}(D_{\eta,\tilde{\eta}}(X)\in G)\in(0,1)$, which in turn, implies the existence of a set $H=\mathbb{R}^q\setminus G\in\mathcal{B}(\mathbb{R}^q)$ which also satisfies that $\mathbb{P}(D_{\eta,\tilde{\eta}}(X)\in H)=1-\mathbb{P}(D_{\eta,\tilde{\eta}}(X)\in G)\in(0,1)$.

    In addition, let us note the following event equality:
    \begin{align}
        &\{\omega\in\Omega:D_{\eta,\tilde{\eta}}(X(\omega))\in G\}\nonumber\\
        =\:&\{\omega\in\Omega:X(\omega)\in D^{-1}_{\eta,\tilde{\eta}}(G)\};
    \end{align}
    this enables us to state that there exists a set ${A=D_{\eta,\tilde{\eta}}^{-1}(G)\in\mathcal{B}(\mathbb{R}^p)}$ such that ${\mathbb{P}(X\in A) = \mathbb{P}(D_{\eta,\tilde{\eta}}(X)\in G) > 0}$ and, analogously, that there exists a set ${B=D_{\eta,\tilde{\eta}}^{-1}(H)\in\mathcal{B}(\mathbb{R}^p)}$ such that ${\mathbb{P}(X\in B)=\mathbb{P}(D_{\eta,\tilde{\eta}}(X)\in H)>0}$.

    Let us retrieve set $M$ from (\ref{eqn:pdf-eq-marginal-set}), we can see that $\lambda^p(M^{\mathsf{c}})=0$ implies that $\mathbb{P}(X\in M^{\mathsf{c}})=0$. In consequence, as $A\cap M^{\mathsf{c}}\subseteq M^{\mathsf{c}}$, it holds that ${\mathbb{P}(X\in A\cap M^{\mathsf{c}}) = 0}$, and therefore, we can observe that ${\mathbb{P}(X\in A\cap M) =}$ ${\mathbb{P}(X\in A) - \mathbb{P}(X\in A\cap M^{\mathsf{c}}) =}$ $\mathbb{P}(X\in A)>0$; analogously, we have that $\mathbb{P}(X\in B\cap M) = \mathbb{P}(X\in B)>0$.

    Furthermore, let us consider $f_X:\mathbb{R}^p\rightarrow[0,\infty)$ as a pdf induced by the marginalization of $X$ from either a density of $(X,Y-\eta(X))$ or $(X,h(W))$. We can see that $\exists\mathbf{a}\in A\cap M$ such that $f_X(\mathbf{a})>0$.\footnote{$\exists\mathbf{a}\in A\cap M:f_X(\mathbf{a})>0$ is true, because otherwise, we would have that $\forall\mathbf{x}\in A\cap M,f_X(\mathbf{x})=0$, which would imply that ${\mathbb{P}(X\in A\cap M) = \int_{A\cap M}f_X(\mathbf{x})\:\mathrm{d}\mathbf{x} = \int_{A\cap M}0\:\mathrm{d}\mathbf{x} = 0}$; i.e., it would contradict that $\mathbb{P}(X\in A\cap M)>0$.} Moreover, it is possible to see that evaluating $\mathbf{x}=\mathbf{a}$ in (\ref{eqn:proof-th2-pdf-gen-after-ind}) and dividing the terms of the equality by $f_X(\mathbf{a})>0$, we get that
    \begin{equation}
    \label{eqn:marginal-eq-a}
        \forall\mathbf{r}\in E_{\mathbf{a}},f_R(\mathbf{r}) = f_{\tilde{h}(W)}(\mathbf{r}+D_{\eta,\tilde{\eta}}(\mathbf{a})).
    \end{equation}
    Analogously, we can see that $\exists\mathbf{b}\in B\cap M$, such that $f_X(\mathbf{b})>0$ and
    \begin{equation}
        \label{eqn:marginal-eq-b}
        \forall\mathbf{r}\in E_{\mathbf{b}},f_R(\mathbf{r}) = f_{\tilde{h}(W)}(\mathbf{r}+D_{\eta,\tilde{\eta}}(\mathbf{b})).
    \end{equation}
    From (\ref{eqn:marginal-eq-a}) and (\ref{eqn:marginal-eq-b}), we get the following:
    \begin{equation}
        \label{eqn:pdf-marginals}
        \forall\mathbf{r}\in E_{\mathbf{a},\mathbf{b}},f_{\tilde{h}(W)}(\mathbf{r}+D_{\eta,\tilde{\eta}}(\mathbf{a})) = f_{\tilde{h}(W)}(\mathbf{r}+D_{\eta,\tilde{\eta}}(\mathbf{b})),
    \end{equation}
    where $E_{\mathbf{a},\mathbf{b}}\equiv E_{\mathbf{a}}\cap E_{\mathbf{b}}\in\mathcal{B}(\mathbb{R}^q)$. Since ${\mathbf{a}\in M}$ and ${\mathbf{b}\in M}$, we get that ${\lambda^q(E_{\mathbf{a}}^{\mathsf{c}})=\lambda^q(E_{\mathbf{b}}^{\mathsf{c}})=0}$; then, for ${E_{\mathbf{a},\mathbf{b}}^{\mathsf{c}}\equiv\mathbb{R}^q\setminus E_{\mathbf{a},\mathbf{b}} = E_{\mathbf{a}}^{\mathsf{c}}\cup E_{\mathbf{b}}^{\mathsf{c}}\in\mathcal{B}(\mathbb{R}^q)}$, we get that  $\lambda^q(E_{\mathbf{a},\mathbf{b}}^{\mathsf{c}})=0$ due to the inequality ${\lambda^q(E_{\mathbf{a},\mathbf{b}}^{\mathsf{c}}) \leq \lambda^q(E_{\mathbf{a}}^{\mathsf{c}})+\lambda^q(E_{\mathbf{b}}^{\mathsf{c}}) = 0}$. In consequence, the equality shown in (\ref{eqn:pdf-marginals}), i.e., a periodicity for the pdf $f_{\tilde{h}(W)}(\cdot)$, is true $\lambda^q$-almost everywhere.

    Let us define $\mathbf{p}=(p_j)_{j=1}^q\in\mathbb{R}^q$ as $\mathbf{p}\equiv D_{\eta,\tilde{\eta}}(\mathbf{a})-D_{\eta,\tilde{\eta}}(\mathbf{b})$. Due to $\mathbf{a}\in A=D^{-1}_{\eta,\tilde{\eta}}(G)$ and $\mathbf{b}\in B=D^{-1}_{\eta,\tilde{\eta}}(H)$, we have that $D_{\eta,\tilde{\eta}}(\mathbf{a})\in G$ and $D_{\eta,\tilde{\eta}}(\mathbf{b})\in H$; then, as ${G\cap H=\emptyset}$, we have that $D_{\eta,\tilde{\eta}}(\mathbf{a})\neq D_{\eta,\tilde{\eta}}(\mathbf{b})$ and, in consequence, ${\mathbf{p}\neq\mathbf{0}\in\mathbb{R}^q}$. If we define the set of points $\tilde{E}\equiv\{\mathbf{h}\in\mathbb{R}^q:\mathbf{h}-D_{\eta,\tilde{\eta}}(\mathbf{b})\in E_{\mathbf{a},\mathbf{b}}\}\in\mathcal{B}(\mathbb{R}^q),$ we get that $\forall\mathbf{h}\in\tilde{E},\mathbf{h}-D_{\eta,\tilde{\eta}}(\mathbf{b})\in E_{\mathbf{a},\mathbf{b}}$; hence,
    \begin{equation}
        \forall\mathbf{h}\in\tilde{E},f_{\tilde{h}(W)}(\mathbf{h}) = f_{\tilde{h}(W)}(\mathbf{h}+\mathbf{p}).
    \end{equation}
    We can observe for ${\tilde{E}^{\mathsf{c}}\equiv\mathbb{R}^q\setminus\tilde{E}=}$ ${\{\mathbf{h}\in\mathbb{R}^q:\mathbf{h}-D_{\eta,\tilde{\eta}}(\mathbf{b})\in E_{\mathbf{a},\mathbf{b}}^{\mathsf{c}}\}\in\mathcal{B}(\mathbb{R}^q)}$ that its $\lambda^q$-measure is $\lambda^q(\tilde{E}^{\mathsf{c}}) = \lambda^q(E_{\mathbf{a},\mathbf{b}}^{\mathsf{c}})=0$.

    In addition, if for an arbitrary $j\in\mathbb{Z}$ we define ${\tilde{E}_j\equiv\{\mathbf{h}\in\mathbb{R}^q:\mathbf{h}+j\mathbf{p}\in\tilde{E}\}\in\mathcal{B}(\mathbb{R}^{q})}$, we can see that $\tilde{E}_j^{\mathsf{c}}\equiv\mathbb{R}^q\setminus\tilde{E}_j=\{\mathbf{h}\in\mathbb{R}^q:\mathbf{h}+j\mathbf{p}\in\tilde{E}^{\mathsf{c}}\}\in\mathcal{B}(\mathbb{R}^q)$ has a $\lambda^q$-measure of $\lambda^q(\tilde{E}_j^{\mathsf{c}})=\lambda^q(\tilde{E}^{\mathsf{c}})=0$ and that $\forall\mathbf{h}\in\tilde{E}_j,\mathbf{h}+j\mathbf{p}\in\tilde{E}$; hence,
    \begin{equation}
        \label{eqn:periodicity-inductive}
        \forall\mathbf{h}\in\tilde{E}_j,f_{\tilde{h}(W)}(\mathbf{h}+j\mathbf{p})=f_{\tilde{h}(W)}(\mathbf{h}+(j+1)\mathbf{p}).
    \end{equation}
    Moreover, if we define $\tilde{E}_{\mathbb{Z}}\equiv\bigcap_{k\in\mathbb{Z}}\tilde{E}_k\in\mathcal{B}(\mathbb{R}^q)$, its complement is $\tilde{E}_{\mathbb{Z}}^{\mathsf{c}}\equiv\mathbb{R}^q\setminus\tilde{E}_{\mathbb{Z}} = \bigcup_{k\in\mathbb{Z}}\tilde{E}_k^{\mathsf{c}}\in\mathcal{B}(\mathbb{R}^q)$. As $\tilde{E}_{\mathbb{Z}}^{\mathsf{c}}$ is the result of a countable union, we get that its $\lambda^q$-measure is $\lambda^q(\tilde{E}_{\mathbb{Z}}^{\mathsf{c}})=0$ due to the inequality ${\lambda^q(\tilde{E}_{\mathbb{Z}}^{\mathsf{c}}) \leq \sum_{k\in\mathbb{Z}}\lambda^q(\tilde{E}_k^{\mathsf{c}})=\sum_{k\in\mathbb{Z}}0=0}$. Let us note that we can apply (\ref{eqn:periodicity-inductive}) inductively to obtain the following property:
    \begin{equation}
        \forall\mathbf{h}\in\tilde{E}_{\mathbb{Z}},[\forall k\in\mathbb{Z},f_{\tilde{h}(W)}(\mathbf{h})=f_{\tilde{h}(W)}(\mathbf{h}+k\mathbf{p})].
    \end{equation}

    Let us note that $\mathbf{p}=(p_j)_{j=1}^q\neq\mathbf{0}\in\mathbb{R}^q$ implies the existence of $\ell\in\{1,2,\ldots,q\}$, a dimension index, such that $p_{\ell}\neq0\in\mathbb{R}$. Then, according to $\ell$, it is possible to define a partition $\mathcal{R}\equiv\{r_k\}_{k\in\mathbb{Z}}\subseteq\mathcal{B}(\mathbb{R}^q)$ of $\mathbb{R}^q$ such that $\forall k\in\mathbb{Z}$, $r_k\equiv\{\mathbf{h}=(h_j)_{j=1}^q\in\mathbb{R}^q:h_{\ell}\in[kp_{\ell},(k+1)p_{\ell})\}$. We can observe that $\forall k\in\mathbb{Z}$, $[\forall\mathbf{h}\in r_0,\mathbf{h}+k\mathbf{p}\in r_k]$; hence, we obtain the following property for every $k\in\mathbb{Z}$:
    \begin{align}
        &\mathbb{P}(\tilde{h}(W)\in r_k)\nonumber\\
        =\:&\int_{r_k}f_{\tilde{h}(W)}(\tilde{\mathbf{h}})\:\mathrm{d}\tilde{\mathbf{h}}\label{eqn:proof-th2-ii-patch-eq-prev-1}\\
        =\:&\int_{r_0}f_{\tilde{h}(W)}(\mathbf{h}+k\mathbf{p})\:\mathrm{d}\mathbf{h}\label{eqn:proof-th2-ii-patch-eq-prev-2}\\
        =\:&\int_{r_0\cap\tilde{E}_{\mathbf{Z}}^{\mathsf{c}}}f_{\tilde{h}(W)}(\mathbf{h}+k\mathbf{p})\:\mathrm{d}\mathbf{h} + \int_{r_0\cap\tilde{E}_{\mathbb{Z}}}f_{\tilde{h}(W)}(\mathbf{h}+k\mathbf{p})\:\mathrm{d}\mathbf{h}\label{eqn:proof-th2-ii-patch-eq-prev-3}\\
        =\:&0 + \int_{r_0\cap\tilde{E}_{\mathbb{Z}}}f_{\tilde{h}(W)}(\mathbf{h})\:\mathrm{d}\mathbf{h}\label{eqn:proof-th2-ii-patch-eq-prev-4}\\
        =\:&\int_{r_0\cap\tilde{E}_{\mathbb{Z}}^{\mathsf{c}}}f_{\tilde{h}(W)}(\mathbf{h})\:\mathrm{d}\mathbf{h} + \int_{r_0\cap\tilde{E}_{\mathbb{Z}}}f_{\tilde{h}(W)}(\mathbf{h})\:\mathrm{d}\mathbf{h}\label{eqn:proof-th2-ii-patch-eq-prev-5}\\
        =\:&\int_{r_0}f_{\tilde{h}(W)}(\mathbf{h})\:\mathrm{d}\mathbf{h}\label{eqn:proof-th2-ii-patch-eq-prev-6}\\
        =\:&\mathbb{P}(\tilde{h}(W)\in r_0)\label{eqn:proof-th2-ii-patch-eq}
    \end{align}
    where (\ref{eqn:proof-th2-ii-patch-eq-prev-1}) and (\ref{eqn:proof-th2-ii-patch-eq}) come from $f_{\tilde{h}(W)}(\cdot)$ being a pdf of $\tilde{h}(W)$, (\ref{eqn:proof-th2-ii-patch-eq-prev-3}) and (\ref{eqn:proof-th2-ii-patch-eq-prev-6}) come from the disjoint union $r_0=(r_0\cap\tilde{E}_{\mathbb{Z}})\cup(r_0\cap\tilde{E}_{\mathbb{Z}}^{\mathsf{c}})$, (\ref{eqn:proof-th2-ii-patch-eq-prev-4}) and (\ref{eqn:proof-th2-ii-patch-eq-prev-5}) come from $\lambda^q(\tilde{E}_{\mathbb{Z}}^{\mathsf{c}})=0$, and (\ref{eqn:proof-th2-ii-patch-eq-prev-2}) comes from a change of variable.
    
    Equation~(\ref{eqn:proof-th2-ii-patch-eq}) implies that ${\mathbb{P}(\tilde{h}(W)\in\mathbb{R}^q) =}$ ${\sum_{k\in\mathbb{Z}}\mathbb{P}(\tilde{h}(W)\in r_k) = \sum_{k\in\mathbb{Z}}\mathbb{P}(\tilde{h}(W)\in r_0)}$. Then, there are only two possible cases, and both lead to a contradiction.

    In the first case, $\mathbb{P}(\tilde{h}(W)\in r_0)=0$; this implies that $\sum_{k\in\mathbb{Z}}\mathbb{P}(\tilde{h}(W)\in r_0) = 0$, which contradicts ${\mathbb{P}(\tilde{h}(W)\in\mathbb{R}^q)=1}$. In the second case $\mathbb{P}(\tilde{h}(W)\in r_0)>0$; this implies that $\sum_{k\in\mathbb{Z}}\mathbb{P}(\tilde{h}(W)\in r_k)$ diverges, which also contradicts $\mathbb{P}(\tilde{h}(W)\in\mathbb{R}^q)=1$.

    In conclusion, we have shown that under the hypotheses of property (ii) of Theorem~\ref{the:gen-mdd-dependency}, assuming both $\tilde{\eta}\not\approx_{\mu}\eta$ and the independence between $X$ and $Y-\eta(X)$ leads to a contradiction. This leads us to prove that under the mentioned hypotheses, $\tilde{\eta}\not\approx_{\mu}\eta$ implies that $X$ and $Y-\eta(X)$ are not independent, proving in this way the property (ii) of Theorem~\ref{the:gen-mdd-dependency}.

\end{itemize}

Having demonstrated both properties (i) and (ii) of Theorem~\ref{the:gen-mdd-dependency}, we conclude our proof for Theorem~\ref{the:gen-mdd-dependency}.\hfill$\square$
\end{prfthfive}

\section{Proof of Theorem~\ref{the:mismodelling-independence-link} --- Fault Existence and Input-Residual Dependency Equivalence}
\label{sec:proof-mismodelling-independence-link}

In this appendix, we prove Theorem~\ref{the:mismodelling-independence-link}. Within this proof, we can observe that Theorem~\ref{the:mismodelling-independence-link} is a corollary of Theorem~{\ref{the:gen-mdd-dependency}}, which corresponds to a broader version of Theorem~\ref{the:mismodelling-independence-link}.

\begin{prfthone}
As Theorem~\ref{the:mismodelling-independence-link} states an equivalence relationship, in this proof, we will prove both implications that constitute the equivalence.

For the first implication (which corresponds to ${[\tilde{\eta}\simeq_{\mu}\eta] \Rightarrow [\text{$X$ and $Y-\eta(X)$ are independent}]}
$), the proof is straightforward. We can see that $\tilde{\eta}\simeq_{\mu}\eta$ (Def.~\ref{def:model-almost-equivalence}) is equivalent to $\tilde{\eta}\approx_{\mu}^{\mathbf{0}}\eta$ (Def.~\ref{def:model-b-equivalence}); then, there exists $\mathbf{c}=\mathbf{0}\in\mathbb{R}^q$ such that $\tilde{\eta}\approx_{\mu}^{\mathbf{c}}\eta$ and, in consequence, due to Theorem~\ref{the:gen-mdd-dependency} point (i), $X$ and $Y-\eta(X)$ are independent.

For the second implication (which corresponds to $[\text{$X$ and $Y-\eta(X)$ are independent}] \Rightarrow [\tilde{\eta}\simeq_{\mu}\eta]$), we need to look at the contrapositive of Theorem~\ref{the:gen-mdd-dependency} point (ii); this is, that the independence between $X$ and $Y-\eta(X)$ implies that either $\tilde{\eta}\not\approx_{\mu}\eta$ is false or that the existence of the densities of $(X,Y-\eta(X))$ and $(X,\tilde{h}(W))$ is false. As the existence of the densities is ensured by our \textit{standard assumptions} --~point (iii)~-- we are only left with the fact that $\tilde{\eta}\not\approx_{\mu}\eta$ is false.

The falsehood of $\tilde{\eta}\not\approx_{\mu}\eta$ implies that there exists a value $\mathbf{c}\in\mathbb{R}^q$ such that $\tilde{\eta}\approx_{\mu}^\mathbf{c}\eta$. Then, as $X\sim\mu$, we have that $\tilde{\eta}(X)\overset{\text{a.s.}}{=}\eta(X)+\mathbf{c}$. Moreover, as $(X,Y)\sim\text{add}(\tilde{\eta},\tilde{h};\mu)$, we have that $(X,Y)\overset{\text{a.s.}}{=}(X,\tilde{\eta}(X)+\tilde{h}(W))$; hence
\begin{align}
    &[\tilde{\eta}(X)\overset{\text{a.s.}}{=} \eta(X)+\mathbf{c}] \:\wedge\:[(X,Y)\overset{\text{a.s.}}{=}(X,\tilde{\eta}(X)+\tilde{h}(W))]\nonumber\\
    \Rightarrow\:&(X,Y)\overset{\text{a.s.}}{=}(X,\eta(X)+\tilde{h}(W)+\mathbf{c}).\label{eqn:th2-prof-biased-aeq}
\end{align}
Then, let us note the following:
\begin{align}
    &\mathbb{E}[Y-\eta(X)]\nonumber\\
    =\:&\mathbb{E}[\eta(X)+\tilde{h}(W)+\mathbf{c}-\eta(X)]\\
    =\:&\mathbb{E}[\tilde{h}(W)] + \mathbf{c}.\label{eqn:bias-equivalence}
\end{align}
Since our \textit{standard assumptions} consider $\mathbb{E}[\tilde{h}(W)]=\mathbf{0}\in\mathbb{R}^q$ and $\mathbb{E}[Y-\eta(X)]=\mathbf{0}\in\mathbb{R}^q$ (points (i) and (ii), respectively), by replacing these values in (\ref{eqn:bias-equivalence}), we get that $\mathbf{c}=\mathbf{0}$. This, in turn, implies that $\tilde{\eta}\approx_{\mu}^{\mathbf{0}}\eta$ (Def.~\ref{def:model-b-equivalence}) or, equivalently, that $\tilde{\eta}\simeq_{\mu}\eta$ (Def.~\ref{def:model-almost-equivalence}).

Finally, we can see that our proof concludes at this point, as we have proven the equivalence relationship stated in Theorem~\ref{the:mismodelling-independence-link} by proving the two implications that constitute it. \hfill $\square$
\end{prfthone}

\section{Proof of Theorem~\ref{the:strong-consistency} --- Strong Consistency}
\label{sec:proof-strong-consistency}

\begin{prfthtwo}
Let us note that every decision scheme ${\Psi_{\mathbf{b},\mathbf{d},\mathbf{a}}^{\lambda} = \left(\psi_{b_n,d_n,a_n}^{\lambda, n}(\cdot)\right)_{n\in\mathbb{N}}\in\Psi_{\text{SC}}}$ follows the construction shown in Sec.~\ref{sec:the_detector}; in consequence, $\Psi_{\mathbf{b},\mathbf{d},\mathbf{a}}^{\lambda}$ is a sequence of decision rules such that $\forall n\in\mathbb{N}$, ${\forall \mathbf{z}_n=(x_j,y_j)_{j=1}^n\in(\mathbb{R}^{p+q})^n}$,
\begin{equation}
    \psi_{b_n,d_n,a_n}^{\lambda, n}(\mathbf{z}_n) = 1_{[a_n,\infty)}\left(I_{b_n,d_n}^{\lambda, n}(\mathbf{j}_n)\right),
\end{equation}
where $\mathbf{j}_n\equiv(x_j,y_j-\eta(x_j))_{j=1}^n$. Let us recall that $I_{b_n,d_n}^{\lambda,n}(\mathbf{j}_n)$ is an estimation of the MI between $X$ and $Y-\eta(X)$ as $\mathbf{j}_n$ consists of $n$ i.i.d. realizations of $(X,Y-\eta(X))$. Hence, $\psi_{b_n,d_n,a_n}^{\lambda,n}(\mathbf{z}_n)$ is the output of a decision rule of length $n$ for testing independence between $X$ and $Y-\eta(X)$, and consequently, $\Psi_{\mathbf{b},\mathbf{d},\mathbf{a}}^{\lambda}$ is a decision scheme for testing independence between the mentioned r.v.s. The usage of the estimator presented in \cite{silva2012complexity} for testing independence was studied by \cite{gonzalez2021indtest}.

From (\ref{eqn:strong_family}), let us notice that any $\Psi_{\mathbf{b},\mathbf{d},\mathbf{a}}^{\lambda}\in\Psi_{\text{SC}}$ satisfies that $\mathbf{b}\approx(n^{-l})_{n\in\mathbb{N}}$ for $l\in(0,1/3)$, $\mathbf{d}\in\ell_1(\mathbb{N})$, $1/\mathbf{d} \in O(\text{exp}(n^{1/3}))$, $\mathbf{a}\in o(1)$, and $\lambda\in(0,\infty)$. Hence, from \cite[Th.~3]{gonzalez2021indtest}, $\Psi_{\mathbf{b},\mathbf{d},\mathbf{a}}^{\lambda}$ is strongly consistent (see \cite[Def.~1]{gonzalez2021indtest}) for detecting independence between $X$ and $Y-\eta(X)$, which means that
\begin{align}
    &[\text{$X$ and $Y-\eta(X)$ are independent}]\nonumber\\
    \Rightarrow\:&\mathbb{P}\left(\lim_{n\rightarrow\infty}\psi_{b_n,d_n,a_n}^{\lambda,n}(\mathbf{Z}_n)=0\right)=1,\label{eqn:proof-ii-imp-i}\\
    &[\text{$X$ and $Y-\eta(X)$ are \text{not} independent}]\nonumber\\
    \Rightarrow\:&\mathbb{P}\left(\lim_{n\rightarrow\infty}\psi_{b_n,d_n,a_n}^{\lambda,n}(\mathbf{Z}_n)=1\right)=1.\label{eqn:proof-ii-imp-ii}
\end{align}
Moreover, as we are under our \textit{standard assumptions} and $(X,Y)\sim\text{add}(\tilde{\eta},\tilde{h};\mu)$, we get from Theorem~\ref{the:mismodelling-independence-link} the equivalence
\begin{equation}
    [\tilde{\eta}\simeq_{\mu}\eta] \Leftrightarrow [\text{$X$ and $Y-\eta(X)$ are independent}];\label{eqn:proof-ii-eq}
\end{equation}
hence, as we can see in (\ref{eqn:hypothesis-test}), Sec.~\ref{sec:mismodelling-test-def}, our null hypothesis ($\mathcal{H}_0:\tilde{\eta}\simeq_{\mu}\eta$) is equivalent to the input-residual independence hypothesis, and our alternate hypothesis ($\mathcal{H}_1:\tilde{\eta}\not\simeq_{\mu}\eta$) is equivalent to the input-residual non-independence hypothesis. In consequence, from (\ref{eqn:proof-ii-imp-i}), (\ref{eqn:proof-ii-imp-ii}), and (\ref{eqn:proof-ii-eq}),
\begin{equation}
    \label{eqn:proof-th3-strong-consistency}
    \mathbb{P}\left(\lim_{n\rightarrow\infty}\left.\psi_{b_n,d_n,a_n}^{\lambda,n}(\mathbf{Z}_n) = i\,\right|\,\mathcal{H}_i\right) = 1,\forall i\in\{0,1\}.
\end{equation}

Equation (\ref{eqn:proof-th3-strong-consistency}) shows that any decision scheme ${\Psi_{\mathbf{b},\mathbf{d},\mathbf{a}}^{\lambda}\in\Psi_{\text{SC}}}$ satisfies (\ref{eqn:strong-consistency}), i.e., Def.~\ref{def:strong-consistency}; hence, any decision scheme in $\Psi_{\text{SC}}$ is strongly consistent in detecting faults, concluding our proof for Theorem~\ref{the:strong-consistency}. \hfill $\square$
\end{prfthtwo}

\section{Proof of Theorem~\ref{the:finite-nominal-time} --- Exponentially-Fast Decision under \texorpdfstring{$\mathcal{H}_0$}{H0}}
\label{sec:proof-finite-nominal-time}

\begin{prfththree}
Every decision scheme $\Psi\in\Psi_{\text{FL}}$ follows the construction shown in Sec.~\ref{sec:the_detector}; consequently, its parameters can be written explicitly as $\Psi\equiv\Psi_{\mathbf{b},\mathbf{d},\mathbf{a}}^{\lambda}=(\psi_{b_n,d_n,a_n}^{\lambda,n}(\cdot))_{n\in\mathbb{N}}$, and as shown in the proof of Theorem~\ref{the:strong-consistency}, it corresponds to a decision scheme for testing independence between $X$ and $Y-\eta(X)$.

From (\ref{eqn:finite-nominal-time}), let us notice that any $\Psi\equiv\Psi_{\mathbf{b},\mathbf{d},\mathbf{a}}^{\lambda}\in\Psi_{\text{FL}}$ satisfies that $\mathbf{b}\approx(n^{-l})_{n\in\mathbb{N}}$ for $l\in(0,1/3)$,$\mathbf{d}\approx(\exp(n^{-1/3}))_{n\in\mathbb{N}}$, $\mathbf{a}\in o(1)$, and $\lambda\in(0,\infty)$. Hence, from \cite[Th.~4]{gonzalez2021indtest}, we can see that
\begin{align}
    &\text{$X$ and $Y-\eta(X)$ are independent}\nonumber\\
    \Rightarrow\:&\forall m\in\mathbb{N},\mathbb{P}(\mathcal{C}_0^{\Psi}(\mathbf{J}_{\infty})\geq m)\leq K\text{exp}(-m^{1/3}),\label{eqn:proof-th4-implication}
\end{align}
for some universal constant $K\in(0,\infty)$, where $\mathbf{J}_{\infty}$ denotes $(X_j,Y_j-\eta(X_j))_{j=1}^{\infty}$ and, from \cite[Def.~3]{gonzalez2021indtest}, $\mathcal{C}_0^{\Psi}(\mathbf{J}_{\infty})$ corresponds to the random value that expresses when the tree-structured partition used for estimating the MI between $X$ and $Y-\eta(X)$ using the parameters of $\Psi\equiv\Psi_{\mathbf{b},\mathbf{d},\mathbf{a}}^{\lambda}$ ($\mathcal{A}=\{A_{\ell}\}_{\ell\in\Lambda}$; see Sec.~\ref{sec:mi-estimator}, Stage~one) collapses to the trivial partition $\{\mathbb{R}^{p+q}\}$,\footnote{This collapsing time functional we denote as $\mathcal{C}_0^{\Phi}(\cdot)$ is denoted as $\mathcal{T}_0(\cdot)$ in \cite{gonzalez2021indtest} (let us recall that we are denoting arbitrary decision schemes as $\Phi$ and the decision schemes induced by our pipeline introduced in Sec.~\ref{sec:the_detector} as $\Psi$). We use the notation $\mathcal{C}_0^{\Phi}(\cdot)$ to avoid confusion with our functional $\mathcal{T}_0^{\Phi}$ (Def.~\ref{def:detection-time}).}\textsuperscript{,}\footnote{Let us note that $[\mathcal{C}_0(\mathbf{J}_{\infty})\geq m]$ denotes an event in which either $\mathcal{C}_0(\mathbf{J}_{\infty})$ exists and is greater or equal than $m$, or $\mathcal{C}_0(\mathbf{J}_{\infty})$ does not exist; this same idea applies to the notation $[\mathcal{T}_0^{\Phi}(\mathbf{Z}_{\infty})\geq m]$.} which in turn implies an MI estimation of $0$.

As we have shown in the proof of Theorem~\ref{the:strong-consistency}, both $(X,Y)\sim\text{add}(\tilde{\eta},\tilde{h};\mu)$ and being under our \textit{standard assumptions}, imply from Theorem~\ref{the:mismodelling-independence-link} that ${[\tilde{\eta}\simeq_{\mu}\eta]\Leftrightarrow[\text{$X$ and $Y-\eta(X)$ are independent}]}$. Hence, as our null hypothesis ($\mathcal{H}_0$) corresponds to $\tilde{\eta}\simeq_{\mu}\eta$ (see (\ref{eqn:hypothesis-test}), Sec.~\ref{sec:mismodelling-test-def}), the consequence of the implication shown in (\ref{eqn:proof-th4-implication}) reduces to
\begin{equation}
    \label{eqn:proof-th4-exp-bound}
    \forall m\in\mathbb{N},\mathbb{P}(\mathcal{C}_0^{\Psi}(\mathbf{J}_{\infty})\geq m\,|\,\mathcal{H}_0) \leq K\text{exp}(-m^{1/3}).
\end{equation}

Let us consider an arbitrary $m\in\mathbb{N}$, then we have that $\mathcal{C}_0^{\Psi}(\mathbf{j}_{\infty})=m\Rightarrow \forall n>m,I_{b_n,d_n}^{\lambda,n}(\mathbf{j}_n) = 0$;\footnote{Here, we denote by $\mathbf{z}_n$ and $\mathbf{j}_n$, the realizations of $\mathbf{Z}_n = (X_j,Y_j)_{j=1}^n$ and $\mathbf{J}_n = (X_j,Y_j-\eta(X_j))_{j=1}^n$, respectively. We extend this notation to $n=\infty$.} hence, ${\forall n>m,\forall \mathbf{a}=(a_n)_{n\in\mathbb{N}}\in o(1),I_{b_n,d_n}^{\lambda,n}(\mathbf{j}_n)<a_n}$. Furthermore, if we observe the sequence of outputs from the decision rule, we have that $\forall n>m=\mathcal{C}_0^{\Psi}(\mathbf{j}_{\infty}),\psi_{b_n,d_n,a_n}^{\lambda,n}(\mathbf{z}_n)=0$. As this last statement is satisfied $\forall m\in\mathbb{N}$, we get the following relationship:
\begin{equation}
    \label{eqn:proof-th4-ineq}
    \mathcal{T}_0^{\Psi}(\mathbf{z}_{\infty}) = \sup\left\{n\in\mathbb{N}:\psi_{b_n,d_b,a_n}^{\lambda,n}(\mathbf{z}_n)=1\right\} \leq \mathcal{C}_0^{\Psi}(\mathbf{j}_{\infty}).
\end{equation}
From (\ref{eqn:proof-th4-ineq}), we get the following relationship between events $\forall m\in\mathbb{N}:$
\begin{align}
    &\{\omega\in\Omega:\mathcal{C}_0^{\Psi}(\mathbf{J}_{\infty}(\omega))<m\}\nonumber\\
    \subseteq\:&\{\omega\in\Omega:\mathcal{T}_0^{\Psi}(\mathbf{Z}_{\infty}(\omega))<m\},
\end{align}
and consequently, ${\mathbb{P}(\mathcal{C}_0^{\Psi}(\mathbf{J}_{\infty})<m\,|\,\mathcal{H}_0)\leq\mathbb{P}(\mathcal{T}_0^{\Psi}(\mathbf{Z}_{\infty})<m\,|\,\mathcal{H}_0)}$, or equivalently,
\begin{equation}
    \label{eqn:proof-th4-prob-ineq}
    \mathbb{P}(\mathcal{T}_0^{\Psi}(\mathbf{Z}_{\infty})\geq m\,|\,\mathcal{H}_0)\leq\mathbb{P}(\mathcal{C}_0(\mathbf{J}_{\infty})\geq m\,|\,\mathcal{H}_0).
\end{equation}
From (\ref{eqn:proof-th4-exp-bound}) and (\ref{eqn:proof-th4-prob-ineq}), we get that for every $\Psi\in\Psi_{\text{FL}}$, it is satisfied that
\begin{equation}
    \forall m\in\mathbb{N},\mathbb{P}(\mathcal{T}_0^{\Psi}(\mathbf{Z}_{\infty})\geq m\,|\,\mathcal{H}_0) \leq K\text{exp}(-m^{1/3}),
\end{equation}
or which is equivalent, that
\begin{equation}
    \label{eqn:proof-iii-conclusion}
    \forall m\in\mathbb{N},\mathbb{P}(\mathcal{T}_0^{\Psi}(\mathbf{Z}_{\infty})<m\,|\,\mathcal{H}_0) \geq 1-K\text{exp}(-m^{1/3}),
\end{equation}
for some universal constant $K\in(0,\infty)$. Equation~(\ref{eqn:proof-iii-conclusion}) shows that $\forall\Psi\in\Psi_{\text{FL}}$, (\ref{eqn:finite-nominal-time}) is satisfied; hence, it concludes our proof for Theorem~\ref{the:finite-nominal-time}. \hfill $\square$
\end{prfththree}

\section{Proof of Theorem~\ref{the:hypothesis-test-metrics} --- Power and Significance Level}
\label{sec:proof-hypothesis-test-metrics}

\begin{prfthfour}
Let us recall that every decision scheme $\Psi\in\Psi_{\text{FL}}$ follows the construction shown in Sec.~\ref{sec:the_detector}; consequently, the parameters of the decision scheme and each decision rule can be written explicitly, i.e., $\Psi\equiv\Psi_{\textbf{b},\mathbf{d},\mathbf{a}}^{\lambda}$ and $\forall n\in\mathbb{N}$, $\psi_n(\cdot)\equiv\psi_{b_n,d_n,a_n}^{\lambda,n}(\cdot)$. Theorem~\ref{the:hypothesis-test-metrics} states two assertions expressed in (\ref{eqn:power-convergence}) and (\ref{eqn:significance-vanish}); hence, we divide this proof into two parts, one for each equation.

\begin{itemize}
    \item \textsc{Proof of Eq.~(\ref{eqn:power-convergence}):} Let us observe that ${\forall\Psi\equiv\Psi_{\mathbf{b},\mathbf{d},\mathbf{a}}^{\lambda}\in\Psi_{\text{FL}}}$, it is satisfied that $\mathbf{b}\approx(n^{-l})_{n\in\mathbb{N}}$ with $l\in(0,1/3)$, $\mathbf{a}\in o(1)$, $\lambda\in(0,\infty)$, and $\mathbf{d}\approx(\exp(n^{-1/3}))_{n\in\mathbb{N}}$. The latter asymptotic equivalence implies that $1/\mathbf{d}\approx\exp(n^{1/3})$, and consequently that $1/\mathbf{d}\in O(\exp(n^{1/3}))$, and on the other hand, it implies that $\sum_{n=1}^{\infty}d_n$ converges due to the convergence of $\sum_{n=1}^{\infty}\exp(n^{-1/3})$, which in turn, due to the non-negativity of $\mathbf{d}$, implies that $\sum_{n=1}^{\infty}|d_n|<\infty$; this last property is equivalent to $\mathbf{d}\in\ell_1(\mathbb{N})$. Consequently, $\Psi\in\Psi_{\text{FL}}\Rightarrow\Psi\in\Psi_{\text{SC}}$, i.e.,
    \begin{equation}
        \label{eqn:family-subsets}
        \Psi_{\text{FL}} \subseteq \Psi_{\text{SC}}.
    \end{equation}

    From (\ref{eqn:family-subsets}) and as we are under the \textit{standard assumptions}, we get, from Theorem~\ref{the:strong-consistency}, that any $\Psi\in\Psi_{\text{FL}}\subseteq\Psi_{\text{SC}}$ is strongly consistent for detecting faults, i.e.,
    \begin{equation}
        \mathbb{P}\left(\left.\lim_{n\rightarrow\infty}\psi_n(\mathbf{Z}_n)=i\:\right|\:\mathcal{H}_i\right) = 1,\forall i\in\{0,1\},
    \end{equation}
    which, in particular, implies that
    \begin{equation}
        \label{eqn:proof-th5-lim-power}
        \mathbb{P}\left(\left.\lim_{n\rightarrow\infty}\psi_n(\mathbf{Z}_n)=0\:\right|\:\mathcal{H}_1\right) = 0.
    \end{equation}

    Moreover, we can develop the following sequence of equalities:
    \begin{align}
        &\mathbb{P}\left(\lim_{n\rightarrow\infty}\psi_n(\mathbf{Z}_n)=0\right)\nonumber\\
        =\:&\mathbb{P}\left(\lim_{n\rightarrow\infty}\{\omega\in\Omega : \psi_n(\mathbf{Z}_n(\omega))=0\}\right)\label{eqn:proof-th5-eq-chain-i}\\
        =\:&\mathbb{P}\left(\limsup\limits_{n\rightarrow\infty}\,\{\omega\in\Omega:\psi_n(\mathbf{Z}_n(\omega))=0\}\right)\label{eqn:proof-th5-eq-chain-ii}\\
        =\:&\mathbb{P}\left(\{\omega\in\Omega:[\forall m\in\mathbb{N},\exists n\geq m:\psi_n(\mathbf{Z}_n(\omega))=0]\}\right)\label{eqn:proof-th5-eq-chain-iii}\\
        =\:&\mathbb{P}\left(\lim_{k\rightarrow\infty}\{\omega\in\Omega:[\forall m\leq k,\exists n\geq m:\psi_n(\mathbf{Z}_n(\omega))=0]\}\right)\label{eqn:proof-th5-eq-chain-iv}\\
        =\:&\mathbb{P}\left(\lim_{k\rightarrow\infty}\{\omega\in\Omega:[\exists n\geq k:\psi_n(\mathbf{Z}_n(\omega))=0]\}\right)\label{eqn:proof-th5-eq-chain-v}\\
        =\:&\lim_{k\rightarrow\infty}\mathbb{P}(\{\omega\in\Omega:[\exists n\geq k:\psi_n(\mathbf{Z}_n(\omega))=0]\}).\label{eqn:proof-th5-eq-chain}
    \end{align}
    Equation (\ref{eqn:proof-th5-eq-chain-i}) comes from the definition of the event subject to the limit [$\psi_n(\mathbf{Z}_n)=0$], and  (\ref{eqn:proof-th5-eq-chain-iii}) and (\ref{eqn:proof-th5-eq-chain-ii}) come from the definition of the limit superior (limsup) of a set sequence and the existence of its limit superior given the existence of its limit, respectively \cite{gut2006probability}. To obtain (\ref{eqn:proof-th5-eq-chain-iv}) and (\ref{eqn:proof-th5-eq-chain-v}), if we define ${\mathcal{E}_n\equiv\{\omega\in\Omega:\psi_n(\mathbf{Z}_n(\omega))=0\}}$, we can see that (\ref{eqn:proof-th5-eq-chain-iv}) and (\ref{eqn:proof-th5-eq-chain-v}) come from ${\bigcap_{m=1}^{\infty}\bigcup_{n=m}^{\infty}\mathcal{E}_n = \lim_{k\rightarrow\infty}\bigcap_{m=1}^k\bigcup_{n=m}^{\infty}\mathcal{E}_n}$ and ${\bigcap_{m=1}^k\bigcup_{n=m}^{\infty}\mathcal{E}_n = \bigcup_{n=k}^{\infty}\mathcal{E}_n}$, respectively. The last equality, (\ref{eqn:proof-th5-eq-chain}), is a consequence of the continuity of the probability measure $\mathbb{P}$.

    Let us observe that for any $k\in\mathbb{N}$ and sampling realization $\mathbf{z}_{k}\in(\mathbb{R}^{p+q})^{k}$,\footnote{For more details on the notation of random samplings and their realizations, we refer the reader to the proof of Theorem~\ref{the:finite-nominal-time} in \ref{sec:proof-finite-nominal-time}.} it is satisfied that ${\psi_k(\mathbf{z}_k)=0\Rightarrow\exists n\geq k:\psi_n(\mathbf{z}_n)=0}$; then, we get that ${\forall k\in\mathbb{N}}$,
    \begin{align}
    &\{\omega\in\Omega:\psi_k(\mathbf{Z}_k(\omega))=0\}\nonumber\\
    \subseteq\:&\{\omega\in\Omega:[\exists n\geq k:\psi_n(\mathbf{Z}_n)=0]\},
    \end{align}
    which implies that $\forall k\in\mathbb{N}$,
    \begin{align}
        &\mathbb{P}(\psi_k(\mathbf{Z}_k)=0\,|\,\mathcal{H}_1)\nonumber\\
        \leq\:&\mathbb{P}(\exists n\geq k:\psi_n(\mathbf{Z}_n)=0\,|\,\mathcal{H}_1).\label{eqn:proof-th5-final-ineq}
    \end{align}
    If we take the limit when $k$ tends to infinity, we get that
    \begin{align}
        &\lim_{k\rightarrow\infty}\mathbb{P}(\psi_k(\mathbf{Z}_k)=0\,|\,\mathcal{H}_1)\nonumber\\
        \leq\:&\lim_{k\rightarrow\infty}\mathbb{P}(\exists n\geq k:\psi_n(\mathbf{Z}_n)=0\,|\,\mathcal{H}_1)\label{eqn:limit-inequal-proof-test-i}\\
        =\:&\mathbb{P}\left(\left.\lim_{n\rightarrow\infty}\psi_n(\mathbf{Z}_n)=0\,\right|\,\mathcal{H}_1\right)\label{eqn:limit-inequal-proof-test-ii}\\
        =\:&0,\label{eqn:limit-inequal-proof-test}
    \end{align}
    where (\ref{eqn:limit-inequal-proof-test-i}) comes from taking the limit of $k$ to infinity on (\ref{eqn:proof-th5-final-ineq}), (\ref{eqn:limit-inequal-proof-test-ii}) is a consequence of conditioning (\ref{eqn:proof-th5-eq-chain}) on $\mathcal{H}_1$, and (\ref{eqn:limit-inequal-proof-test}) comes from (\ref{eqn:proof-th5-lim-power}). Moreover, from the non-negativity of the probability measure, we get that
    \begin{equation}
        \lim_{k\rightarrow\infty}\mathbb{P}(\psi_{k}(\mathbf{Z}_k)=0\,|\,\mathcal{H}_1) = 0.
    \end{equation}
    In consequence, recalling that $\beta_{\psi_n} = \mathbb{P}(\psi_n(\mathbf{Z}_n)=0\,|\,\mathcal{H}_1)$ from Def.~\ref{def:hypothesis-test-metrics}, we can see that
    \begin{align}
        &\lim_{n\rightarrow\infty}(1-\beta_{\psi_n}) = 1-\lim_{n\rightarrow\infty}\beta_{\psi_n}\nonumber\\
        =\:&1-\lim_{n\rightarrow\infty}\mathbb{P}(\psi_n(\mathbf{Z}_n)=0\,|\,\mathcal{H}_1) = 1,
    \end{align}
    which proves (\ref{eqn:power-convergence}), concluding this part of the proof.

    \item \textsc{Proof of Eq. (\ref{eqn:significance-vanish}):} Let us observe that under the \textit{standard assumptions}, it is satisfied due to Theorem~\ref{the:finite-nominal-time}, that ${\forall\Psi\in\Psi_{\text{FL}}}$,
    \begin{align}
        &\mathbb{P}(\mathcal{T}_0^{\Psi}((Z_n)_{n=1}^{\infty})<m\,|\,\mathcal{H}_0)\nonumber\\
        \geq\:&1-K\text{exp}(-m^{-1/3}),\forall m\in\mathbb{N},
    \end{align}
    for a universal constant $K>0$, and in consequence,
    \begin{equation}
        \label{eqn:proof-5-abs-inequality}
        \mathbb{P}(\mathcal{T}_0^{\Psi}((Z_n)_{n=1}^{\infty})\geq m\,|\,\mathcal{H}_0)\leq K\text{exp}(-m^{1/3}).
    \end{equation}
    In addition, let us note that for an arbitrary sample size $n\in\mathbb{N}$ and an arbitrary sampling realization ${\mathbf{z}_{\infty}\in(\mathbb{R}^{p+q})^{\infty}}$, we have that $\psi_n(\mathbf{z}_n)=1\Rightarrow\mathcal{T}_0^{\Psi}(\mathbf{z}_{\infty})\geq n$, and in consequence
    \begin{align}
        &\{\omega\in\Omega:\psi_n(\mathbf{Z}_n(\omega))=1\}\nonumber\\
        \subseteq\:&\{\omega\in\Omega:\mathcal{T}_0^{\Psi}(\mathbf{Z}_{\infty}(\omega))\geq n\},
    \end{align}
    which in turn implies that
    \begin{equation}
        \label{eqn:proof-5-def-inequality}
        \mathbb{P}(\psi_n(\mathbf{Z}_n)=1\,|\,\mathcal{H}_0) \leq \mathbb{P}(\mathcal{T}_0^{\Psi}(\mathbf{Z}_{\infty})\geq n\,|\,\mathcal{H}_0).
    \end{equation}
    Recalling $\alpha_{\psi_n}$ from Def.~\ref{def:hypothesis-test-metrics}, we can see, from (\ref{eqn:proof-5-def-inequality}) and (\ref{eqn:proof-5-abs-inequality}), that $\forall n\in\mathbb{N}$,
    \begin{align}
        &\alpha_{\psi_n} = \mathbb{P}(\psi_n(\mathbf{Z}_n)=1\,|\,\mathcal{H}_0)\nonumber\\
        \leq\:&\mathbb{P}(\mathcal{T}_0^{\Psi}(\mathbf{Z}_{\infty})\geq n\,|\,\mathcal{H}_0) \leq K\text{exp}(-n^{1/3}),
    \end{align}
    which proves (\ref{eqn:significance-vanish}), concluding this part of the proof.
\end{itemize}
As we have proven both (\ref{eqn:power-convergence}) and (\ref{eqn:significance-vanish}), we conclude our proof for Theorem~\ref{the:hypothesis-test-metrics}. \hfill $\square$
\end{prfthfour}

\section{Generality of the Non-Bias Assumptions and the Uniform Universal Noise}
\label{sec:generality}

In this appendix, we show how the assumptions (i) and (ii) of the \textit{standard assumptions} (see Sec.~\ref{sec:md-ir-dependency}) and the uniform distribution for the (\textit{universal}) noise term ($W$) in Def.~\ref{def:additive-model} do not imply a loss of generality.

\subsection{Assumption of Unbiased Noise}
\label{sec:unbias-noise}
The assumption of unbiased noise corresponds to assumption (i) of our \textit{standard assumptions}, this is that ${\mathbb{E}[\tilde{h}(W)] = \mathbf{0} \in\mathbb{R}^q}$. This does not imply a loss of generality, as if ${(X,Y)\sim\text{add}(\tilde{\eta},\tilde{h};\mu)}$ with $\mathbb{E}[\tilde{h}(W)]=\mathbf{b}\in\mathbb{R}^q$ and $\mathbf{b}\neq\mathbf{0}\in\mathbb{R}^q$, it is possible to embed the bias ($\mathbf{b}$) into $\tilde{\eta}(\cdot)$. Embedding the noise bias into $\tilde{\eta}(\cdot)$ can be done by considering the mappings $\bar{\eta}:\mathbb{R}^p\rightarrow\mathbb{R}^q$ and $\bar{h}:[0,1]\rightarrow\mathbb{R}^q$ such that $\forall x\in\mathbb{R}^p,\bar{\eta}(x)=\tilde{\eta}(x)+\mathbf{b}$ and $\forall w\in[0,1],\bar{h}(w)=\tilde{h}(w)-\mathbf{b}$, respectively. Then, we have that
\begin{align}
    &(X,Y) \overset{\text{a.s.}}{=} (X,\tilde{\eta}(X)+\tilde{h}(W))\nonumber\\
    =\:&(X,(\tilde{\eta}(X)+\mathbf{b})+(\tilde{h}(W)-\mathbf{b})) = (X,\bar{\eta}(X)+\bar{h}(W)).\label{eqn:noise-unbias-equality}
\end{align}
This implies that 
\begin{equation}
\label{eqn:noise-unbias-ii-equivalence}
    (X,Y)\overset{\text{a.s.}}{=}(X,\tilde{\eta}(X)+\tilde{h}(W)) \Leftrightarrow (X,Y)\overset{\text{a.s.}}{=}(X,\bar{\eta}(X)+\bar{h}(W)).
\end{equation}
Hence, the following equivalence holds
\begin{equation}
    \label{eqn:noise-unbias-equivalence}
    [(X,Y)\sim\text{add}(\tilde{\eta},\tilde{h};\mu)] \Leftrightarrow [(X,Y)\sim\text{add}(\bar{\eta},\bar{h};\mu)]
\end{equation}
as a consequence of (\ref{eqn:noise-unbias-ii-equivalence}) --~i.e., equivalence on point (ii) of Def.~\ref{def:additive-model}~-- and the fact that $X\sim\mu$ holds for both terms equivalent in (\ref{eqn:noise-unbias-equivalence}) --~i.e., equivalence on point (i) of Def.~\ref{def:additive-model}.\footnote{$X\sim\mu$ denotes that $\mathbb{P}(X\in A)=\mu(A),\forall A\in\mathcal{B}(\mathbb{R}^p)$.} The equivalence stated in (\ref{eqn:noise-unbias-equivalence}) lets us know that no generality is lost when assuming $\mathbb{E}[\tilde{h}(W)]=\mathbf{0}\in\mathbb{R}^q$.

\subsection{Assumption of Unbiased Estimator}
\label{sec:unbias-model}
The assumption of unbiased estimator corresponds to assumption (ii) of our \textit{standard assumptions}, this is that ${\mathbb{E}[Y-\eta(X)]=\mathbf{0}\in\mathbb{R}^q}$ for a system $(X,Y)\sim\text{add}(\tilde{\eta},\tilde{h};\mu)$. This does not imply a loss of generality, as if $\eta(\cdot)$ is a biased estimator such that $\mathbb{E}[Y-\eta(X)]=\mathbf{b}\in\mathbb{R}^q$ with $\mathbf{b}\neq\mathbf{0}\in\mathbb{R}^q$, it is possible to correct it to build an unbiased one. This correction can be made by considering a mapping $\bar{\eta}:\mathbb{R}^q\rightarrow\mathbb{R}^q$ such that $\bar{\eta}(x)=\eta(x)+\mathbf{b},\forall x\in\mathbb{R}^p$. Then, we have that ${\mathbb{E}[Y-\bar{\eta}(X)] =}$ ${\mathbb{E}[Y-\eta(X)-\mathbf{b}] = \mathbb{E}[Y-\eta(X)]-\mathbf{b} = \mathbf{b}-\mathbf{b}=\mathbf{0}}$.\footnote{Even though $\mathbf{b}$ may be unknown, it can be easily estimated from a sampling of the system: $\mathbf{Z}_n = (X_j,Y_j)_{j=1}^n$, as $\hat{\mathbf{b}} = \frac{1}{n}\sum_{j=1}^nY_j-\eta(X_j)$.} In consequence, no generality is lost when assuming $\mathbb{E}[Y-\eta(X)]=\mathbf{0}$.

\subsection{Generality of the Uniform Universal Noise}
Here, we clarify the generality of Def.~\ref{def:additive-model} in terms of using a universal noise with uniform distribution, i.e., the usage of $W\sim\text{Uniform}([0,1])$; this generality is a consequence of noise outsourcing (see Lemma~\ref{lem:noise-outsourcing}).

Let us consider a r.v. -- i.e., a system -- $(X,Y)$ such that $X\sim\mu$ and $(X,Y)\overset{\text{a.s.}}{=}(X,\eta(X)+h(\tilde{W}))$ with $\eta:\mathbb{R}^p\rightarrow\mathbb{R}^q$ and $h:\mathbb{R}^s\rightarrow\mathbb{R}^q$, where $\tilde{W}$ takes values in $(\mathbb{R}^s,\mathcal{B}(\mathbb{R}^s))$ with an arbitrary $s\in\mathbb{N}$ and an arbitrary distribution. If we consider $K$ as a constant random variable, i.e., $K:\Omega\rightarrow\mathbb{R}$, where $\forall \omega\in\Omega,K(\omega)=0$; then, for the r.v. $(K,\tilde{W})$, Lemma~\ref{lem:noise-outsourcing} ensures the existence of the mapping $f:\mathbb{R}\times[0,1]\rightarrow\mathbb{R}^s$ such that $(K,\tilde{W})\overset{\text{a.s.}}{=}(K,f(K,W))$ with $W\sim\text{Uniform}([0,1])$ independent of $K$. Moreover, as [$\forall\omega\in\Omega, K(\omega)=0$] implies that $K\overset{\text{a.s.}}{=}0$, we obtain the following sequence of implications:
\begin{align}
    &(K,\tilde{W})\overset{\text{a.s.}}{=}(K,f(K,W)) \:\wedge\: K\overset{\text{a.s.}}{=}0\nonumber\\
    \Rightarrow\:&\mathbb{P}((K,\tilde{W})=(K,f(K,W)),K=0) = 1\label{eqn:u_noise_implications-i}\\
    \Rightarrow\:&\mathbb{P}((0,\tilde{W})=(0,f(0,W))) = 1\label{eqn:u_noise_implications-ii}\\
    \Rightarrow\:&\tilde{W}\overset{\text{a.s.}}{=} f(0,W),\label{eqn:u_noise_implications}
\end{align}
where (\ref{eqn:u_noise_implications-i}) is a consequence of intersecting almost-surely events --~see (\ref{eqn:as-event-property}) in \ref{sec:general-theorem}~-- (\ref{eqn:u_noise_implications-ii}) comes from $0$ being the only possible value for the r.v. $K$, and (\ref{eqn:u_noise_implications}) is a consequence of decoupling the a.s. r.v.s in (\ref{eqn:u_noise_implications-ii}) into their two constitutive coordinates and considering the second ones.

Let us note that it is possible to define a mapping ${\bar{h}:[0,1]\rightarrow\mathbb{R}^q}$ such that $\forall w\in[0,1]$, $\bar{h}(w) = h(f(0,w))$. From (\ref{eqn:u_noise_implications}), we get that ${\bar{h}(W) = h(f(0,W)) \overset{\text{a.s.}}{=} h(\tilde{W})}$. This last relationship implies ${(X,Y) \overset{\text{a.s.}}{=} (X,\eta(X)+h(\tilde{W})) \overset{\text{a.s.}}{=} (X,\eta(X)+\bar{h}(W))}$; then, recalling Def.~\ref{def:additive-model} (and that $X\sim\mu$), we get that ${(X,Y)\sim\text{add}(\eta,\bar{h};\mu)}$.

The fact that for any system $(X,Y)$ with $X\sim\mu$ and ${(X,Y)\overset{\text{a.s.}}{=}(X,\eta(X)+h(\tilde{W}))}$ where $\tilde{W}$ follows an arbitrary distribution, there exists a function $\bar{h}(\cdot)$ such that $(X,Y)\sim\text{add}(\eta,\bar{h};\mu)$ shows the generality of Def.~\ref{def:additive-model}.

\section{Experiment Description --- Synthetic Distribution Drifts}
\label{sec:exp-desc}

This appendix extends the experiment description presented in Sec.~\ref{sec:experimental} regarding the numerical analysis performed on synthetic distributions to make our results reproducible. The core idea of these experiments is to show how our method and baselines behave when ranging a parametrized fault -- i.e., a parametrized drift -- $\delta\in\mathbb{R}^2$. For this purpose, we consider a nominal model $\eta_{\theta}(\cdot)$ and a system with unknown drift $(X^{[\delta]},Y^{[\delta]})\sim\text{add}(\eta_{\theta,\delta},h;\mu)$.\footnote{Although in Sec.~\ref{sec:experimental}, a potentially drifted system is expressed as ${(X,Y)\sim\text{add}(\eta_{\theta,\delta},h;\mu)}$, in this description, we express it as ${(X^{[\delta]},Y^{[\delta]})\sim\text{add}(\eta_{\theta,\delta},h;\mu)}$ to make explicit the effects of $\delta$ in the random variables.} Let us note that the nominal model is the MMSE estimator of $Y$ given $X$ assuming a nominal system distributes as $\text{add}(\eta_{\theta,\mathbf{0}},h;\mu)$.

Our numeric analysis consists of, first, obtaining a sampling $\mathbf{Z}^{[\delta]}_n = (X^{[\delta]}_j,Y^{[\delta]}_j)_{j=1}^n$ of $n$ samples from the 
$\delta$-drifted system $(X^{[\delta]},Y^{[\delta]})\sim\text{add}(\eta_{\theta,\delta},h;\mu)$, then, building a joint input-residual sampling as expressed in points~1~and~2 of Sec.~\ref{sec:the_detector}: ${\mathbf{J}_{n}^{[\delta]} = (X_{j}^{[\delta]},Y_{j}^{[\delta]}-\eta_{\theta}(X_{j}^{[\delta]}))_{j=1}^n}$, and finally, obtaining a quantification of the fault using our method and two baseline methods.

\subsection{Description of Synthetic Systems}
\label{sec:model-desc}
Here, we describe the six systems used in our experiments: four forward and two autoregressive systems. Before going into the details of each system, we show that all of them (independent of their drift) share the following common random variables:
\begin{align}
    U &\sim \text{Uniform}([a_\mathrm{U},b_\mathrm{U}]),\\
    S &\sim \mathcal{N}(\mu_\mathrm{S},\sigma_\mathrm{S}^2),\\
    H &\sim \mathcal{N}(\mu_\mathrm{H},\sigma_\mathrm{H}^2),\\
    W &\sim \text{Uniform}([a_\mathrm{W},b_\mathrm{W}]),
\end{align}
where all $U$, $S$, $H$, and $W$ are independent with each other. We denote the samplings of each r.v. as $(U_j)_{j=1}^n$, $(S_j)_{j=1}^n$, $(H_j)_{j=1}^n$, and $(W_j)_{j=1}^n$, respectively. In Table~\ref{tab:coefficients} (at the end of this subsection), we show the values of all constant coefficients used in our analysis, including the distribution parameters of the mentioned r.v.s.

\subsubsection{Forward Systems}

The input of these models corresponds to $X^{[\delta]} \equiv X=(U,S)$, and in consequence, a sampling of the input corresponds to $(X_j)_{j=1}^n = (U_j,S_j)_{j=1}^n$.\footnote{Note that $X$ does not depend on $\delta$; according to our notation, this can be written as $X_{j}^{[\delta]} = X_j,\forall\delta\in\mathbb{R}^2$.} The generative model of the output corresponds $\forall j\in\{1,2,\ldots,n\}$~to
\begin{equation}
    Y_{j}^{[\delta]} = \eta_{\theta,\delta}(X_j) + h(W_j).
\end{equation}
The nominal model corresponds to $\eta_{\theta}(\cdot) = \eta_{\theta,\mathbf{0}}(\cdot)$; hence, the residual is as follows $\forall j\in\{1,2,\ldots,n\}$:
\begin{equation}
\label{eqn:forward-residual}
    R_{j}^{[\delta]} \equiv Y_{j}^{[\delta]}-\eta_{\theta,\mathbf{0}}(X_j) = \eta_{\theta,\delta}(X_j) - \eta_{\theta}(X_j) + h(W_j).
\end{equation}

\paragraph{Linear System}

The linear systems are determined by a simple linear combination of the input coordinates. The expressions for $\eta_{\theta,\delta}(\cdot)$ and $h(\cdot)$ are as follows $\forall(u,s)\in\mathbb{R}^2$ and $\forall w\in\mathbb{R}$:
\begin{align}
    \eta_{\theta,\delta}(u,s) &= (c_1+\delta_1)u + (c_2+\delta_2)s,\\
    h(w) &= kw.    
\end{align}
In consequence, the (actual) output (\ref{eqn:lin-act-out}), nominal output (\ref{eqn:lin-nom-out}), and residual (\ref{eqn:lin-res}), correspond $\forall j\in\{1,2,\ldots,n\}$ to
\begin{align}
    Y_{j}^{[\delta]} &= (c_1+\delta_1)U_j + (c_2+\delta_2)S_j + kW_j,\label{eqn:lin-act-out}\\
    \eta_{\theta}(X_j) &= c_1U_j + c_2S_j,\label{eqn:lin-nom-out}\\
    R_{j}^{[\delta]} &= \delta_1U_j + \delta_2S_j + kW_j.\label{eqn:lin-res}
\end{align}

\paragraph{Polynomial System}

The polynomial systems are determined by a linear combination of non-linear transformations of the input coordinates. The expressions for $\eta_{\theta,\delta}(\cdot)$ and $h(\cdot)$ are as follows $\forall(u,s)\in\mathbb{R}^2$ and $\forall w\in\mathbb{R}$:
\begin{align}
    \eta_{\theta,\delta}(u,s) &= (c_1+\delta_1)u^2 + (c_2+\delta_2)s^3,\\
    h(w) &= kw.    
\end{align}
In consequence, the (actual) output (\ref{eqn:pol-act-out}), nominal output (\ref{eqn:pol-nom-out}), and residual (\ref{eqn:pol-res}), correspond $\forall j\in\{1,2,\ldots,n\}$ to
\begin{align}
    Y_{j}^{[\delta]} &= (c_1+\delta_1)U_j^2 + (c_2+\delta_2)S_j^3 + kW_j,\label{eqn:pol-act-out}\\
    \eta_{\theta}(X_j) &= c_1U_j^2 + c_2S_j^3,\label{eqn:pol-nom-out}\\
    R_{j}^{[\delta]} &= \delta_1U_j^2 + \delta_2S_j^3 + kW_j.\label{eqn:pol-res}
\end{align}

\paragraph{Trigonometric System}

The trigonometric systems are determined by an input-output relationship expressed by trigonometric functions. The expressions for $\eta_{\theta,\delta}(\cdot)$ and $h(\cdot)$ are as follows $\forall(u,s)\in\mathbb{R}^2$ and $\forall w\in\mathbb{R}$:
\begin{align}
    \eta_{\theta,\delta}(u,s) &= (A+\delta_1)\sin(u\cdot s + \phi + \delta_2),\\
    h(w) &= A_{\mathrm{W}}\sin(f_{\mathrm{W}}\cdot w+\phi_{\mathrm{W}}).
\end{align}
In consequence, the (actual) output (\ref{eqn:tri-act-out}) and nominal output (\ref{eqn:tri-nom-out}), correspond $\forall j\in\{1,2,\ldots,n\}$~to
\begin{align}
    Y_{j}^{[\delta]} &= (A+\delta_1)\sin(U_j\cdot S_j+\phi+\delta_2)\nonumber\\
    &\hspace{3mm}+ A_{\mathrm{W}}\sin(f_{\mathrm{W}}\cdot W_j+\phi_{\mathrm{W}}),\label{eqn:tri-act-out}\\
    \eta_{\theta}(X_j) &= A\sin(U_j\cdot S_j+\phi).\label{eqn:tri-nom-out}
\end{align}
The residual expression ($R_{j}^{[\delta]}$) cannot be simplified further than shown in (\ref{eqn:forward-residual}).

\paragraph{MLP System}

The MLP systems are built upon an MLP-based deterministic input-output relationship affected by additive noise. In this case, the MLP consists of a single hidden layer with two hidden units; the nominal parameters were chosen randomly from a fixed random seed. The expression for the function induced by the $\delta$-disturbed MLP, $\eta_{\theta,\delta}(\cdot)$, is $\forall\mathbf{x}\in\mathbb{R}^2$,
\begin{equation}
    \label{eqn:mlp-eta}
    \eta_{\theta,\delta}(\mathbf{x}) = \mathbf{w}_{\text{hidden}}^{\mathsf{T}}\cdot \sigma((\mathbf{W}_{\text{in}}+\delta_{\mathbf{W}})\cdot\mathbf{x} + \mathbf{b}_{\text{in}}) + b_{\text{hidden}},
\end{equation}
where $\sigma(\cdot)$ is a LeakyReLU activation function with positive and negative slope of $1$ and $0.01$, respectively \citep{maas2013rectifier}, and $\mathbf{W}_{\text{in}}$, $\delta_{\mathbf{W}}$, $\mathbf{b}_{\text{in}}$, and $\mathbf{w}_{\text{hidden}}$ are such that
\begin{align}
    \mathbf{W}_{\text{in}} &= \left(\begin{array}{cc}w_{11} & w_{12}\\w_{21} & w_{22}\end{array}\right),\label{eqn:mlp-param-eqn-i}\\
    \delta_{\mathbf{W}} &= \left(\begin{array}{cc}\delta_1 & \delta_2\\0&0\end{array}\right),\label{eqn:mlp-param-eqn-ii}\\
    \mathbf{b}_{\text{in}} &= \left(\begin{array}{cc}b_1&b_2\end{array}\right)^{\mathsf{T}},\label{eqn:mlp-param-eqn-iii}\\
    \mathbf{w}_{\text{hidden}} &= \left(\begin{array}{cc}w_1^{\text{h}}&w_2^{\text{h}}\end{array}\right)^{\mathsf{T}}.\label{eqn:mlp-param-eqn-iv}
\end{align}
The values of the nominal parameters shown in (\ref{eqn:mlp-eta})--(\ref{eqn:mlp-param-eqn-iv}) for the fixed random seed we used in our numeric analysis can be found (up to five decimals) in Table~\ref{tab:mlp-parameters}.

\begin{table}[ht!]
    \caption{Nominal parameters of the MLP system.}
    \label{tab:mlp-parameters}
    \begin{center}
    \begin{tabular}{crcr}\toprule
         \textbf{Parameter} & \textbf{Value} & \textbf{Parameter} & \textbf{Value} \\\cmidrule(lr){1-2}\cmidrule(lr){3-4}
         \rule{0pt}{2.5ex}$w_{11}$ & $-0.66612$ & $w_{12}$ & $-0.13874$\\
         \rule{0pt}{2.5ex}$w_{21}$ & $-0.33963$ & $w_{22}$ & $-0.18860$\\
         \rule{0pt}{2.5ex}$b_1$ & $-0.62466$ & $b_2$ & $0.28375$\\
         \rule{0pt}{2.5ex}$w_1^{\text{h}}$ & $-0.63385$ & $w_2^{\text{h}}$ & $-0.04506$\\
         \rule{0pt}{2.5ex}$b_{\text{hidden}}$ & $0.24580$ & & \\\bottomrule
    \end{tabular}
    \end{center}
\end{table}

With respect to the noise model, $h(\cdot)$ is the identity function, i.e., $\forall w\in\mathbb{R},h(w) = w$. In conclusion, the (actual) output (\ref{eqn:mlp-act-out}), nominal output (\ref{eqn:mlp-nom-out}) and residual (\ref{eqn:mlp-res}), correspond $\forall j\in\{1,2,\ldots,n\}$ to
\begin{align}
    Y_{j}^{[\delta]} &= \mathbf{w}_{\text{hidden}}^{\textsf{T}}\cdot\sigma((\mathbf{W}_{\text{in}}+\delta_{\mathbf{W}})\cdot X_j+\mathbf{b}_{\text{in}}) + b_{\text{hidden}} + W_j,\label{eqn:mlp-act-out}\\
    \eta_{\theta}(X_j) &= \mathbf{w}_{\text{hidden}}^{\mathsf{T}}\cdot\sigma(\mathbf{W}_{\text{in}}\cdot X_j+\mathbf{b}_{\text{in}}) + b_{\text{hidden}},\label{eqn:mlp-nom-out}\\
    R_{j}^{[\delta]} &= \mathbf{w}_{\text{hidden}}^{\mathsf{T}}\cdot\left[\sigma((\mathbf{W}_{\text{in}}+\delta_{\mathbf{W}})\cdot X_j+\mathbf{b}_{\text{in}})\right.\nonumber\\
    &\hspace{18mm}\left.-\sigma(\mathbf{W}_{\text{in}}\cdot X_j+\mathbf{b}_{\text{in}})\right]+W_j.\label{eqn:mlp-res}
\end{align}

\subsubsection{Autoregressive Systems}
Autoregressive (AR) systems treated in our experiments are a modification of Def.~\ref{def:additive-model} that is described by the following recursive formula $\forall j\in\{1,2,\ldots,n\}$:\footnote{We consider the special cases $D_0\equiv d_0$ (see Table~\ref{tab:coefficients}) and $W_0\equiv0$.}
\begin{align}
    D_{j}^{[\delta]} &= \eta_{\theta,\delta}(D_{j-1}^{[\delta]},U_j) + H_j,\\
    Y_{j}^{[\delta]} &= D_{j}^{[\delta]} + W_j,
\end{align}
where the r.v.s $U$, $H$, and $W$ are the exogenous input, model noise, and measurement noise, respectively. Let us note that the system output is $Y_{j}^{[\delta]}$, so the inner state of the system ($D_{j}^{[\delta]}$) is not accessible without measurement noise contamination; hence, our input for the nominal model, $\eta_{\theta}(\cdot)$, corresponds to $X_{j}^{[\delta]} = (Y_{j-1}^{[\delta]},U_j)$. This said, the nominal output corresponds to $\eta_{\theta}(X_{j}^{[\delta]})$; hence, the residual corresponds to
\begin{equation}
    \label{eqn:autoregressive-residual}
    R_{j}^{[\delta]} \equiv Y_{j}^{[\delta]}-\eta_{\theta}(X_{j}^{[\delta]}) = \eta_{\theta,\delta}(D_{j-1}^{[\delta]},U_j)-\eta_{\theta}(Y_{j-1}^{[\delta]},U_j) + H_j + W_j.
\end{equation}

\paragraph{ARX System}

The linear autoregressive system with exogenous input (ARX) is similar to the Linear system, but rather than a full exogenous input, one component of its input is a previous step of the system state. The expression of $\eta_{\theta,\delta}(\cdot)$ is as follows $\forall(d,u)\in\mathbb{R}^2$:
\begin{equation}
    \eta_{\theta,\delta}(d,u) = (c_1+\delta_1)d + (c_2+\delta_2)u.
\end{equation}
In consequence, the (actual) output (\ref{eqn:arx-act-out}), nominal output (\ref{eqn:arx-nom-out}), and residual (\ref{eqn:arx-res}) correspond $\forall j\in\{1,2,\ldots,n\}$ to
\begin{align}
    Y_{j}^{[\delta]} &= (c_1+\delta_1)D_{j-1}^{[\delta]} + (c_2+\delta_2)U_{j} + H_j + W_j,\label{eqn:arx-act-out}\\
    \eta_{\theta}(X_{j}^{[\delta]}) &= c_1(D_{j-1}^{[\delta]} + W_{j-1}) + c_2U_{j},\label{eqn:arx-nom-out}\\
    R_{j}^{[\delta]} &= \delta_1D_{j-1}^{[\delta]} + \delta_2U_{j} - c_1W_{j-1} + H_j + W_j.\label{eqn:arx-res}
\end{align}

\paragraph{NARX System}

The non-linear autoregressive system with exogenous input (NARX) is a system with a non-linear relationship between the current output and the AR input (previous output and exogenous input). The expression of $\eta_{\theta,\delta}(\cdot)$ is as follows $\forall(d,u)\in\mathbb{R}^2$:
\begin{equation}
    \eta_{\delta}(d,u) = (c_3+\delta_1+c_4\exp(-d^2))\cdot d + (c_5+\delta_2)\cdot u^2.
\end{equation}
In consequence, the (actual) output (\ref{eqn:narx-act-out}) and nominal output (\ref{eqn:narx-nom-out}), correspond $\forall j\in\{1,2,\ldots,n\}$~to
\begin{align}
    Y_{j}^{[\delta]} &= (c_3+\delta_1+c_4\exp(-(D_{j-1}^{[\delta]})^2))\cdot D_{j-1}^{[\delta]}\nonumber\\
    &\hspace{3mm}+ (c_5+\delta_2)\cdot U_j^2 + H_j + W_j,\label{eqn:narx-act-out}\\
    \eta_{\theta}(X_{j}^{[\delta]}) &= (c_3+c_4\exp(-(D_{j-1}^{[\delta]}+W_{j-1})^2))\cdot(D_{j-1}^{[\delta]}+W_{j-1})\nonumber\\
    &\hspace{3mm}+ c_5\cdot U_j^2.\label{eqn:narx-nom-out}
\end{align}
The residual expression ($R_j^{[\delta]}$) can be obtained by replacing (\ref{eqn:narx-act-out}) and (\ref{eqn:narx-nom-out}) in (\ref{eqn:autoregressive-residual}).

\begin{table}[ht!]
\caption{Constant coefficient values for the numerical analysis.}
    \begin{center}
    \begin{tabular}{crcr}\toprule
    \textbf{Coefficient} & \textbf{Value} & \textbf{Coefficient} & \textbf{Value}\\\cmidrule(lr){1-2}\cmidrule(lr){3-4}
    $a_\mathrm{U}$ & $-2.0$ & $c_3$ & $0.8$\\
    $b_\mathrm{U}$ & $2.0$ & $c_4$ & $-0.5$\\
    $\mu_\mathrm{S}$ & $0.5$ & $c_5$ & $1.0$\\
    $\sigma_\mathrm{S}$ & $2\sqrt{3}/3$ & $A$ & $1.0$\\
    $\mu_\mathrm{H}$ & $0.0$ & $\phi$ & $0.0$\\
    $\sigma_\mathrm{H}$ & $0.1$ & $d_0$ & $0.0$\\
    $a_\mathrm{W}$ & $-0.1$ & $k$ & $1.0$\\
    $b_\mathrm{W}$ & $0.1$ & $A_{\mathrm{W}}$ & 1.5\\
    $c_1$ & $0.6$ & $f_{\mathrm{W}}$ & 1.0\\
    $c_2$ & $-0.4$ & $\phi_{\mathrm{W}}$ & 0.0\\\bottomrule
    \end{tabular}
    \end{center}
\label{tab:coefficients}
\end{table}

\subsection{Fault Quantification Methods}
\label{sec:method-desc}
Here, we describe the methods we use for quantifying faults; these methods were applied to all systems, as seen in Figures~\ref{fig:experiment} and \ref{fig:ar_experiment}. One of these methods corresponds to our MI-based RIV method, and the other two are baselines we use to compare our method. To quantify faults, the RIV and Correlation methods have the input-residual sampling ($\mathbf{J}_{n}^{[\delta]}$) as their own input, while the RMSE method only considers the residual sampling ($\mathbf{R}_{n}^{[\delta]}$) as its input.

\subsubsection{RIV Method}

This method is our methodological contribution, and as detailed in Sec.~\ref{sec:md_method}, a RIV corresponds to the MI estimation between the input and the residual with the MI estimator presented in \cite{silva2012complexity}. An explanation of how this MI estimation is computed is shown in Sec.~\ref{sec:mi-estimator}. Formally, given a sampling $\mathbf{J}_{n}^{[\delta]}$, we compute $I_{b_n,d_n}^{\lambda,n}(\mathbf{J}_{n}^{[\delta]})$, where $\lambda=2.3\cdot10^{-5}$, $d_n=\exp(n^{-1/3})$, $b_n=wn^{-l}$ with $w=5\cdot10^{-2}$ and $l=0.167$, and $n=2000$.

\subsubsection{Correlation Method}

This method corresponds to the maximum absolute value of the Pearson correlation coefficient of the residual with both coordinates of the input; we denote this as
\begin{equation}
\text{MAPC}(\mathbf{J}_{n}^{[\delta]}) \equiv \max\{|\text{corr}(\mathbf{X}_{n,(1)}^{[\delta]},\mathbf{R}_{n}^{[\delta]})|,|\text{corr}(\mathbf{X}_{n,(2)}^{[\delta]},\mathbf{R}_{n}^{[\delta]})|\},
\end{equation}
where we denote $\forall j\in\{1,2,\ldots,n\}$, ${X_{j}^{[\delta]} \equiv (X_{j,(1)}^{[\delta]},X_{j,(2)}^{[\delta]})}$, $\mathbf{X}_{n,(1)}^{[\delta]} \equiv (X_{j,(1)}^{[\delta]})_{j=1}^n$, and $\mathbf{X}_{n,(2)}^{[\delta]} \equiv (X_{j,(2)}^{[\delta]})_{j=1}^n$; and $\text{corr}(\cdot,\cdot)$ is a mapping such that for two arbitrary vectors, $\mathbf{x}=(x_j)_{j=1}^n\in\mathbb{R}^n$ and ${\mathbf{r}=(r_j)_{j=1}^n\in\mathbb{R}^n}$, it provides their Pearson correlation coefficient, which is expressed by
\begin{equation}
    \text{corr}(\mathbf{x},\mathbf{r}) \equiv \frac{\sum_{j=1}^n(x_j-\bar{x})(r_j-\bar{r})}{\left[\left(\sum_{j=1}^n(x_j-\bar{x})^2\right)\cdot\left(\sum_{j=1}^n(r_j-\bar{r})^2\right)\right]^{1/2}},
\end{equation}
where $\bar{x} = \frac{1}{n}\sum_{j=1}^nx_j$ and $\bar{r} = \frac{1}{n}\sum_{j=1}^nr_j$.

\subsubsection{RMSE Method}
This method corresponds to the empirical root mean squared error computed from the residual; we denote this as $\text{RMSE}(\mathbf{R}_{n}^{[\delta]})$, which can be expressed as
\begin{equation}
    \text{RMSE}(\mathbf{R}_{n}^{[\delta]}) \equiv \sqrt{\frac{1}{n}\sum_{j=1}^n(R_{j}^{[\delta]})^2}.
\end{equation}

\subsection{Summary of Figure~\ref{fig:experiment} and Figure~\ref{fig:ar_experiment}}
With all the elements described in this section, we summarize the description of Figures~\ref{fig:experiment} and \ref{fig:ar_experiment}. These figures contain 12 and 6 colormaps, respectively. Each figure is divided into 3 rows; Figures~\ref{fig:experiment} and \ref{fig:ar_experiment} are divided into 4 and 2 columns, respectively. The rows correspond to results for each one of the methods described in \ref{sec:method-desc}, and the columns correspond to results for each one of the systems described in \ref{sec:model-desc}. The correspondence is made clear in the labels of each row and column.

Each colormap is a visualization of a matrix. Each cell of these matrices is related to a single $\delta=(\delta_1,\delta_2)$ value where both $\delta_1$ and $\delta_2$ range from $-0.15$ to $0.15$ with a step of $0.0015$; i.e., each matrix has $201\cdot201 = 40\:401$~cells. The value of the cell corresponds to the average value of applying the corresponding method to a sampling $\mathbf{J}_{n}^{[\delta]}$ or $\mathbf{R}_n^{[\delta]}$, respectively, of the corresponding $\delta$-drifted system; this average value is obtained over 10 different random seeds.

\section{Error-Bar Analysis of Model Drift Experiments on Synthetic Distributions}
\label{sec:synthetic-error-bar}

\begin{figure*}[ht!]
    \centering
    \subfloat[Forward Systems.]{\includegraphics[height=6cm]{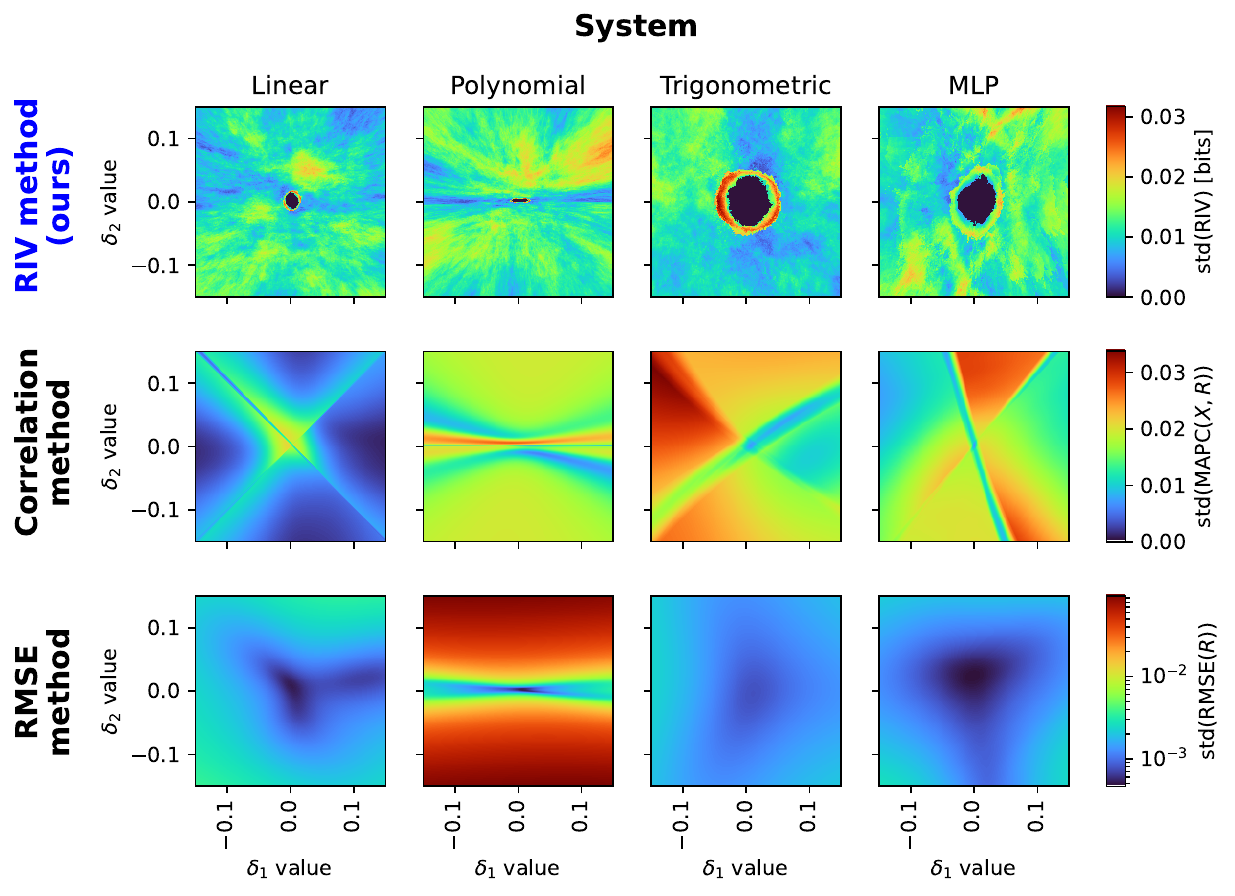}\label{sfig:error-forward}}\hspace{1cm}
    \subfloat[Autorergessive Systems.]{\includegraphics[height=6cm]{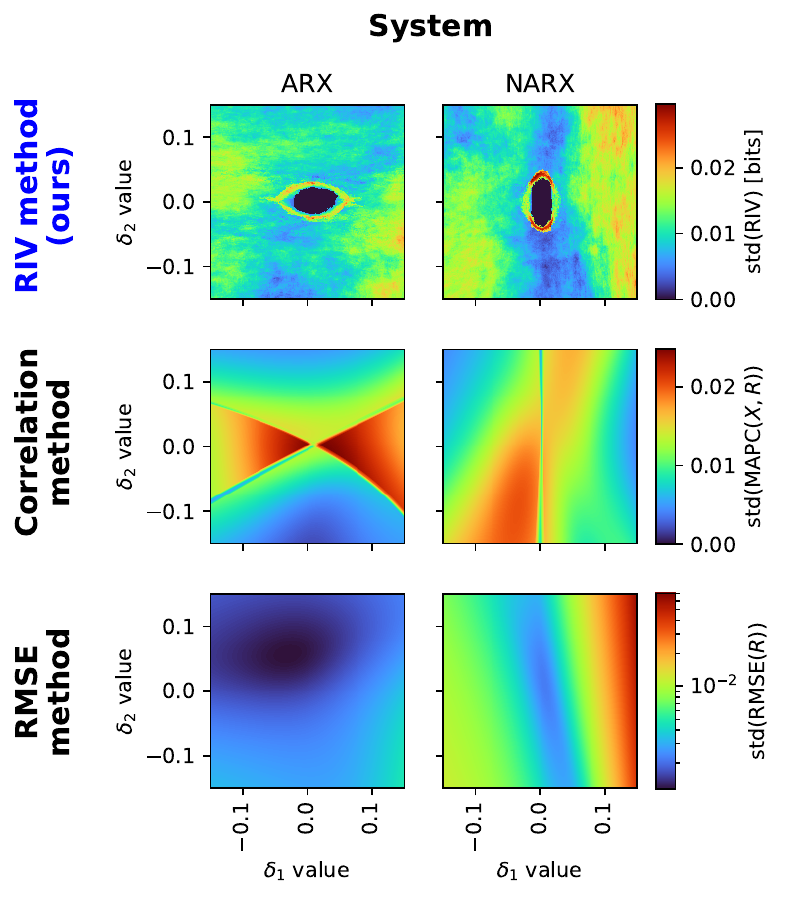}\label{sfig:error-ar}}
    \caption{Standard deviations for the numerical results of our RIV method and baselines on parametrized faults.}
    \label{fig:error}
\end{figure*}

In this section, we perform an error-bar analysis for the experiment conducted in Sec.~\ref{sec:experimental}. This error analysis is summarized in Figure~\ref{fig:error}, which contains figures analogous to Figures~\ref{fig:experiment} and \ref{fig:ar_experiment} (Figures~\ref{sfig:error-forward} and \ref{sfig:error-ar}, respectively) with the only difference of reporting the standard deviation instead of the average value.

First, we can see that all the colormaps shown in Figure~\ref{fig:error} display standard deviation values of at least one order of magnitude smaller than the average values reported in Figures~\ref{fig:experiment} and \ref{fig:ar_experiment}; this validates the consistency of our results over different seeds, and in consequence, validates our discussion and interpretation of the results. Moreover, this consistency strengthens the numerical validation of our RIV method for detecting faults.

We remark on two elements seen in Figure~\ref{fig:error} regarding our method. First, we can see regions of zero empirical standard deviation at the regions where zero average estimated MI was reported in Figures~\ref{fig:experiment} and \ref{fig:ar_experiment}; this has to do with the exponentially fast decision convergence of our method on the null hypothesis (Theorem~\ref{the:finite-nominal-time}). Second, we highlight that outside the regions with zero estimated MI, there is no clear increase or decrease of standard deviation as we go farther from $\delta=\mathbf{0}$; we understand this as a display of uniform consistency of our method regardless of the fault characteristics.

\section{Supporting Result --- Proof of Lemma~\ref{lem:degenerate-rv}}
\label{sec:lemma-proof}

\begin{prflemma}
Before starting this proof, we need to state some minor definitions. Let us consider a pair of arbitrary vectors $\mathbf{a}=(a_j)_{j=1}^r\in\mathbb{R}^r$ and $\mathbf{b}=(b_j)_{j=1}^r\in\mathbb{R}^r$ such that $\forall j\in\{1,2,\ldots,r\}$, $a_j<b_j$ for an arbitrary $r\in\mathbb{N}$, and let us define the rectangle whose main opposite vertices are $\mathbf{a}$ and $\mathbf{b}$ as $I_{\mathbf{a,b}}\equiv\times_{j=1}^r[a_j,b_j)\in\mathcal{B}(\mathbb{R}^r)$.

The idea of this proof is to build a sequence of nested rectangle-based partitions and note that we can define a sequence of sets that converges to a singleton such that each of its elements has a probability measure of 1. The value in the singleton of convergence of the sequence will correspond to the single deterministic point to which the r.v. is almost surely equal.

We start by partitioning $I_{\mathbf{a},\mathbf{b}}$ in dyadic rectangles. For this purpose, let us consider any pair $(a,b)\in\mathbb{R}^2$ which defines an interval such that $a<b$; then, we can get the set that contains both pair of values that define each dyadic sub-intervals of the original interval by using a mapping ${d_{\text{y}}:\mathbb{R}^2\rightarrow\mathcal{B}(\mathbb{R}^2)}$ such that $d_{\text{y}}(a,b)=\{(a,(a+b)/2),((a+b)/2,b)\}$;\footnote{We remark, to avoid confusion, that both $(a,(a+b)/2)$ and $((a+b)/2,b)$ in $d_{\text{y}}(a,b)$ are coordinate pairs in $\mathbb{R}^2$ and not open intervals. Moreover, it is easy to see that any interval defined by $(a,b)$ has length $b-a$, and any interval defined by $(x,y)\in d_{\text{y}}(a,b)$ has length $(b-a)/2$.} using this mapping, we can define the partition of $I_{\mathbf{a},\mathbf{b}}$ as ${P_{\mathbf{a},\mathbf{b}}\equiv\left\{\times_{j=1}^r[x_j,y_j):\forall j\in\{1,2,\ldots,r\},(x_j,y_j)\in d_{\text{y}}(a_j,b_j)\right\}}$. Let us note that $P_{\mathbf{a},\mathbf{b}}$ partitions the rectangle $I_{\mathbf{a},\mathbf{b}}$ into $2^r$ smaller rectangles with half the side of the original.

Another definition we are interested in for operating in our sequence of sets is their \textit{generalized diagonal}. If we consider $d_{\text{E}}$ as the Euclidean distance, then $(\mathbb{R}^r,d_{\text{E}})$ is a metric space, and we can define the generalized diagonal of every non-empty set $S\subseteq\mathbb{R}^r$ as $\text{diag}(S)\equiv \sup_{(x,y)\in S^2}d_{\text{E}}(x,y)$. Let us note that $\text{diag}(I_{\mathbf{a},\mathbf{b}}) = \sqrt{\sum_{j=1}^r(b_j-a_j)^2}$, and that for every $S\in P_{\mathbf{a},\mathbf{b}}$, its generalized diagonal corresponds to ${\text{diag}(S) = \sqrt{\sum_{j=1}^r((b_j-a_j)/2)^2} = \text{diag}(I_{\mathbf{a},\mathbf{b}})/2}$.

As we are dealing with a r.v. ($U$) such that ${\forall A\in\mathcal{B}(\mathbb{R}^r)}$, $\mathbb{P}(U\in A)\in\{0,1\}$; we need to note that if we have an arbitrary set $I$ and a partition $P$ of it, then\footnote{Equation~(\ref{eqn:lemma-partition}) can be proven by noting that $\mathbb{P}(U\in I) = \sum_{S\in P}\mathbb{P}(U\in S)$. Existence of $B$ can be proven easily, as else it would be satisfied that ${\forall S\in P}$, $\mathbb{P}(S\in U)=0$, which in turn, would imply that $\mathbb{P}(U\in I)=0$, and consequently, contradicting $\mathbb{P}(U\in I)=1$. Uniqueness can be proven by contradiction; if we consider $B_1\in P$ and $B_2\in P$ such that $\mathbb{P}(U\in B_1)=\mathbb{P}(U\in B_2)=1$, this would imply that $\mathbb{P}(U\in I) \geq \mathbb{P}(U\in B_1)+\mathbb{P}(U\in B_2)=2$, which is a contradiction.}
\begin{equation}
    \label{eqn:lemma-partition}
    [\mathbb{P}(U\in I)=1]\Rightarrow[\exists!B\in P:\mathbb{P}(U\in B)=1].
\end{equation}

Moving now to the core of our proof, if we denote ${\mathbf{1}\equiv (1)_{j=1}^r\in\mathbb{R}^r}$, we can see that $\{I_{\mathbf{k},\mathbf{k}+\mathbf{1}}\}_{\mathbf{k}\in\mathbb{Z}^r}$ is a partition of $\mathbb{R}^r$, and from (\ref{eqn:lemma-partition}), there exists a unique $\mathbf{m}\in\mathbb{Z}^r$ such that $\mathbb{P}(I_{\mathbf{m},\mathbf{m}+\mathbf{1}})=1$. Then, we can build a sequence of rectangles $(I_{\mathbf{a}_n,\mathbf{b}_n})_{n\in\mathbb{N}}$ such that $(\mathbf{a}_1,\mathbf{b}_1) = (\mathbf{m},\mathbf{m}+\mathbf{1})$ and $(\mathbf{a}_{n+1},\mathbf{b}_{n+1})$ corresponds to the unique element in the singleton ${\{(\mathbf{x},\mathbf{y})\in\mathbb{R}^{r+r}:[I_{\mathbf{x},\mathbf{y}}\in P_{\mathbf{a}_n,\mathbf{b}_n}\:\wedge\:\mathbb{P}(U\in I_{\mathbf{x},\mathbf{y}})=1]\}}$. This means that the $(n+1)$-th term of the sequence corresponds to the unique rectangle with a probability measure of $1$ within the ones generated by the dyadic partition of the rectangle that corresponds to the $n$-th term.

We are interested in defining a sequence of closed sets such that each element has a probability measure of $1$. The idea for this sequence is to converge to a singleton and use the continuity of the probability measure to conclude that this singleton has a probability measure of~$1$. To deal with closed sets, we will consider the closed rectangle as the closure of an (open) rectangle, i.e., $\text{cl}(I_{\mathbf{a},\mathbf{b}}) \equiv \times_{j=1}^r[a_j,b_j]\in\mathcal{B}(\mathbb{R}^r)$. Then, it is possible to define the sequence of sets $(C_n)_{n\in\mathbb{N}}$, where $\forall n\in\mathbb{N},C_n = \text{cl}(I_{\mathbf{a}_n,\mathbf{b}_n})$; this sequence satisfies the following properties $\forall n\in\mathbb{N}$:
\begin{equation}
    \label{eqn:proof-lemma-seq-prop}
    [\mathbb{P}(U\in C_n)=1] \:\wedge\: [C_{n+1}\subseteq C_n] \:\wedge\: [\text{$C_n$ is a closed set}].
\end{equation}
One additional property of $(C_n)_{n\in\mathbb{N}}$, is that $\text{diag}(C_1) = \sqrt{r}$ and $\text{diag}(C_{n+1}) = \text{diag}(C_n)/2$ as $I_{\mathbf{a}_{n+1},\mathbf{b}_{n+1}}\in P_{\mathbf{a}_n,\mathbf{b}_n}$. In consequence, $\forall n\in\mathbb{N},\text{diag}(C_n) = \sqrt{r}/2^{n-1}$, and in particular, we have that $\lim_{n\rightarrow\infty}\text{diag}(C_n) = 0$. This last limit, in addition to $(C_n)_{n\in\mathbb{N}}$ being a sequence of nested closed non-empty sets --~see (\ref{eqn:proof-lemma-seq-prop})~-- enables us to infer from Cantor's Intersection Theorem \cite[Theorem~C,~Sec.~2.12]{simmons1963introduction} that there exists $\mathbf{c}\in\mathbb{R}^r$ such that $\bigcap_{i=1}^{\infty}C_i = \{\mathbf{c}\}$. Then, we can see that
\begin{equation}
    \label{eqn:proof-lemma-sing-prob}
    \mathbb{P}(U=\mathbf{c}) = \mathbb{P}(U\in\{\mathbf{c}\}) = \mathbb{P}\left(U\in\bigcap_{k=1}^{\infty}C_k\right),
\end{equation}
where the last term, due to the continuity of the probability measure $\mathbb{P}$ and the set-theoretic definition of limit \cite{gut2006probability}, can be expressed as
\begin{equation}
    \label{eqn:proof-lemma-prob-continuity}
    \mathbb{P}\left(U\in\bigcap_{k=1}^{\infty}C_k\right) = \mathbb{P}\left(U\in\lim_{n\rightarrow\infty}\bigcap_{k=1}^nC_k\right) = \lim_{n\rightarrow\infty}\mathbb{P}\left(U\in\bigcap_{k=1}^nC_k\right).
\end{equation}
From the nested property of $(C_n)_{n\in\mathbb{N}}$ shown in (\ref{eqn:proof-lemma-seq-prop}), it is implied that $\forall n\in\mathbb{N},$
\begin{equation}
    \label{eqn:proof-lemma-nested-intersection}
    \bigcap_{k=1}^nC_k = C_n.
\end{equation}
As a last step, let us note that
\begin{align}
    &\mathbb{P}(U=\mathbf{c})\nonumber\\
    =\:&\mathbb{P}\left(U\in\bigcap_{k=1}^{\infty}C_k\right)\label{eqn:proof-lemma-last-eq-i}\\
    =\:&\lim_{n\rightarrow\infty}\mathbb{P}\left(U\in\bigcap_{k=1}^nC_k\right)\label{eqn:proof-lemma-last-eq-ii}\\
    =\:&\lim_{n\rightarrow\infty}\mathbb{P}(U\in C_n)\label{eqn:proof-lemma-last-eq-iii}\\
    =\:&\lim_{n\rightarrow\infty}1 = 1,\label{eqn:proof-lemma-last-eq}
\end{align}
where (\ref{eqn:proof-lemma-last-eq-i}), (\ref{eqn:proof-lemma-last-eq-ii}), (\ref{eqn:proof-lemma-last-eq-iii}), and (\ref{eqn:proof-lemma-last-eq}) come from (\ref{eqn:proof-lemma-sing-prob}), (\ref{eqn:proof-lemma-prob-continuity}), (\ref{eqn:proof-lemma-nested-intersection}) and the first property shown in (\ref{eqn:proof-lemma-seq-prop}), respectively.

At this point, we have shown the existence of $\mathbf{c}\in\mathbb{R}^r$ such that $\mathbb{P}(U=\mathbf{c})=1$, which was precisely what is stated in Lemma~\ref{lem:degenerate-rv}. Hence, concluding our proof. \hfill $\square$
\end{prflemma}

\section*{Acknowledgements}
This material is based on work supported by grants of CONICYT-Chile, Fondecyt 1210315, and the Advanced Center for Electrical and Electronic Engineering, Basal Project AFB240002. Camilo Ramírez is supported by ANID-Subdirección de Capital Humano/Magíster-Nacional/2023 - 22230232 master’s scholarship. The work of Marcos Orchard is supported by grants of CONICYT-Chile , Fondecyt 1210031. We thank Diane Greenstein and Sebastián Espinosa for editing and proofreading all this material.

\bibliographystyle{elsarticle-num}
\bibliography{references}

\end{document}